\documentclass[manuscript,useAMS,usenatbib]{mnras}

\usepackage{times,epsfig,rotating}
\usepackage{lscape}
\usepackage{hyperref}
\usepackage{float}

\usepackage[T1]{fontenc}
\usepackage{ae,aecompl}

\usepackage{graphicx}	% Including figure files
\usepackage{amsmath}	% Advanced maths commands
\usepackage{amssymb}	% Extra maths symbols

\usepackage{color}

\def\mnras{MNRAS}
\def\aj{AJ}
\def\aap{A\&A}
\def\apj{ApJ}
\def\apjl{ApJ}
\def\apjs{ApJS}
\def\araa{ARA\&A}

\def\nat{Nature}

\def\aaps{A\&AS}

\newcommand{\angstrom}{\mbox{\normalfont\AA}}
\newcommand{\z}{\mbox{$z$}}
\newcommand{\msun}{\mbox{$\mathrm{M_{\sun}}$}}

\newcommand{\sunrise}{\mbox{\sc sunrise}}

\newcommand{\hyperion}{\mbox{\sc hyperion}}
\newcommand{\pd}{\mbox{\sc powderday}}

\newcommand{\ART}{\mbox{\sc ART}}

\newcommand{\gadget}{\mbox{\sc gadget-3}}

\newcommand{\gizmo}{\mbox{\sc gizmo}}
\newcommand{\gasoline}{\mbox{\sc gasoline}}

\newcommand{\fsps}{\mbox{\sc fsps}}

\newcommand{\numpy}{\mbox{\sc numpy}}
\newcommand{\scipy}{\mbox{\sc scipy}}
\newcommand{\matplotlib}{\mbox{\sc matplotlib}}
\newcommand{\cython}{\mbox{\sc cython}}

\newcommand{\caesar}{\mbox{\sc caesar}}

\newcommand{\yt}{\mbox{\sc yt}}
\newcommand{\sextractor}{\mbox{\sc sextractor}}
\newcommand{\sep}{\mbox{\sc sep}}
\newcommand{\astrolib}{\mbox{\sc astrolib}}
\newcommand{\mufasa}{\mbox{\sc mufasa}}

\newcommand{\hst}{\textit{HST}}
\newcommand{\avgSNR}{${\left<SNR\right>}$}

\newcommand{\art}{\mbox{\sc art}}
\newcommand{\music}{\mbox{\sc music}}

\newcommand{\smass}{$M_\ast$}

\title[Quantitative Morphology Measures in Galaxy Simulations]{Identifying Mergers Using Quantitative Morphologies in Zoom Simulations of High-Redshift Galaxies}

\author[M.W. Abruzzo et al.]{Matthew W. Abruzzo$^{1,5}$\thanks{E-mail:
    mwa2113@columbia.edu}
Desika Narayanan$^{2,3,4,5}$,
Romeel Dav\'{e}$^{6,7,8}$
\newauthor and Robert Thompson$^{9}$\\
$^{1}$Columbia University, Department of Astronomy, New York, NY 10025, USA\\
$^{2}$Department of Astronomy, University of Florida, 211 Bryant Space Sciences Center, Gainesville, FL 32611\\
$^{3}$University of Florida Informatics Institute, 432 Newell Drive, CISE Bldg E251, Gainesville, FL 32611\\
$^{4}$Cosmic Dawn Centre (DAWN), University of Copenhagen, Julian Maries vej 30, DK-2100, Copenhagen, Denmark\\
$^{5}$Department of Physics and Astronomy, Haverford College, 370 Lancaster Ave, Haverford, PA 19041\\
$^{6}$Institute for Astronomy, University of Edinburgh, Royal Observatory, Edinburgh EH9 3HJ, UK\\
$^{7}$University of the Western Cape, Bellville, Cape Town 7925, South Africa\\
$^{8}$South African Astronomical Observatory, Cape Town 7925, South Africa\\
$^{9}$Portalarium, 3410 Far West Blvd, Austin, TX 78731}

\date{Submitted to MNRAS}

\begin{document}
\pagerange{\pageref{firstpage}--\pageref{lastpage}} \pubyear{2018}

\maketitle

\label{firstpage}

\begin{abstract}
Non-parametric morphology measures are a powerful tool for identifying galaxy mergers at low redshifts. We employ cosmological zoom simulations using {\sc Gizmo} with the {\sc Mufasa} feedback scheme, post-processed using 3D dust radiative transfer into mock observations, to study whether common morphological measures Gini $G$, $M_{20}$, concentration $C$, and asymmetry $A$ are effective at identifying major galaxy mergers at $z\sim 2-4$, i.e. ``Cosmic Noon''.  Our zoom suite covers galaxies with $10^{8.6} \lesssim M_*\lesssim 10^{11}M_\odot$ at $z\approx 2$, and broadly reproduces key global galaxy observations.  Our primary result is that these morphological measures are unable to robustly pick out galaxies currently undergoing mergers during Cosmic Noon, typically performing no better than a random guess.  This improves only marginally if we consider whether galaxies have undergone a merger within the last Gyr.%, or employ a higher $A$ threshold for merger classification.
When also considering minor mergers, galaxies display no trend of moving towards the merger regime with increasing merger ratio.  From $z=4\rightarrow 2$, galaxies move from the non-merger towards the merger regime in all statistics, but this is primarily an effect of mass: Above a given noise level, higher mass %and thus more luminous 
galaxies display a more complex outer morphology induced by their clustered environment.  We conclude that during Cosmic Noon, these %non-parametric 
morphological statistics are of limited value in identifying galaxy mergers.
\end{abstract}

\section{Introduction}

A major outstanding question in modern day astronomy is how galaxies
form. This question is intimately tied to galaxy morphology. Early on, it was noted that
galaxies could be classified by their visual morphology. The \citet{Hubble1926} morphological classification system remains a useful characterisation of galaxies to this day mainly because its morphological classifications are strongly correlated with physical properties: spiral galaxies typically have ongoing star formation and are 
dominated by the light of bright, bluer, younger stars, whereas 
elliptical galaxies are usually quiescent and dominated by the light of  older, redder, stars \citep{conselice14a}. Moreover, within the $\Lambda$ Cold Dark Matter ($\Lambda$CDM) paradigm, dark matter haloes grow hierarchically via mergers \citep{white78a,lacey93a,guo08a}, and semi-analytic models of galaxy formation successfully tie the resulting merging process to the morphological evolution of galaxies

Theoretical simulations have demonstrated that
major galaxy mergers, canonically defined as having a progenitor mass ratio above 1:4, can have a dramatic effect on the luminosity, structural, and chemical histories of galaxies.  Tidal torques from interacting systems can drive gaseous inflows that subsequently fuel intense nuclear starbursts \citep[e.g.][though see \citet{teyssier10a}]{Barnes1992,Mihos1996,Springel2005b,narayanan10b,hopkins13a,hayward13c}.
This star formation activity can be followed by a period of intense
black hole growth \citep[e.g.][]{Springel2005b,hopkins05a,younger09a,gabor16a},
and eventually result in a "red and dead" elliptical galaxy
\citep[e.g.][]{Springel2005a,cox08a}.  These processes can have a dramatic impact on the structural and thermal properties of the
interstellar medium in galaxies \citep[e.g.][]{narayanan11a,narayanan12a}, the kinematic structure of galaxies \citep[e.g.][]{cox06b,wuyts10a}, and the formation of stellar bulges \citep{hopkins09a}.

In the local Universe, it is clear that the most bolometrically luminous galaxies are
principally comprised of mergers.  For example, analysis of the morphological structure of infrared-bright galaxies in the local
Universe evidenced that the majority of systems forming stars above $\sim 50 \ M_\odot {\rm yr}^{-1}$ (or, with infrared luminosities
greater than $\sim 10^{11.5} \ L_\odot {\rm yr}^{-1}$) owed their origin to major mergers \citep{sanders96a,veilleux02a,casey14a,larson16a}.  Indeed, surveys of a large sample of relatively local ($z<0.1)$ galaxies show
a trend with increasing Luminous Infrared Galaxy ($L_{\rm IR} > 10^{11} M_\odot$) fraction with decreasing pair separation in galaxy mergers \citep{ellison13a}.

What is less clear, however, is whether a similar situation holds at higher redshifts.  At redshifts $z \la 4$, at a fixed stellar mass, galaxy star formation rates intrinsically increase
\citep[e.g.][]{rodighiero11a,elbaz11a,whitaker12a,speagle14a} owing to the strong redshift dependence of the cosmological accretion rate~\citep[e.g.][]{dekel09}.  Of these galaxies, dusty infrared-luminous (and often submillimetre-selected) galaxies appear to play a substantial role in
contributing to the cosmic star formation rate density (SFRD) \citep{lefloch05a,dunlop17a,michalowski17a,smith17a,koprowski17a}, contributing at
least $\sim 40\%$ of the SFRD through $z\sim 4$.  Given the strong correlation between infrared luminosity in galaxies in the local
Universe and galaxy mergers, a natural extrapolation to the high-redshift Universe would suggest a strong impact of galaxy mergers
on the cosmic star formation rate density.  At the same time, the role of mergers in driving the luminosity of heavily star-forming galaxies
at high-redshift is under vigorous debate in the community \citep[e.g.][]{casey09a,dekel08a,dave10a,engel10a,hayward11a,hayward13a,narayanan10a,narayanan10b,narayanan15a,tacconi08a}. Identifying mergers, therefore, is of significant value in understanding their relative role in the growth and evolution of
galaxies over cosmic time.  Redshifts $z\sim2-4$ in particular, so-called "Cosmic Noon", represent an important phase in cosmological
galaxy formation where the black hole accretion rate and star formation rate density both peak \citep[see][for recent reviews]{Shapley2011,Madau2014}.

%Galaxies grow their baryonic components via three major processes: (1)
%in situ star formation; (2) mergers with other galaxies (major or
%minor); and (3) the accretion of gas from the intergalactic medium.
%Quantifying the relative significance of these as a function of galaxy
%mass and redshift remains an outstanding question in galaxy formation
%theory and observations \citep{Conselice2014}.

%There are two main processes through which galaxies are assembled: (i)
%major galaxy mergers, where two galaxies of similar mass merge to
%form a larger galaxy, and (ii) the accretion of cold gas and dark
%matter from the environment. Major galaxy mergers are typically
%associated with bursts of star formation
%\citep{Mihos1996,Narayanan2010} followed by the remnants of the
%mergers taking on the appearance of red spheroidal galaxies
%\citep{cox08a}. Smooth accretion of cold gas has been shown to
%correlate with the growth of galactic disks \citep{Dekel2013}.  In
%reality, galaxies are assembled by some complex combination of the
%processes. However the relative significance of each process in %galaxy
%assembly is not well known over all parts of cosmic history.

There are two primary methods for identifying galaxy mergers: (1) identifying close pairs as galaxies {\it yet} to merge
\citep[e.g.][]{barton00a,lin08a}, and (2) utilising irregular or disturbed morphologies as an identifier of an {\it ongoing} or
recently elasped merger \citep[for a recent review see][]{conselice14a}.  

Within the latter category, there are two broad methods for identifying mergers.  The most common approach in using galaxy morphologies to identify ongoing mergers involves visual inspection.  In an era of deep HST surveys alongside massive citizen science campaigns, visual inspection has shown great utility
in understanding merger fractions through $z \approx 2$ \citep{lintott11a,kocevski2012,Kartaltepe2015}   The second major method, developed principally over the last two decades, involves non-parametric quantitative morphological measures.  These have the advantage that they are less subjective and do not require a priori assumptions about morphological characteristic of mergers, but must still be calibrated via visually-identified samples typically at low-$z$.  This paper focuses on studying whether such non-parametric measures are successful at identifying mergers at higher redshifts.

The predominant quantitative galaxy morphology measures utilise the galaxy's concentration ($C$), asymmetry ($A$) and clumpiness ($S$)
\citep[formally known as the CAS system][]{conselice03a}, as well as Gini and $M_{20}$ \citep[][]{lotz04a} (to be described quantitatively shortly).  More recently, some authors have additionally begun to
explore multimode (M), intensity (I) and deviation(D) ($MID$) statistics \citep{Freeman2013}.  Regions of the parameter space of each set of measures, which were empirically identified using observations of local galaxies, classify morphologies as ``normal'' or ``disturbed'' \citep[e.g][]{lotz08b, Freeman2013}, where those with
``disturbed'' morphologies are often considered galaxy mergers.  The CAS system tends to identify mostly major mergers while Gini and
$M_{20}$ often identify both major and minor mergers \citep{conselice14a}. These morphological measures are powerful tools;
they have been shown to effectively identify local idealised galaxy mergers \citep{lotz08b,lotz08a}, are free of human bias, and scale to large galaxy surveys \citep[e.g.][]{Cassata2005,Grogin2011}.

%In light of their utility, additional work has been done to try to
%better understand the behavior of these measures. These works have
%used 2 primary tools: surveys and numerical simulations.  Since their
%conception, these quantitative measures have been employed in numerous
%surveys \citep[e.g.][]{Cassata2005,Grogin2011}. By comparing the
%morphological measures to other observed quantities, our understanding
%of their behavior has been improved. 

Calibrations of quantitative morphological techniques have been done against low-redshift galaxies, where visual classification of
mergers is relatively straight forward.  However, at high-redshift, the relative lack of spatial resolution as compared to present-epoch
galaxies complicates calibrations.  Moreover, galaxies at $\z \ga 2$ are generally less organized than their lower-redshift counterparts, with rather complex distributions of gas, dust, and young stars \citep[e.g.][]{finlator06a,dave10a,ivison13a,geach16a,rujopakarn16a,koprowski16a}.  It is at present unclear how non-parametric morphological indicators perform in this regime of complex environments in high-$z$ galaxies.

In this regard, numerical simulations of galaxies in evolution provide a promising way forward.  By coupling realistic simulations of galaxy evolution with a methodology for creating mock observables, one can calibrate observational techniques against known quantities, and in effect 'ground-truth' non-parametric quantitative galaxy morphology
indicators.

%The other method of studying the
%behavior of these measures uses of numerical galaxy formation
%simulations. The measures are computed from mock observations of
%simulated galaxies, which are comparable to how an observer would
%observe a real galaxy, and their behavior is directly compared to the
%physical properties and detailed history of the galaxy. The advantage
%of using simulations over surveys is that simulations allow for the
%direct comparison of the morphological parameters against galaxies'
%physical properties, rather than the comparison to properties derived
%from their appearance.

The method of calibrating quantitative morphology measures via numerical simulations of galaxy evolution was pioneered by
\citet{lotz08b}, who studied the behaviour of morphological measures in idealised simulations of gas-rich major galaxy mergers.  By
coupling \gadget \ simulations of galaxy mergers with \sunrise \ 3D dust radiative transfer, these authors focused on understanding the dependence of the morphological measures on the observer's viewing angle, the total mass of the merging galaxies, gas properties, supernova feedback, and the initial orbit of the merging
galaxies. Subsequently, the same group used similar methodologies to study how the morphology measures are affected by the mass ratio between merging galaxies \citep{lotz10a} and the gas fractions of the merging galaxies \citep{lotz10b}.

In recent years, the methodology has evolved to utilise bona fide cosmological hydrodynamic simulations of galaxies in evolution.  While
computationally demanding, cosmological simulations offer the notable advantage of modelling the full cosmic environment of galaxies, and therefore may be advantageous over idealised simulations in studying the characteristically complex environments of high-redshift galaxies.

For example, \citet{torrey15a}, \citet{snyder15b} and \citet{bignone17a} utilised the large-box Illustris cosmological
simulation to develop mock catalogs and study galaxy morphologies, with the latter authors focusing specifically on non-parametric
morphological indicators.  Owing both to particle mass resolution, as well as the computational expense, these authors were unable to employ dust radiative transfer, and instead utilised attenuation calculations in order to generate their mock images.  More
recent work, therefore, has focused on the cosmological zoom technique in order to better resolve galaxy morphologies at high-redshift; this technique encodes the attractive aspects of both large-scale cosmological simulations and idealised galaxy evolution simulations at
the expense of significant computational cost \citep[see][for a summary]{somerville14a}.  For example, \citet{snyder15a} and \citet{thompson15b} employed \ART\ and \gadget \ zoomed in simulations (respectively) of galaxies at $z \ga 1$ to understand quantitative morphological indicators via mock observations.  The results have been less conclusive than in the idealised case, with the efficacy of non-parametric statistics in quantifying morphologies and identifying mergers being less clear.

In this paper, we present a systematic study of the non-parametric morphological indicators Gini-$M_{20}$ and CAS of galaxies at redshifts $z \sim 2-4$.  To do this, we employ high-resolution cosmological zoom galaxy formation simulations that span 2.5 decades in mass, bracketing the mass range that encapsulates proto-Milky Way galaxies through massive
submillimetre-luminous systems.  These simulations are run with the same star formation feedback modules as the {\sc Mufasa} cosmological simulations, which have shown to be successful at reproducing a broad range of global galaxy observables~\citep{Dave2016}.  We couple these simulations to 3D dust radiative transfer simulations in order to model the emergent optical morphologies, and employ image analysis techniques analogous to what is done in observations.  The combination of high-resolution zooms using a state-of-the-art galaxy formation model, sophisticated 3D dust radiative transfer, and a careful accounting of instrumental effects makes this study a step forward with respect to  previous efforts.

In \S~\ref{section:zooms} we present our galaxy formation simulations.  In
\S~\ref{section:mock_observations} we detail our conversion of these
simulations to observations, discussing both our radiative transfer
calculations, as well as our image analysis techniques.  We present
our results in \S~\ref{section:results}, provide discussion in
\S~\ref{section:discussion}, and conclude in
\S~\ref{section:conclusions}.

\section{Cosmological Zoom Simulations}
\label{section:zooms}

\begin{table*}
	\centering
	\caption{Descriptions of the simulated Galaxies. }
	\label{tab:galaxies}
	\begin{tabular}{lccccccc}
		\hline
		Name & Marker Color & Marker Shape & $M_{\rm *,central}^a$ & $M_{\rm *,halo}^a$ &$M_{\rm DM}^a$ & ${\rm sSFR}_{\rm halo}^b$ & $z_{\rm final}$\\
		\hline
		mz0   & green   & thin diamond  & $8.4 \times 10^{10}$ & $4.1 \times 10^{11}$&$4.1 \times 10^{13}$ & 0.0042 & 2.15\\
		mz5   & blue    & circle   & $6.9 \times 10^{10}$ & $8.3 \times 10^{11}$&$6.3 \times 10^{13}$ & 0.16 & 2\\ 
		mz10  & magenta & square   & $6.8 \times 10^{10}$ & $1.6 \times 10^{11}$&$1.1 \times 10^{13}$ & 0.077 & 2\\
		mz45  & red     & down triangle & $1.3 \times 10^{10}$ &$1.3 \times 10^{11}$& $3.7 \times 10^{13}$ & 0.62 & 2\\
      	z0mz287 & brown & hexagon & $1.0 \times 10^{9}$ & $2.3 \times 10^{9}$&$2.9 \times 10^{11}$ & 0.26 & 2\\
        z0mz352 & grey & star & $2.7 \times 10^9$ &$7.2 \times 10^{9}$& $9.2 \times 10^{11}$ & 0.84 & 2\\
        z0mz374 & pink & diamond & $1.0 \times 10^{8}$& $2.9 \times 10^{8}$& $1.8 \times 10^{11}$ & 1.4 & 2\\
        z0mz401 & yellow  & pentagon & $2.4 \times 10^9$ &$3.8 \times 10^{9}$& $5.8 \times 10^{11}$ & 1.2 & 2\\
        z0mz1500 & cyan & left triangle & $4.5 \times 10^{8}$& $7.6 \times 10^{8}$& $1.7 \times 10^{11}$ & 1.5 & 2\\
		\hline
	\end{tabular}
    \\$^a$ Masses given in ${\rm M}_\odot$ at $z=z_{\rm final}$.
    \\$^b$ 50 Myr averaged Specific Star Formation Rate given in ${\rm Gyr}^{-1}$ at $z=z_{\rm final}$.
\end{table*}

\subsection{Simulation Details}

To run our galaxy formation simulations, we use a modified version of
the hydrodynamic code \gizmo\ \citep{hopkins15a}, which draws heavily
from the framework of \gadget\ \citep{Springel2005}. With a cosmology
$\Omega_\Lambda =0.7$, $\Omega_b =0.048$, $H_0=68$ km
$\mathrm{s^{-1}}$ $\mathrm{Mpc^{-1}}$ and $\sigma_8=0.82$ we generate
initial conditions for a $50h^{-1}$ Mpc box at $z=249$ using
\music\ \citep{Hahn2011}. We run our initial dark matter only
simulation, which includes $512^3$ particles with a dark matter mass
resolution of $7.8\times 10^8h^{-1}$ \msun , down to $z=0$. At $z_{\rm{sim}}$,
we use \caesar
\footnote{\url{http://caesar.readthedocs.io/en/latest/}}
\citep{thompson15a} to identify halos to re-simulate at a higher
resolution.  Specifically, we select nine halos to re-simulate at
higher resolution.  Four of these are selected at $z_{\rm sim}=2$, while the
latter five are selected at $z_{\rm sim}=0$ (though only analysed down to $z=2$
for the purposes of this paper).

%$z_{\rm{sim}}=2$ with halo masses of $M_{DM}\ga 10^{12}$ \msun\. Likewise, we also
%re-simulate five halos selected at $z_{\rm{sim}}=0$, though for the purposes of this paper only analyse thee.  The number at the
%end of each simulation's name refers to the mass of the halo at
%$z_{\rm{sim}}$ being re-simulated relative to all other halos in the
%volume (e.g. mz0 re-simulates the largest halo and mz5 re-simulates
%the sixth largest halo at $z=2$). From this point on, we drop the
%``z0'' prefix when referring to the names of the simulations
%re-simulated from halos chosen at $z_{\rm{sim}}=0$ (e.g. we refer to
%z0mz287 as mz287).

Following the procedure outlined in \citet{Hahn2011}, we set the
Lagrangian high-resolution region to be re-simulated for each halo to
be the region enclosed by the distance to the farthest dark matter
particle included in the dark matter halo at $z_{\rm sim}$ multiplied
by a factor of 2.5.
\begin{figure}
% Make sure the figure is centered:
\centering
\includegraphics[width = 3.15 in]{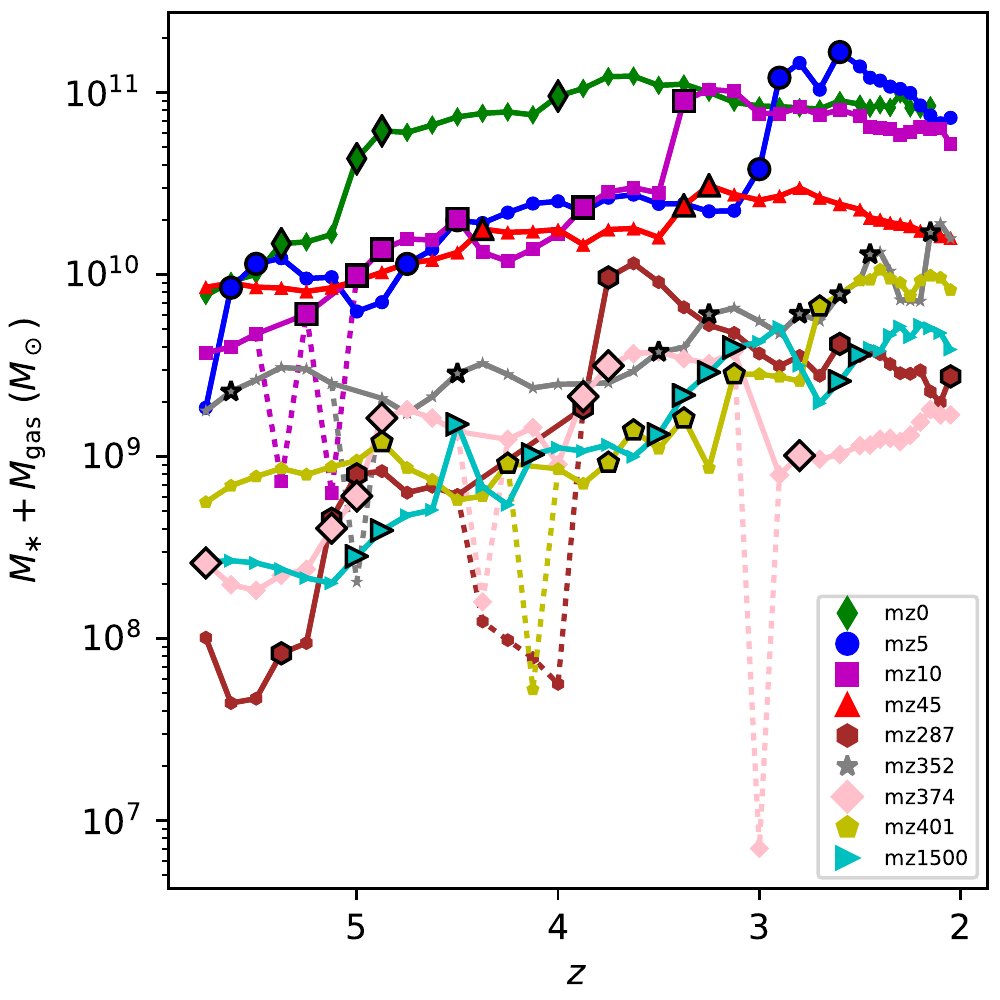}
% Give the caption for the Figure here. 
\caption{\label{figure:mass_evo} Baryonic mass evolution of the
  central galaxy from each simulation as a function of
  redshift. Dashed lines join the baryonic masses of the central 
  galaxy in snapshots that have been omitted from our analysis 
  due to the unphysical baryonic mass drops from previous 
  snapshots. Larger markers with a black outline indicate a
  baryonic mass increase of at least 25\%, from the most recent 
  snapshot connected by a solid line, which corresponds to a major 
  merger. The physical properties of these galaxies are detailed in 
  Table~\ref{tab:galaxies}.}
\end{figure}
We run these zoomed galaxy formation simulations (with baryons) with
the same sub-resolution physics employed in the \mufasa\ cosmological hydrodynamic
simulations \citep{Dave2016,Dave2017}. Like in the
\mufasa\ simulations, we run \gizmo\ using its meshless finite mass
(MFM) method, which evolves fluid in a way that conserves mass within
fluid elements, with a cubic spline kernel with 64 neighbours
\citep{Dave2016}. The initial conditions of the \mufasa\ simulation
are the same as those used in our original dark matter simulation.
 
In the simulations, decoupled two-phase winds are used to model
feedback from young stars. The probability of ejection for these winds
is given by some fraction, $\eta$, of the star formation rate
probability. The value of $\eta$, which is a best-fit relationship of
the mass outflow rate in the Feedback in Realistic Environments (FIRE)
simulations \citep{Muratov2015}, is given by
\begin{equation}
\label{mass_outflow}
\eta = 3.55\left(\frac{M_*}{10^{10} \mathrm{M_\odot}}\right)^{-0.351}
\end{equation}
where the galaxy stellar mass $M_*$ comes from an on-the-fly friends
of friends finder \citep{Dave2016}. The velocity of the ejections
depends on the galaxy circular velocity using the 
\citet{Muratov2015} relationship, with an increased amplitude as discussed in \citet{Dave2016}. The wind fluid elements remain decoupled until
its velocity, relative to surrounding gas, drops below half of the
local sound speed.  Alternatively, the wind fluid elements will also
recouple if either the wind is ejected into gas below a threshold
density of 1\% of the critical density required for star formation, or
if the wind has been decoupled for more than 2\% of the Hubble time
when it was ejected. See \citet{Dave2016} for a more in-depth
discussion of the stellar winds.

Star formation occurs within
molecular gas, where the $\mathrm{H_2}$ abundance is given by the
prescription detailed in \citet{Krumholz2009}, assuming a minimum metallicity of $10^{-3}$
$\mathrm{Z_\odot}$. We also track the evolution of 11 elements, which
consist of H, He, C, N, O, Ne, Mg, Si, S, Ca, and Fe.  Specifically,
we consider the feedback mechanisms from Type Ia supernovae, Type II
supernovae, and Asymptotic Giant Branch stars.  See \citet{Dave2016}
for specifics about the yields. The only deviation between the physics
in our simulations and that of the \mufasa\ simulations is that we do
not include the on-the-fly, heuristic quenching mechanism \citep{Dave2016}.

We run each of the cosmological zoom simulations, with dark matter
particle masses of $M_{DM}=1\times 10^6h^{-1}$ \msun\ and baryon
particle masses of $M_b =1.9\times 10^5h^{-1}$ \msun , down to
$z\approx2$\footnote{The simulations selected at $z_{\rm sim}=0$ were
  in practice typically run past $z=2$, though these later time
  snapshots are not included in this paper, whose focus is galaxies
  during cosmic noon.}. While running the simulations, we used
adaptive gravitational softening \citep{hopkins15a}; for dark matter,
gas, and stars the minimum softening lengths are 280 pc, 7 pc and 2.8
pc, respectively.  For each of the simulations, we record 85 snapshots
spanning $z\sim30-2$ (except mz0 for which we record 82 snapshots
spanning $z\approx30-2.15$), though focus our efforts here on the
redshift range $z=2-4$.

\subsection{Galaxy Identification}
To identify galaxies and halos in each run, we employ \caesar , which uses a Friends of Friends algorithm with a linking length of 0.2 times the mean inter-particle distance to identify halos and constructs a merger history by linking halos to each other (galaxies are identified using a linking length of $0.2 \times$ the halo linking length). 
We make use of \caesar 's ability to track galaxies between snapshots and calculate intrinsic properties
of the galaxies such as their centres of mass, baryonic masses, etc. For the purposes of our analysis, we identify the central galaxy in a given simulation as the most massive galaxy in the high-resolution halo at the lowest $z$ where it has been simulated.\footnote{Note that there are two galaxies in the high-resolution halo of mz45 of comparable mass. For this analysis, we have identified the central galaxy with slightly higher stellar mass, and slightly smaller baryonic mass as the central galaxy.}
See Table \ref{tab:galaxies} for a list of the central galaxies that we follow from each simulation. Figure
\ref{figure:mass_evo}\footnote{The sizeable dips in baryonic mass shown by dashed lines in Figure~\ref{figure:mass_evo} owe to a merger temporarily unbinding the majority in the system. Because, during these time frames, the bound systems would be undetectable, we omit these snapshots from our analysis.} shows the evolution of the baryonic mass associated with the central galaxy for each
simulation over cosmic history, and indicates points in time when we identify a major merger has occurred.

\begin{figure}
\begin{tabular}{cc}
\hspace{-0.25cm}
\includegraphics[width=\columnwidth]{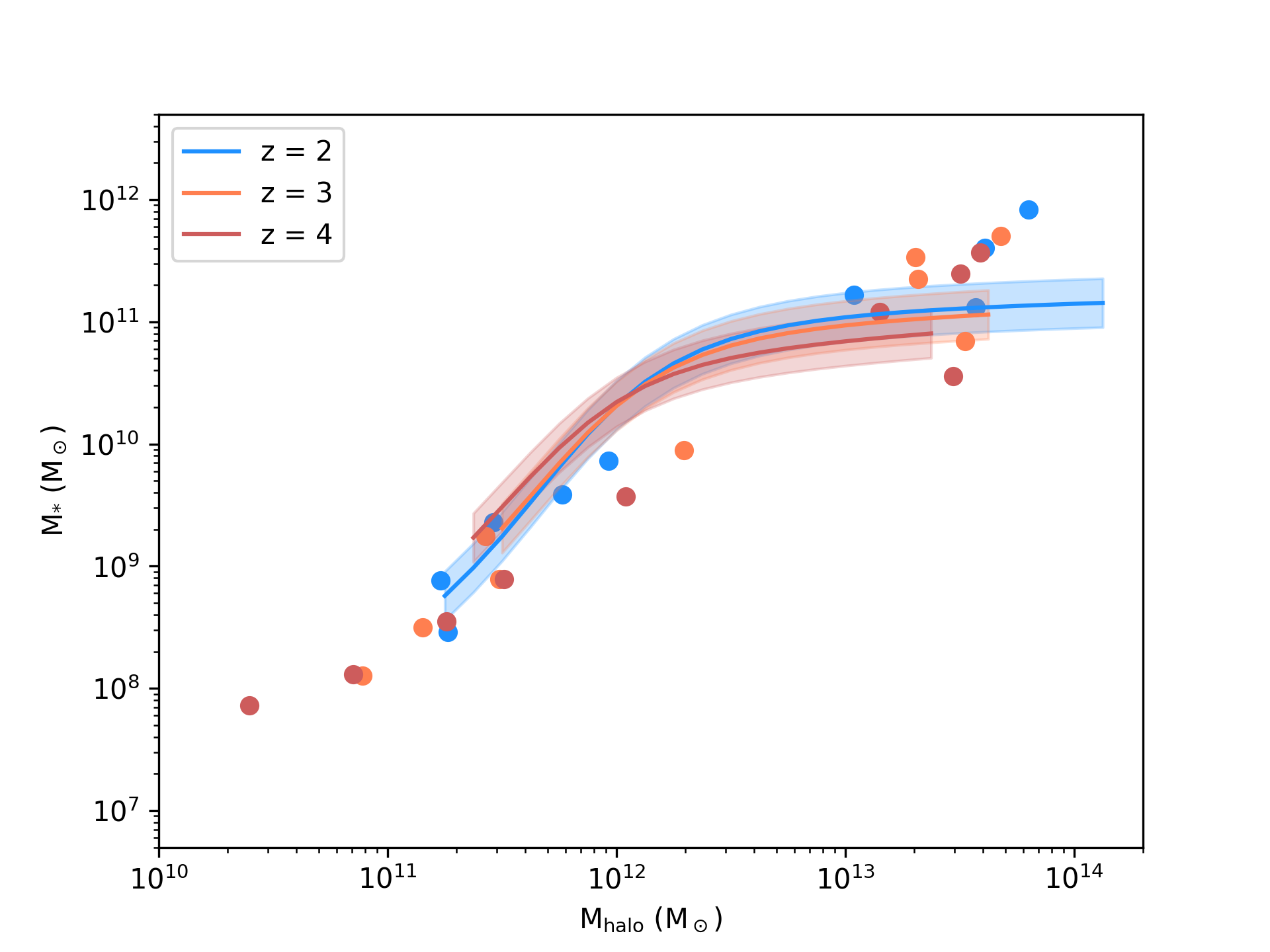}

\end{tabular}
    \vspace{-0.25cm}
    \caption{$M_*$-$M_{\rm halo}$ relation for our model halos at
      integer redshifts between $z=2-4$.  The shaded lines come from
      the average relations derived by \citet{behroozi13a}, with an
      assumed $0.2$ dex uncertainty.
    \label{figure:mstar_mhalo}}
\end{figure}

\begin{figure}
\begin{tabular}{cc}
\hspace{-0.25cm}
\includegraphics[width=\columnwidth]{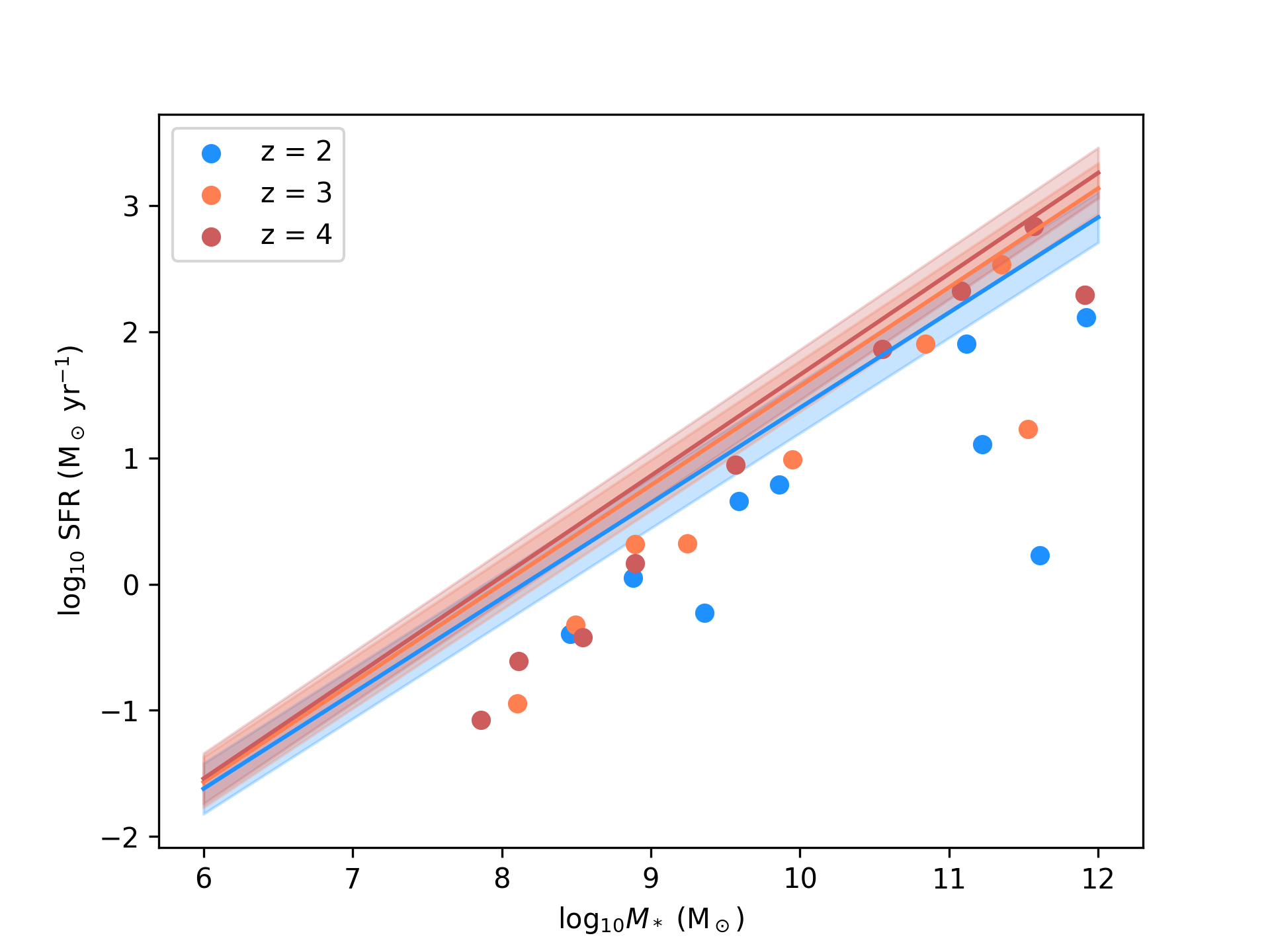}
\end{tabular}
    \vspace{-0.25cm}
    \caption{SFR-$M_*$ relation for our model halos at integer redshifts between $z=2-4$.   The shaded lines represent the average locations of the star forming main sequence derived by \citet{speagle14a}, with an
      assumed $0.2$ dex uncertainty.
    \label{figure:sfr_mstar}}
\end{figure}

In Figure~\ref{figure:mstar_mhalo}, we show the $M_*-M_{\rm halo}$ relation for our model halos at integer redshifts $z=2-4$.  The shaded curves are the best fit abundance matching models from \citet{behroozi13a} with an assumed $0.2$ dex uncertainty.  In Figure~\ref{figure:sfr_mstar}, we show the SFR-$M_*$ plane of our model galaxies, with the shaded regions showing the best fit main-sequence relations from \citet{speagle14a} (again, assuming a $0.2$ dex uncertainty).  Our model galaxies tend to lie on or near the typical SFR-$M_*$ relation, although typically somewhat below as has been commonly found for galaxy formation simulations including {\sc Mufasa}~\citep[e.g.][]{Somerville15,Dave2016}.
A few galaxies at $z\sim 2$ fall closer to the passive region $>1$~dex below the main sequence. These galaxies have exhausted the bulk of their star formation owing to gas consumption and have relatively low gas fractions~\citep[see e.g.][]{feldmann16a}.

Similar to \citet{thompson15b}, we consider a major merger to be at least a 4:1 merger and a minor merger to be at least a 10:1 merger. We use the ratio, $R$, of the increase in baryonic mass between two snapshots to the baryonic mass of the snapshot before the mass increase to identify galaxy mergers. For each snapshot, we calculate the ratio given by:
\begin{equation}
\label{merger ratio}
R=\frac{M_t-M_{\left(t-1\right)}}{M_{\left(t-1\right)}}
\end{equation}
where $M_t$ is the baryonic mass of a galaxy at one time and
$M_{\left(t-1\right)}$ is the baryonic mass of the central galaxy in
the preceding snapshot. Values of $R$ that are $\geq0.25$ or
$\geq0.10$ indicate that between the snapshot at $t$ and snapshot at
$t-1$, a galaxy that will participate in a major or minor merger has
gotten close enough to the central galaxy that the FoF algorithm binds them as a single galaxy.

There are of course some ambiguities in utilising baryonic mass
increases as a signature for mergers.  First, a rapid succession of
minor mergers can mimic a major merger if enough mass is bound to the
central galaxy.  Similarly, a first passage of a galaxy during a
merger followed by a delayed second approach can trigger two merger
'events' if the first passage is close enough.  Without visual
inspection, it is difficult to remove these technical ambiguities.
This said, both of these points are somewhat academic; in principle quantitative morphology measures should be blind to these issues.

%the first approach of a major merger, the second approach of a major merger or several simultaneous minor mergers. There is similar uncertainty when $R\geq0.10$.

%\textbf{Left over from old version (Not necessarily true now):}

%Due to these caveats, we mainly study major galaxy
%mergers in this paper. While it is challenging to disentangle
%specifics about the merger history from visual inspection for the more
%massive galaxies, we observe at least one case for the central galaxy in 
%mz5 where $R\geq0.25$ indicates the second approach of a merging galaxy. 
%It is important to realize that from a single observation of the case 
%where $R$ indicates the second approach of a merger, visual inspection 
%of this observation alone (without knowledge of the detailed merger 
%history) would probably not be able to differentiate whether a first 
%%or second pass is being observed. For this reason and the fact that 
%we only notice one obvious scenario like this, we are satisfied with the
%performance of $R$.

\begin{figure*}
% Make sure the figure is centered:
\centering
\includegraphics[width = 7 in]{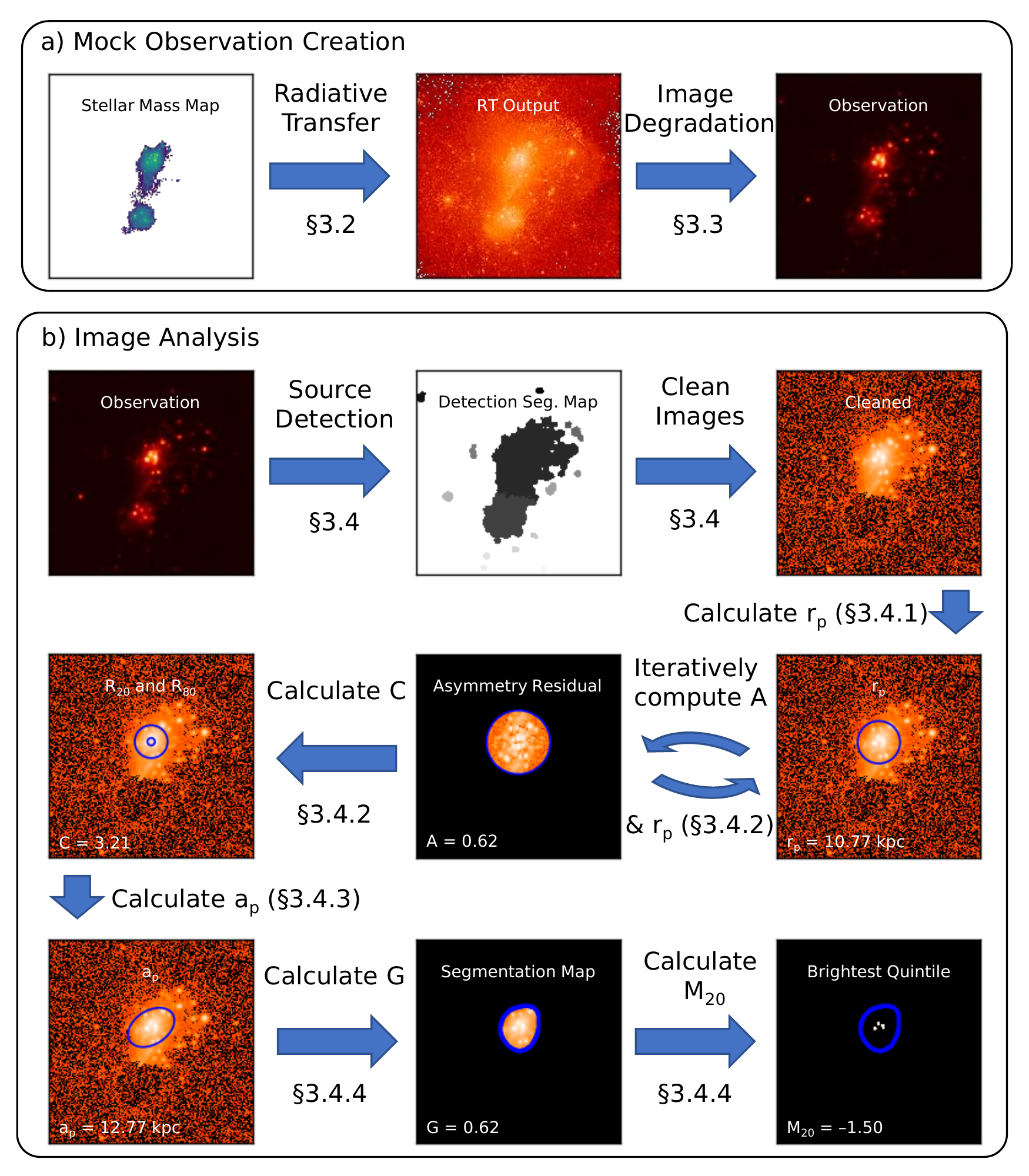}
%\includegraphics[width = 4 in]{figures/mass_evo.pdf}
% Give the caption for the Figure here. 
\caption{\label{figure:flowchart} Illustration of the steps involved
  with the creation of a mock observation (section a)) and the
  analysis of the observation (section b)) for a single line of sight
  of the central galaxy of mz10 at $z\approx2.8$.  The panel labelled
  ``Stellar Mass Map'' shows the projection of all of the stellar mass
  particles assigned by \caesar\ to the central galaxy and represents
  the galaxy model we start with. The panel titled ``RT Output'' shows
  the image produced by \pd\ and the mock observation created by
  degrading the initial image is shown in the panels titled
  ``Observation''. The panel labelled ``Detection Seg. Map''
  illustrates all of the detected galaxies in the image. ``Cleaned
  Image'' illustrates the observation after masking and
  sky-subtraction while ``$\mathrm{r_p}$'' shows the same image with
  the Petrosian radius overplotted. The panel called ``Asymmetry
  Residuals'' shows the absolute values of the residuals used to
  measure $A$. The panel labelled ``$\mathrm{r_{20}}$ and
  $\mathrm{r_{80}}$'' shows the radii used to compute $C$ whereas the
  panel labelled ``$\mathrm{a_p}$'' illustrates an ellipse with the
  ellipticity and position angle of the best-fit ellipse and the
  Petrosian semi-majors. The outline of the segmentation map used to
  calculate $G-M_{20}$ is shown in ``Segmentation Map'' and
  ``Brightest Quintile.'' In the former, the contents of the
  segmentation map is shown while in the latter, only the brightest
  quintile is shown.}
\end{figure*}

\section{Mock Observations and Morphological Measures}
\label{section:mock_observations}
\subsection{Overview}

In this Section, we describe how we generate realistic mock
observations of the our model galaxies and how we calculate the
quantitative morphology measures from that observation. In
Figure~\ref{figure:flowchart}, we provide a schematic that summarises
our methods that is cross-linked with the section number describing
those methods quantitatively.

In section (a) Figure~\ref{figure:flowchart}, the panel labelled
``Stellar Mass Map'' illustrates the galaxy model that we start
with. Although in this panel only the stellar particles assigned to
the central galaxy are shown (for simplicity), in reality we perform
radiative transfer on all stellar particles in the surrounding region
(c.f. \S~\ref{section:radiative_transfer}).  The result of radiative
transfer is an image like the panel labelled ``RT Output'' in Figure
\ref{figure:flowchart}. The specifics of the radiative transfer
simulations are detailed in \S \ref{section:radiative_transfer}. We
then degrade the radiative transfer output to make it comparable to
bona fide observations of galaxies; the produced image resembles the
panel labelled ``Observation'' from Figure \ref{figure:flowchart} and
the details of this process are detailed in \S
\ref{section:image_degredation}. Once we finish degrading the image,
we start image analysis which is explained in \S
\ref{section:image_analysis}.

Section (b) of Figure~\ref{figure:flowchart} demonstrates our image
analysis techniques, which can be subdivided into several steps. The
first of which consists of detecting galaxies in the image; the panel
labelled ``Detection Seg.  Map'' shows the regions of the mock
observation that are assigned to different galaxies. Although it has
been omitted from Figure~\ref{figure:flowchart} for space and clarity,
we use the stellar mass map to determine which of the detected
galaxies corresponds to the central galaxy. Afterwards we ``clean''
the mock observation by subtracting the noise and masking the light
associated with all detected galaxies other than the central galaxy to
produce an image like the panel labelled ``Cleaned.''

At this point, we move onto the task of calculating asymmetry ($A$)
and concentration ($C$). To do this we calculate the Petrosian radius
($r_p$) and find the centre of the galaxy such that the region
enclosed within $1.5r_p$ has minimized $A$. We recalculate $r_p$ and
again search for the centre at which $A$ is minimized using the new
$r_p$. The panel titled ``$\mathrm{r_p}$'' illustrates the size of the
final calculated $r_p$ while the panel titled ``Asymmetry Residual''
shows the absolute value of the difference of the region used to
calculate $A$ and the region rotated by 180$\degr$.  We
calculate the radii that enclose 20 per cent and 80 per cent of the
galaxy's light in order to determine $C$.  These radii are shown in
the panel labelled ``$\mathrm{r_{20}}$ and $\mathrm{r_{80}}$''. The
final stage of image analysis is made up of the calculation of the
Petrosian semi-major axis ($a_p$), Gini ($G$) and $M_{20}$. We
calculate $a_p$ at the centre where $A$ was minimized; it is
illustrated in the panel called ``$\mathrm{a_p}$.''  The panels from
figure \ref{figure:flowchart} named ``Segmentation Map'' and
``Brightest Quintile'' show intermediary results critical to the
calculation of $G$ and $M_{20}$.

In what follows, we describe these methods in greater detail.  This
said, the reader principally interested in the main results may skip
the remainder of this section without loss of continuity.  Finally, we
note that in the Appendices \S~\ref{section:diggss} and \S~\ref{section:idealized_comparison}, we validate our methods
  against test simulation data sets from the literature.

\subsection{Radiative Transfer Simulation} \label{section:radiative_transfer}

We employ the \pd\ dust radiative transfer software
\citep{narayanan15a,narayanan17a}, which is built on
\yt\ \citep{turk11a}, \hyperion\ \citep{robitaille11a}, and
\fsps\ \citep{conroy10a} in order to generate raw mock images of the
galaxies from the \gizmo\ simulations.  In short, \pd \ generates
stellar SEDs from the stars formed in the cosmological simulations,
and propagates these through the dusty interstellar medium (ISM) in a
Monte Carlo and iterative fashion until the radiation field and dust
temperatures are converged.

The stellar SEDs are calculated as simple stellar populations
generated in \fsps\footnote{Functionally, we use the python \fsps
  \ hooks located at \url{https://github.com/dfm/python-fsps}}, with
the ages and metallicities taken directly from the galaxy formation
simulation.  We assume a \citet{kroupa02a} stellar initial mass
function, and the Padova isochrones \citep{marigo07a,marigo08a}.  We
calculate the attenuation these stars see via \hyperion \ dust
radiative transfer.  We construct an octree grid from the hydrodynamic
simulation upon which to perform the radiative transfer by projecting
the metal mass using a spline smoothing kernel.  The octree is
constructed by placing the entire simulated region onto a grid with a
single cell, and then recursively refining until a maximum number of
gas particles (here, $64$) are contained within a cell.  Functionally,
the octree is constructed within \yt \ \citep{turk11a}.

The radiative transfer is propagated via a Monte Carlo method
\citep{robitaille11a}, and the radiative equilibrium calculation uses
the \citet{lucy99a} algorithm.  We determine convergence when the dust
temperature in $99\%$ of the cells have changed by less the $1\%$
between iterations.  We assume a $R_v = 3.15$ \citet{weingartner01a}
dust size distribution.  The dust mass is assumed to be $40\%$ of the
projected metal mass, following constraints from both local and
high-$z$ galaxies \citep{dwek98a,vladilo98a,watson11a}.

We centre the observations on the most massive progenitors of the most
massive galaxy in the simulation at $z=2$ and image 23 snapshots from
$z=2$ to $z=4$. Individual images made in 16 cameras oriented with
respect to the simulation axes. The cameras each view the centre of
mass of the central galaxy and are positioned at all combinations of
$\theta$ and $\phi$, within the sets $\theta \in \left[ 0\degr,
  30\degr, 60\degr, 90\degr \right]$ and $\phi \in \left[ 0\degr,
  30\degr, 60\degr, 90\degr \right]$. The images are produced at
$\lambda_{\rm obs} = 4325$ \angstrom\ to simulate an observation in
the rest-frame $B$ filter.  Our final model images are $512\times512$ pixels.

The physical size of the imaged region is dependent on
the angular diameter distance at for the \z\ of the snapshot
is chosen such that the resulting image has a pixel scale of
$\sim0.05 '' \mathrm{pix}^{-1}$, comparable to that of \hst 's
Wide Field Camera 3 (WFC3). The brightness of the image is
finally scaled to the luminosity distance at the redshift of the
snapshot.

\subsection{Image Degradation} \label{section:image_degredation}

%We then add sky noise by adding random Poisson
%noise such that the average signal to noise ratio, \avgSNR , of each
%of the central galaxy's pixels is 25.

%\red{(Might be better to explain this when we describe adding noise) Address how while adding noise we targetted $\left<SNR\right>\ge25$, why it's ok that our cutoff is now lower $\left<SNR\right>\ge20$ - 2 items: (i) \citet{lotz04a} showed that the dependence of the variations in morphological measures are negligible at $\left<SNR\right>\sim10-15$; (ii) in our method of adding noise, we were just approximating the locations of the galaxy pixels used to compute $G-M_{20}$, we expect the light to get slightly smeared out so it's not surprising that some of the edge pixels that are not aligned with projected stellar mass will have smaller $SNR$ and will lower the $\left<SNR\right>$ slightly}

To best compare our model images to observations, we roughly follow
the procedures employed by \citet{lotz08b} and \citet{snyder15b} to
degrade the images produced by the radiative transfer simulations. We
convolve the image with a Gaussian beam with full width at half
maximum (FWHM) corresponding to the Rayleigh criterion appropriate for
the \hst\ mirror size. We then add sky noise targeting an average signal 
to noise ratio, \avgSNR , of at least 20 to minimize the effects of 
\avgSNR\ on our analysis. We do this by adding random Poisson
noise such that the \avgSNR\ of the central galaxy's pixels is 
25. We determine the central galaxy's pixels by identifying the non-zero 
pixels when the galaxy's stellar mass is projected onto an array of equal 
resolution to the image. We deliberately aim for adding noise such that 
the galaxy's pixels have an \avgSNR\ of 25, to increase the number of 
galaxies with $\left<SNR\right>\ge 20$ because we expect the central galaxy's 
light to be slightly smeared out relative to the pixels where it has 
non-zero projected stellar mass. Thus when a central galaxy is detected 
from the observation, it may include additional pixels not included in 
its projected stellar mass, which may cause it to have an \avgSNR\ below 25.

\begin{table}
	\centering
	\caption{Values of the essential parameters for the \sep\ Background class 
    and extract function and the equivalent \sextractor\ parameters. The \sep\ 
    parameters that lack an equivalent \sextractor\ parameter have been omitted 
    from the table.}
	\label{tab:sep}
	\begin{tabular}{lcc}
		\hline
		\sep\ param & \sextractor\ param & value\\
		\hline
		bw, bh & BACK\textunderscore FILTERSIZE & 128, 128 \\
		fw, fh & BACK\textunderscore SIZE & 1, 1 \\ 
		minarea & DETECT\textunderscore MINAREA & 50 \\
                deblend\textunderscore nthresh & DEBLEND\textunderscore NTHRESH & 16\\
		deblend\textunderscore cont & DEBLEND\textunderscore MINCONT& 0.05 \\
		clean & CLEAN & True \\
		clean\textunderscore param & CLEAN\textunderscore PARAM & 1.0\\
		\hline
	\end{tabular}
\end{table}

\subsection{Image Analysis} \label{section:image_analysis}

We make use of the python library, \sep\ \citep{Barbary2016}, to 
detect sources in the mock images and generate initial segmentation 
maps. This library applies many of the algorithms from \sextractor\ 
\citep{Bertin1996}, on images stored in memory. See Table~\ref{tab:sep} 
for a list of parameters used for \sep\ that have equivalent parameters 
in \sextractor . We make a temporary sky subtracted image and detect 
sources with the detection threshold set to the global background RMS 
of the background and use the applicable parameters listed in Table 
\ref{tab:sep}. While performing extraction, we use the filter 
distributed with \sextractor\ in 
``tophat\textunderscore 5.0\textunderscore 5x5.conv'' \citep{Bertin1996} 
and set the filter\textunderscore type parameter to ``conv.'' 
We then identify the central galaxy in the image, using the 
projection of the central galaxy's stellar mass map that we computed 
when adding noise. We sum the stellar mass enclosed in the 
regions assigned to each detected galaxies in the initial detection 
segmentation map produced by \sep . The region enclosing the greatest 
projected stellar mass is identified as the detection of the central 
galaxy. We generate a masked image by copying the original mock observation 
and setting the pixels enclosed by the initial segmentation maps of 
all galaxies other than the central galaxy equal to the noise value we 
previously added to the pixels.

Unless otherwise stated, from this point on we follow the algorithm
employed in the code used in \citet{lotz08b}. Our next step is to
identify the largest square region of the masked image that is no
bigger than 50x50 pixels, is no smaller than 12x12 pixels, does not
overlap with any of the detected galaxies and contains more than 90 per 
cent non-zero pixels. Then, we generate a sky-subtracted image by
subtracting the average flux the square region from the masked
image. At this point, we move onto computing the central galaxy's
Petrosian radius $\left(r_p\right)$, and semi-major axis
$\left(a_p\right)$, concentration $\left(C\right)$ and asymmetry
$\left(A\right)$ of the CAS statistics \citep{conselice03a}, and
Gini and $M_{20}$ \citep{lotz04a}. Rather than following the
algorithms employed by \citet{lotz08b} to calculate the centre of
the galaxy and the parameters for the best-fit ellipse, we use the
values calculated by \sep .

\subsubsection{Petrosian Radius}\label{section:pet_radius}

We adopt the same definition for the Petrosian radius $\left(r_p\right)$ 
as described by \citet{lotz04a}. The Petrosian radius is defined as the 
radius at which the quotient of the the surface brightness enclosed by 
a circular annulus, $\mu (r_p)$, and the mean surface brightness 
enclosed by a circular aperture, $\bar{\mu} (r<r_p)$, is equal to a 
constant, $\eta$, or 
\begin{equation}
\label{r_p}
\eta = \frac{\mu (r_p)}{\bar{\mu}(r<r_p)}
\end{equation}
By convention, $\eta=0.2$. To compute the size of apertures, we use 
the IDL task dist\textunderscore ellipse from the IDL Astronomy Library 
(\astrolib)\footnote{\url{https://idlastro.gsfc.nasa.gov/homepage.html}} 
\citep{Landsman1993} with 
elliptical parameters corresponding to a circle to determine the pixels 
that belong to different apertures. We then employ an iterative algorithm 
that iterates over the radius, $r$, of the circular aperture and 
calculates $\eta$ for each aperture. The algorithm starts with $r=2$ pixels 
and between iteration $r$ is increased by 1. The algorithm terminates when 
$\eta \ge 0.2$ and we determine $r_p$ using $r_p = r(0.8+\eta)$.

\subsubsection{Asymmetry and Concentration}

The asymmetry $(A)$ of a galaxy quantifies the rotational symmetry of 
a galaxy's light \citep{conselice03a}. To determine $A$, the image 
of a galaxy is rotated by $180\degr$ about its centre and is 
subtracted from the original image. The absolute value of the 
residuals summed and divided by the sum of the fluxes in the 
original image and the average asymmetry of the background is 
subtracted:
\begin{equation}\label{Asymmetry}
A = \frac{\sum |I_0-I_{180}|}{\sum I_0}-B_{180}.
\end{equation}
Here, $I_0$ and $I_{180}$ represent the flux values 
of individual pixels in the original image and rotated images, 
respectively, while $B_{180}$ is the average background asymmetry. The 
summations only sum over pixels within $1.5r_p$ of the galaxy's 
centre which is chosen such that $A$ is minimized.

To compute $A$ at a particular centre, we rotate 
the image about the centre using bilinear interpolation and subtract 
the rotated image from the original image. We use the algorithm from 
dist\textunderscore ellipse \citep{Landsman1993} to identify all pixels 
that lie within $1.5r_p$ of the galaxy's centre. Next, we sum the 
absolute value of the residuals that lie within those pixels and divide 
by the sum of the absolute value of the original fluxes at those pixels.
At this point, we have computed the uncorrected asymmetry and all that 
remains is to compute the average background asymmetry, $B_{180}$. We 
take the square background region of the sky-subtracted image, which include 
the same pixels from the square background region we identified to 
perform sky-subtraction, rotate the region about its centre, subtract 
the rotated values from the original values, sum the absolute value of 
the residuals and divide by the number of pixels in the region. Then, 
we multiply the quotient by the number of pixels that lie within 
$1.5r_p$ of the galaxy's centre and divide by the sum of the absolute 
value of the galaxy's fluxes at those pixels, which gives $B_{180}$. 
Subtracting $B_{180}$ from the uncorrected asymmetry, gives the value of 
$A$ at that centre.

In practice, we start trying to calculate the $A$ using the $r_p$ 
calculated when the centre of the galaxy was set to be the centre of 
the best-fit ellipse. We then minimize Equation \ref{Asymmetry} using 
modified Powell's method, which determines the centre at which $A$ is 
minimized starting from the centre of the best-fit ellipse. The result 
is the initial guess for $A$, and the initial guess for the centre. Then, 
we recalculate $r_p$ at this new centre, and again use the modified 
Powell's method to minimize $A$ using the $r_p$ and the new centre as a 
guess. As a result, we determine the galaxy's $A$ and the centre where 
$A$ is minimized.

We use the definition of Concentration $(C)$ given by 
\citet{bershady00a}. It is the ratio of the radii at which 
circular apertures contain 20 and 80 per cent of a galaxy's total 
flux: 
\begin{equation}
\label{Concentration}
C = 5 \log_{10} \left(\frac{r_{80}}{r_{20}}\right).
\end{equation}
Like \citet{conselice03a}, we consider the total flux to be the flux 
contained in $1.5r_p$. The galaxy's centre used in the calculation 
of $C$ is the centre determined while measuring $A$. When computing 
$C$, we start by recalculating $r_p$ at the centre where $A$ is 
minimized. Next, we use a similar iterative method to tabulate the 
fraction of the flux enclosed within an aperture of radius $1.5r_p$ for 
all apertures with positive integer radii less than $1.5r_p$. Then, we 
use linear interpolation to determine the smallest radii to contain 
0.2 and 0.8 of the total flux, which yields $r_{20}$ and $r_{80}$. 
Finally, we determine $C$ with Equation \ref{Concentration}.

\subsubsection{Petrosian semi-major axis} \label{section:pet_semi-major}
After computing the Concentration $C$, we then calculate the Petrosian
semi-major axis ($a_p$), or the elliptical Petrosian radius,
\citep{lotz04a}, which is defined in the same way as $r_p$ except that
elliptical apertures and annuli are used. To calculate it, we use an
algorithm that differs from the one employed by \citet{lotz08b} and draws
significant inspiration from the algorithm employed in the photometric
pipeline of the Sloan Digital Sky Survey (SDSS) to compute $r_p$
\citep{Strauss2002}. Let $\theta$ be the angular distance between the
centre of an object and the edge of an annulus or aperture. Following
\citet{Strauss2002}, we can rewrite Equation \ref{r_p}, the equation
for the Petrosian ratio, as
\begin{equation}
\label{PetRatio}
\eta \left(\theta\right) = \frac{\int_{r_{in}\theta}^{r_{out}\theta}I\left(\theta^\prime\right) dA / \left[A\left(r_{out}\theta\right)-A\left(r_{in}\theta\right)\right]}{\int_{0}^{\theta}I\left(\theta^\prime\right) dA/A\left(\theta\right)}.
\end{equation}
In the above equation $r_{in}$ and $r_{out}$ are constants that represent 
the inner and outer limits of an annulus. Additionally, 
$A\left(\theta\right)$ and $I\left(\theta\right)$ are the area and 
average flux enclosed by an elliptical aperture with a semi-major 
axis $\theta$. For a given galaxy, all apertures have a constant 
ellipticity, $ell = 1 - b/a$. Like with $r_p$, the Petrosian semi-major 
axis, $a_p$, is defined as the value of $\theta$ at which $\eta=0.2$.

Suppose $L\left(\theta\right)$ is the function describing the 
cumulative light profile of a galaxy. It gives the total light 
enclosed within a semi-major axis $\theta$ and is defined as
\begin{equation}
\label{CumProfile}
L\left(\theta\right)=\int_0^\theta I\left(\theta^\prime\right) dA.
\end{equation}
We can substitute Equation \ref{CumProfile} into Equation \ref{PetRatio} to get
\begin{equation}
\label{simpPetRatio}
\eta \left(\theta\right) = \frac{\left[L\left(r_{out}\theta\right)-L\left(r_{in}\theta\right)\right] / \left[A\left(r_{out}\theta\right)-A\left(r_{in}\theta\right)\right]}{L\left(\theta\right)/A\left(\theta\right)}.
\end{equation} 
Because all of our apertures have fixed ellipticity, we know that 
$A\left(\theta\right)=\pi\theta^2(1-ell)$. Using this to 
simplify our equation for $\eta$, we find
\begin{equation}
\label{finalPetRatio}
\eta \left(\theta\right) = \frac{L\left(r_{out}\theta\right)-L\left(r_{in}\theta\right) }{L\left(\theta\right)\left(r_{out}^2-r_{in}^2\right)}.
\end{equation}
Finally, $a_p$ can be found by just solving this equation for when 
$\eta = 0.2$.

To actually calculate $a_p$, we start by determining points along the 
cumulative light profile of the galaxy, $L\left(\theta_i\right)$, for 
56 exponentially spaced semi-major axes $\theta_i$. The minimum 
semi-major axis, $\theta_0$ is the semi-major of an ellipse that 
encloses an area of 1 $\mathrm{pix}^2$ and the values of the remaining 
semi-major axes are given by $\theta_i=(1.057/0.9457)\theta_{i-1}$. 
After we compute the semi-major axes, we compute the flux enclosed 
within the elliptical aperture with a semi-major axis of $\theta_0$ 
and the fluxes enclosed in elliptical apertures extending from 
$\theta_{i-1}$ to $\theta_i$ for the remaining semi-major axis. We 
will refer to these flux measurements as $F\left(\theta_i\right)$. 
The elliptical aperture and elliptical annuli each have ellipticity 
equal to that of the best-fit ellipse and are all centred on the 
centre of the galaxy where $A$ had been minimized. To perform the actual 
photometry we use the sum\textunderscore ellipse and 
sum\textunderscore functions from the python package, \sep\ 
\citep{Barbary2016} and we have the functions calculate the exact 
overlap between the pixels and the apertures. After computing 
$F\left(\theta_i\right)$, we calculate $L\left(\theta_i\right)$ using
$L\left(\theta_i\right)=\sum_{k=0}^{n=i}F\left(\theta_k\right)$.

Like the SDSS photometric pipeline, we construct a cubic spline with 
the ``not-a-knot'' condition to find $\mathrm{asinh}\left(L\right)$ 
as a function $\mathrm{asinh}\left(\theta\right)$ \citep{Strauss2002}.
To construct the spline, we use all values of $\theta_i$ and 
$L\left(\theta_i\right)$, and set $L\left(0\right) = 0$.
Similar to the SDSS photometric pipeline, we take $\mathrm{asinh}$ 
of all of our $\theta_i$ and $L\left(\theta_i\right)$ because of the 
large dynamic range in $L\left(\theta\right)$ and because 
$\mathrm{asinh} x$ is better behaved than $\log x$ when $x$ 
approaches 0 \citep{Strauss2002}. We use Equation 
\ref{finalPetRatio}, our cubic spline, $r_{in}=0.84$, and $r_{out}=1.19$ 
to determine $\eta\left(\theta_i\right)$. Then, we use another 
``not-a-knot'' cubic spline to find $\eta$ as a function of 
$\mathrm{asinh}\left(\theta\right)$. Finally, $a_p$ is given by the 
value of $\theta$ for which $\eta = 0.2$. If there are multiple values 
of $a_p$, we choose the smallest value of that is $\ge5$ pix, if 
applicable; however if all values are smaller than 5 pix, choose the 
maximum value. Like in the algorithm employed by \citet{lotz08b}, 
if $a_p<2$ pix, we set $a_p$ equal to the value of $r_p$.

There are additional differences from the algorithm employed in the SDSS 
photometric pipeline beyond our change to make the algorithm apply to 
$a_p$. These changes include  calculating points of the cumulative 
light profile $L\left(\theta_i\right)$ at more $\theta_i$ values that 
were more closely spaced, not using the ``taut'' condition for our 
cubic splines, setting $L\left(0\right)=0$, and using different values 
of $r_{in}$ and $r_{out}$ \citep{Strauss2002}. See Appendix 
\ref{ap explanation} for an in-depth explanation as to why we employ 
this algorithm to compute $a_{p}$ instead of that which was employed 
by \citet{lotz08b}.

\subsubsection{Gini and $M_{20}$}

\begin{figure*}
% Make sure the figure is centered:
\centering
\includegraphics[width = 6.8 in]{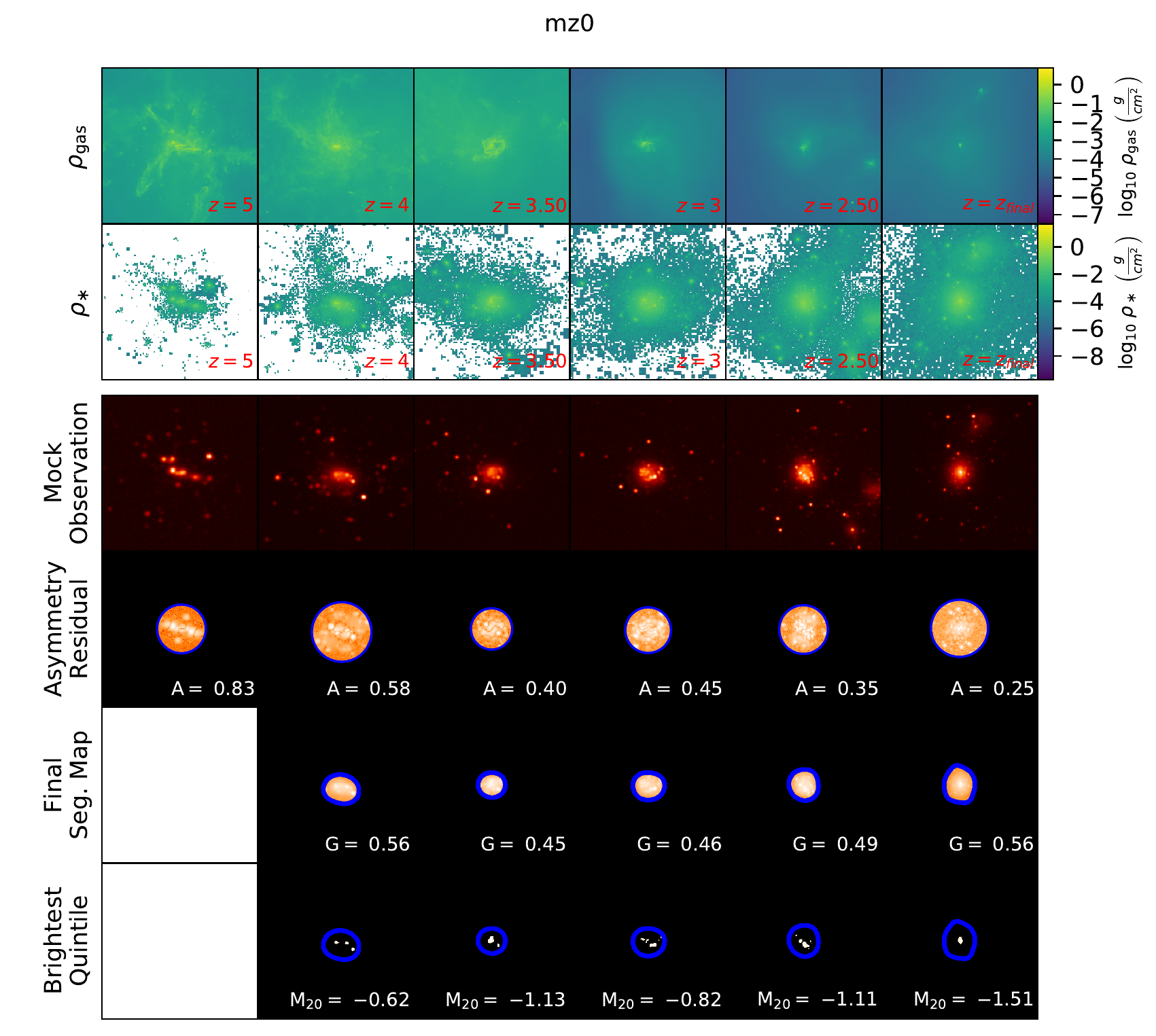}
\caption{\label{figure:mz0_stamps} Gas surface density, stellar surface density, mock $B$-band observations, asymmetry residuals, final segmentation maps, and brightest quintile centred on the centre of mass of the central galaxy of mz0, generated along a single line of sight, at $z\approx$ $5$, $4$, $3.5$, $3$, $2.5$ and $z_{final}$. All gas surface density plots share a single colourmap, as do the stellar surface density plots. Each panels has an angular size of $\approx12.8$ \arcsec . The final segmentation map and brightest quintile has been omitted at the $z$ when the final segmentation map has \avgSNR $< 20$ or is not contiguous.}
\end{figure*}

The Gini coefficient $(G)$ describes the distribution of light among a 
pixels in a galaxy's segmentation map constructed with
procedure described by \citet{lotz04a}. Specifically, the value of $G$
is given by
\begin{equation}
\label{Gini}
G = \frac{1}{\bar{\mid f\mid} n\left(n-1\right)}\sum_{i}^n \left(2i-n-1\right)\mid f_i\mid
\end{equation}
where $n$ is the number of pixels that are in the object's segmentation
map, $\mid f_i\mid$ is the $i$th smallest absolute flux value in the
segmentation map, and
$\bar{\mid f\mid}=\sum_{i}^n \frac{\mid f_i\mid}{n}$.
Higher values of G indicate that the light is less equitably distributed 
among a galaxy's pixels while lower values of G indicate that the light 
is more equitably distributed among a galaxy's pixels \citep{lotz08b}. 
A value of 0 means that all pixels have the same brightness while a value 
of 1 means that all of the light comes from a single pixel \citep{lotz04a}.

Before calculating $G$, we first determine the segmentation map using 
$a_p$. To determine the segmentation map at a given centre we first 
convolve the image with a Gaussian of $\sigma = a_p/5$ to get the 
smoothed image which is only used for determining the segmentation 
map. Following the algorithm used by 
\citet{lotz08b}, we set the FWHM of the Gaussian kernel to the 
maximum of $a_p/10$, 3 times the FWHM used for the point spread 
function, and 1.0. We also set the width of the kernel 
to 5 times the FWHM of the kernel, rounded down to the nearest 
integer. Next, we identify pixels in the smoothed image with at least 
as bright as the average surface brightness $\mu$ at $a_p$ in the smoothed 
image. To measure $\mu (a_p)$, we take the average of all 
pixels in the smoothed image we determine to lie between $a_p-1$ and 
$a_p+1$ using the algorithm from the IDL task dist\textunderscore ellipse 
\citep{Landsman1993}. Then, we determine all pixels at least as bright as 
$\mu (a_p)$. If there are less than 2 pixels brighter than $\mu$, we cut 
$\mu$ in half, and repeat this process of dividing $\mu$ in half until 
we have at least 2 pixels. 

%Following the methodology employed by \citet{lotz08b}, we do not check that the 
%flux of pixels in the segmentation is within 10$\sigma$ of their 
%neighbours as these are simulated images and there is no chance of 
%finding cosmic rays \citep{Lotz2004}.
At this point, we generate an array where all pixels in the
segmentation map have values of 10 and all other pixels have values of
0, and apply the algorithm employed in the IDL task
sigma\textunderscore filter from \astrolib. % when the
%box\textunderscore width is 10 and the ALL\textunderscore PIXELS
%keyword is enabled \citep{Landsman1993}.
All pixels in the resulting array with non-zero values are identified
as pixels in the segmentation map. This procedures removes completely
isolated bright pixels from the segmentation map segmentation map and
includes dimmer pixels that are predominantly surrounded by pixels in
the segmentation map.

% we calculate the mean of all of the other pixels
% For a given pixel in the array, we calculate the average value of all 
% other pixels in an 11 x 11 box centred on the given pixel. Next, we 
% calculate the deviation from the mean of all of the averages at every 
% pixel. Then we use these deviations to calculate a standard deviation 
% of all pixels surrounding the central pixel in the 11x11. Pixels in 
% the array of means are replaced by their original value if the original 
% value is less than 3 sigma of its neighbours. Then the array is returned.
%
% In the simple case where all pixels in an 11x11 box where all pixels 
% have a constant mean this does 2 things to the segmentation map:
% 1. Remove pixels in the map when there are no other pixels in the box 
%    that were also originally in the segmentation map.
% 2. Add pixels to the segmentation map if MORE THAN 90% of the other 
%    pixels in the box were originally in the segmentation map.

We calculate the segmentation map at the centre where the total second 
order moment, $M_{tot}$, is minimized. The total second order moment at 
a given centre, $(x_c, y_c)$, is given by 
\begin{equation}
\label{Mtot}
M_{tot} = \sum_{i}^n M_i =  \sum_{i}^n f_i \left[\left(x_i-x_c\right)^2+\left(y_i-y_c\right)^2\right]
\end{equation}
where $f_i$ is the flux of the pixel at $(x_i, y_i)$. This formula only 
applies to pixels in the segmentation map centred at $(x_c, y_c)$. To 
determine the centre at which $M_{tot}$ is minimized we use the modified 
Powell's method on a function that computes the segmentation map and 
$M_{tot}$ at various centres. We supply the centre at which $A$ is 
minimized as an initial guess. Once we have the segmentation 
map where $M_{tot}$ is minimized, we check to see if all pixels in the 
segmentation map are contiguous. If so, we compute $G$ using Equation 
\ref{Gini}. Otherwise we simply do not calculate $G$ or $M_{20}$ for 
that galaxy.

Additionally, we calculate $M_{20}$, the normalized second-order moment 
of the brightest 20\% of a galaxy's light, from the same segmentation 
map used to calculate $G$. We determine $M_{20}$ by summing $M_i$ for 
the pixels ordered by 
decreasing flux until the total flux of the pixels is 20\% of the 
total flux in the segmentation map, and normalizing the sum by 
$M_{tot}$. This is summarized by
\begin{equation}
\label{M20}
M_{20} = \log_{10} \left(\frac{\sum_{i}^n M_i}{M_{tot}}\right)\mbox{, while} \sum_{i}f_i < 0.2 f_{tot}.
\end{equation}
where $f_{tot}$ is the total flux in the segmentation map. According
\citet{lotz08b} find that more positive values ($M_{20} \ge -1$),
intermediate values ($M_{20} \sim -1$), and more negative values
($M_{20} \le -2$) typically indicate mergers, late type galaxies, and
early type galaxies, respectively.  We calculate $M_{20}$ by applying
Equation \ref{M20} to the segmentation map used to calculate $G$.

After we calculate $M_{20}$, we determine the average signal-to-noise 
ratio per pixel, \avgSNR , for the segmentation map of the galaxy. 
Following \citet{lotz04a}, we compute \avgSNR\ via: 
\begin{equation}
\label{eqn:avgSNR}
\left<SNR\right> = \frac{1}{n}\sum_{i}^n\frac{f_i}{\sqrt{\sigma_{sky}^2+f_i}},
\end{equation}
where $n$ is the number of pixels in the segmentation map, $f_i$ is 
the flux of the $i$th pixel in the segmentation map, and $\sigma_{sky}$ 
is the sky noise. To determine $\sigma_{sky}$, we apply 
a transposed version of the IDL task robust\textunderscore sigma from 
\astrolib\ \citep{Landsman1993} on the square region of the sky 
subtracted image that we use to compute $B_{180}$ in the calculation 
of $A$.

\section{Morphology Analysis}
\label{section:results}
\subsection{Model Sample Selection}
\label{section:sample_selection}
\begin{figure}
  \includegraphics[scale=0.7]{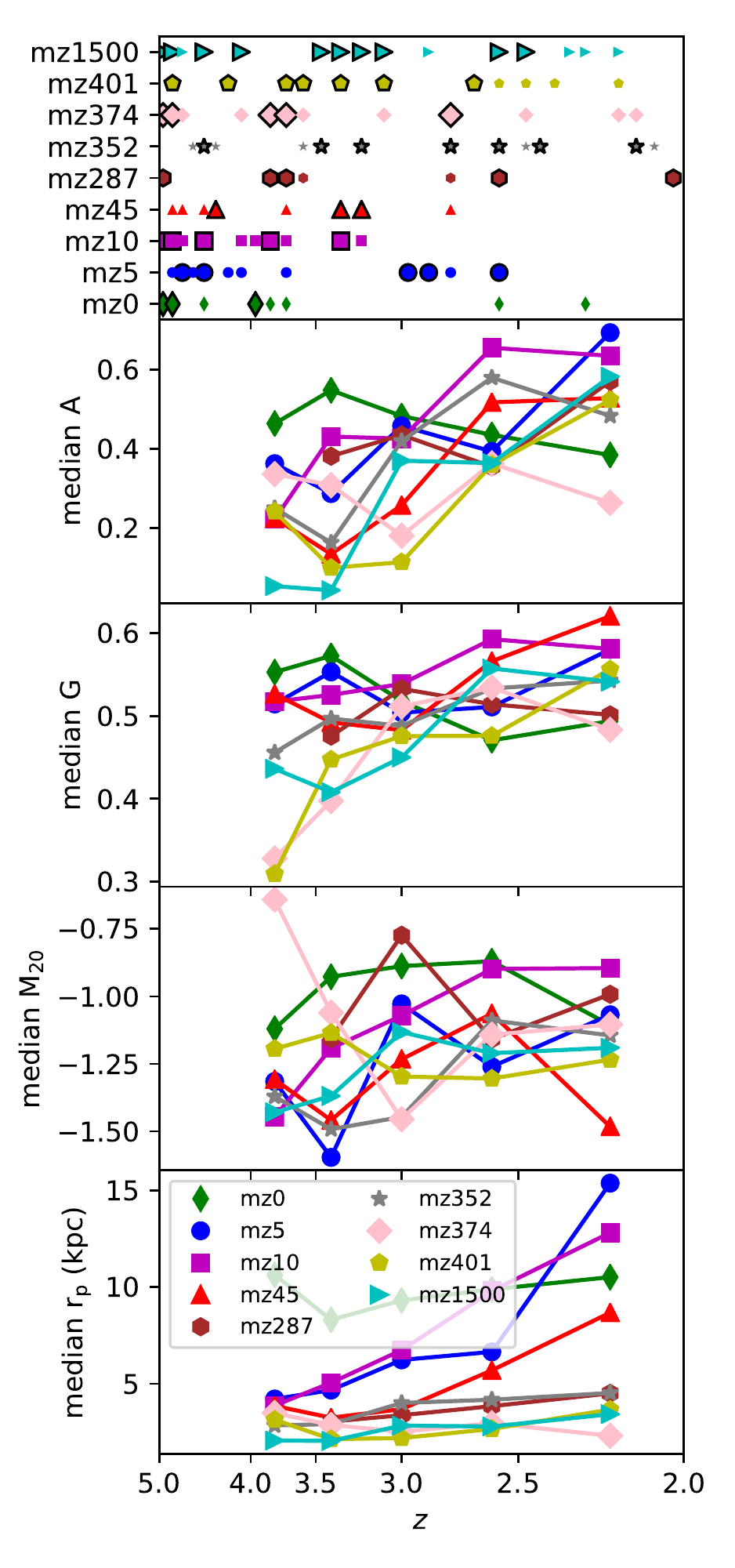}
  \caption{\label{figure:param_evo} Galaxy merger history and 
  	evolution of $A$, $G$, $M_{20}$ and the Petrosian Radius 
    $r_{\rm P}$ as functions of $z$. The top panel illustrates 
    mass increases indicative of galaxy merger; larger markers 
    with black outline correspond to mass increases indicative of 
    major mergers ($R\ge0.25$), whereas smaller markers without 
    outlines indicate mass increases indicative of minor mergers 
    ($0.10\le R <0.25$). In the panels showing the evolution of the 
    morphology measures we only show the region between $z=2-4$ as 
    non-contiguous segmentation maps and low SNR at higher redshifts 
    make the model data somewhat incomplete.}
\end{figure}

We now turn to analysis of the evolution of morphology over
$z\sim2-4$.  We examine morphological parameters along 13 lines of
sight\footnote{While the galaxies are all imaged by 16 cameras in \pd,
  the cameras positioned at $\left(\theta,\phi\right)=$ $(0\degr
  ,0\degr )$, $(0\degr ,30\degr )$, $(0\degr ,60\degr )$, and $(0\degr
  ,90\degr )$ all look along the same line of sight.  The only
  difference in the images produced by these cameras is that they are
  rotated in the plane of the sky, and are therefore not unique. Thus
  of those 4 cameras, we only consider the morphological parameters
  measured with the camera at $(0\degr ,0\degr )$.}, and consider
galaxies which were computed with contiguous segmentation maps that
have $\left<SNR\right>\ge20$. We also manually inspect segmentation
maps and discard 5 additional individual sightlines (out of all sightlines and
all snapshots) in which the segmentation maps enclose the majority of
the image, and are obviously incorrectly computed; these predominantly
come from the lowest mass model. 

For our CAS analysis, we manually inspect images produced for all
observed galaxies, similar to the panel labelled ``Cleaned'' in Figure
\ref{figure:flowchart} and discard values for galaxies when the
circles are very obviously offset from the central galaxy.  In doing
this, we cut measurements for one sightline from mz5, two sightlines 
from mz10, one sightline from mz45, two sightlines from mz287, one 
sightline from mz374, two sightlines from mz401 and three sightlines 
from mz1500.

After imposing these two sets of filtering criteria, we have
$G-M_{20}$ data for 1747 valid sightlines from all of our
galaxy snapshots, and 2631 measurements of $C-A$ data from all of our
galaxy snapshots.

\subsection{Evolution of Non-Parametric Morphology Parameters} \label{section:evo}

In Figure \ref{figure:mz0_stamps} we illustrate the evolution of the
gas and stellar morphology of model mz0 at important redshift
intervals.  We additionally show the mock $B$ band (4325 \angstrom )
observation, the asymmetry residual, final segmentation map, and
brightest quintile measurement.  These latter three quantities are
important for calculating $A$, $G$ and $M_{20}$, respectively.  In
Appendix~\ref{section:sim_stamps}, we show a similar series of postage
stamps for all zoom simulations in our simulation suite\footnote{We note that
we reject images that have either poor SNR or non-contiguous
segmentation maps, and so the redshift stamps are similar, though not
identical from model to model.}.

Figure~\ref{figure:mz0_stamps} and
Figures~\ref{figure:mz5_stamps}-\ref{figure:mz1500_stamps} reveal a
number of salient points.  First, the morphologies of high-redshift
galaxies are highly complex compared to local galaxies.  The rich
accretion histories of satellite galaxies result in complex
environments with extended morphologies, multiple nuclei, and stellar
bridges at nearly all redshifts.  For all galaxies, the morphologies
become more extended at later times ($z \sim 2$) as more subhalos
accrete over cosmic time.

Second, higher mass galaxies tend to have more complex and spatially
distributed optical emission than lower mass systems.  The massive
systems have extended optical morphologies from early times ($z \sim
5$).  Accordingly, aside from the most massive halo in our model
sample (mz0, which we will discuss shortly), this results in elevated
$G$ and $A$ measures as multiple nuclei and distorted morphologies are
ubiqitous over the redshift range considered for the most massive galaxies.  

Third, for the lowest mass galaxies in our simulation sample, the
contribution of satellites to the optical flux at very high ($z \sim
5)$ redshifts is insufficient enough that the optical morphologies
remain compact until later times ($z \sim 2$). Because the regions
used to compute the non-parametric measures $G-M_{20}$ and $A$ are so
compact, they are highly sensitive to small irregularities in the
optical morphology (e.g. if the nucleus is slightly elongated or 
slightly offset from the centre of the envelope). At mid to later
times ($z \sim 3-2$), the low mass galaxies each undergo a series of
mergers in quick succession. Due to the short intervals of time
between mergers ($\la 0.55$ Gyr), the galaxies are unable to relax,
making it easier for non-parametric measures to register ongoing
mergers.

These trends are quantitatively apparent in
Figure~\ref{figure:param_evo}, where we present the evolution with
redshift of non-parametric morphology measures for our model galaxies.
The lowest mass galaxies have the lowest $G$, $A$, and $M_{20}$ values
at early times, due to the difficulty of detecting infalling
satellites.  As the central galaxies grow, mergers impact $G$ and
$M_{20}$ more and these values tend to rise toward later times.  More
massive galaxies (e.g. models mz5, mz10) have relatively elevated
non-parametric morphology measures throughout the redshift range being
considered, though the most massive halo in our model sample (mz0)
deviates from this trend.  The elevated $G$, $A$ and $M_{20}$ values
for these massive galaxies (mz5, mz10) lie in the classical merger
range of these non-parametric indicators, even at times well-separated
from major mergers due to a rich accretion history of subhalos.

This said, while there are broad trends with galaxy/halo mass, it
important to recognise that the diversity in halo accretion history
can cause dispersion or slight deviations in these trends. To see
this, we examine the specific cases of low mass galaxy mz287 and the
most massive system in our sample, mz0.  As a reminder, mz287 is a
proto-Milky Way mass galaxy, while mz0 has a mass comparable to the
most massive galaxies detected at $z\sim 2$.

At early times, low mass galaxy mz287 differs from these trends with
its elevated $A$, $G$, and $M_{20}$, that are mostly indicative of
mergers. For comparison intermediate mass galaxies and other low mass
galaxies all have far lower $A$, $G$, and $M_{20}$.  The elevated $A$,
$G$, and $M_{20}$ can be partially explained by the abundance of
satellites and how it takes on a relatively extended morphology during
these early mergers. Although low mass galaxy mz374 and intermediate
galaxy mz401 also are surrounded by several satellites at $z\la4$, in
comparison to mz287 their interactions with satellites are much
shorter, and they quickly relax following each merger not becoming as
extended.

%It is worth noting
%that at $z=2$ the stellar mass of mz287 is $\sim2-10 \times$ that of
%the other low mass galaxies mz374 and mz1500 and is within a factor of
%$~2$ of the the smallest intermediate mass galaxy mz401 at $z=2$; in
%other words mz287 straddles the boundary between intermediate and low
%mass galaxies. 

%HALO 173 STUFF
%This said, while there are broad trends with galaxy/halo mass, it is
%important to recognise that the diversity in halo accretion histories
%at a given halo mass can cause significant deviations from these
%trends.  To see this, we examine the cases of intermediate mass halo
%z0mz173 and the most massive system in our sample, mz0.

%Intermediate mass galaxy z0mz173 has a low $A$, $G$ and $M_{20}$ for
%the majority of its life, with only modest elevations at times
%associated with major mergers.  The reason for this owes to this
%system's modest merger rate compared to galaxies of a similar mass
%\red{This needs to be checked with 352 and 373}.  Because of this, the
%galaxy remains compact, and the periods of time during which the optical 
%regions of the central galaxy and the merging companions are in close 
%enough proximity to both be included in the segmentation map before 
%the galaxy relaxes are very short. The
%segmentation map, therefore, typically does not include accreting
%satellites, and $G$, $A$, and $M_{20}$ are only elevated for short
%periods following a merger.

The most massive galaxy in our simulation suite, model mz0 also deviates 
slightly from these broad trends in a subtle way.  This galaxy exhibits 
somewhat elevated $G$, $A$ and
$M_{20}$ values early on (as expected given its mass), though these
values plateau, or even decrease at later times.  This is in contrast
to almost all other model halos which have rising $G,A,M_{20}$ values
with time.  The origin of this trend is subtle.  The rapid accretion
history at early times gives rise to significant amounts of extended
light that has a relatively uniform distribution within the final
segmentation map.  The spatial uniformity drives down $G$ and $A$
values, even during periods of heavy bombardment.
Whether or not model mz0 is representative of all galaxies at this
extreme end of the mass function (i.e. that represented, likely, by
high-$z$ dusty star forming galaxies) is unclear.  More simulations in
this mass regime will be revealing, and are deferred to future work.

%\begin{figure}
% Make sure the figure is centered:
%\centering
%\includegraphics[width = \columnwidth]{figures/median_gm20.pdf}
%\includegraphics[width = 3.15 in]{figures/median_gm20.pdf}
% Give the caption for the Figure here. 
%\caption{\label{figure:median_gm20} Location in $G-M_{20}$ space of the
%  central galaxy in each simulation. The $G$ and $M_{20}$ values are
%  sightline averages (over the unique viewing angles) and the
%  different shapes correspond to the different simulations which are
%  given in Table \ref{tab:galaxies}. The labels indicate the Hubble
%  Sequence classifications of local galaxies typically found in each
%  region \citep{lotz08a}.}
%\end{figure}

\begin{figure*}
% Make sure the figure is centered:
\centering
\includegraphics[width = 6.9 in]{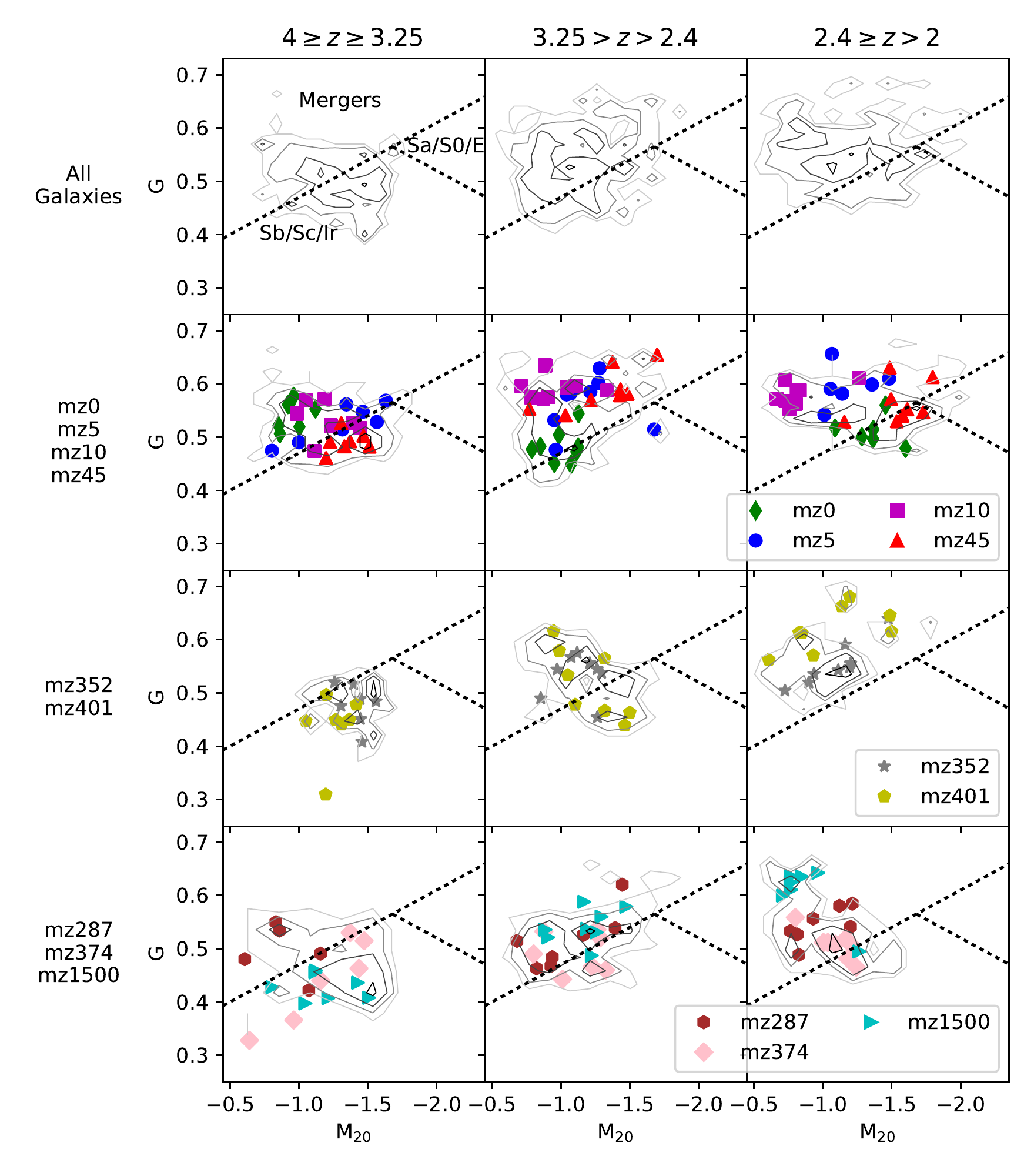}
% Give the caption for the Figure here. 
\caption{\label{figure:median_gm20} Location in $G-M_{20}$ space of the
  central galaxy in each simulation. The contours are spaced 
  logarithmically from the second smallest non-zero number density value 
  to the second largest value. 
  The $G$ and $M_{20}$ markers are
  sightline medians (over the unique viewing angles) and the
  different shapes correspond to the different simulations which are
  given in Table \ref{tab:galaxies}. The labels indicate the Hubble
  Sequence classifications of local galaxies typically found in each
  region \citep{lotz08a}.}
\end{figure*}

%\begin{figure}
% Make sure the figure is centered:
%\centering
%\includegraphics[width = \columnwidth]{figures/median_asym_conc.pdf}
%\includegraphics[width = 3.15 in]{figures/median_asym_conc.pdf}
% Give the caption for the Figure here. 
%\caption{\label{figure:median_conc_asym} Location in $C-A$ space of
%  the central galaxy in each simulation. Akin to
%  Figure~\ref{figure:median_gm20}, the $C$ and $A$ values are sightline averages (over the unique viewing angles), the points
%  are coloured by $z$, and the different shapes correspond to the
%  different simulations. The dashed line indicates a division in $A$
%  values used to identify mergers in local galaxies
%  \citep{conselice03a}; points to the right of the vertical dashed
%  line typically indicate galaxy mergers.}
%\end{figure}

\begin{figure*}
% Make sure the figure is centered:
\centering
\includegraphics[width = 6.9 in]{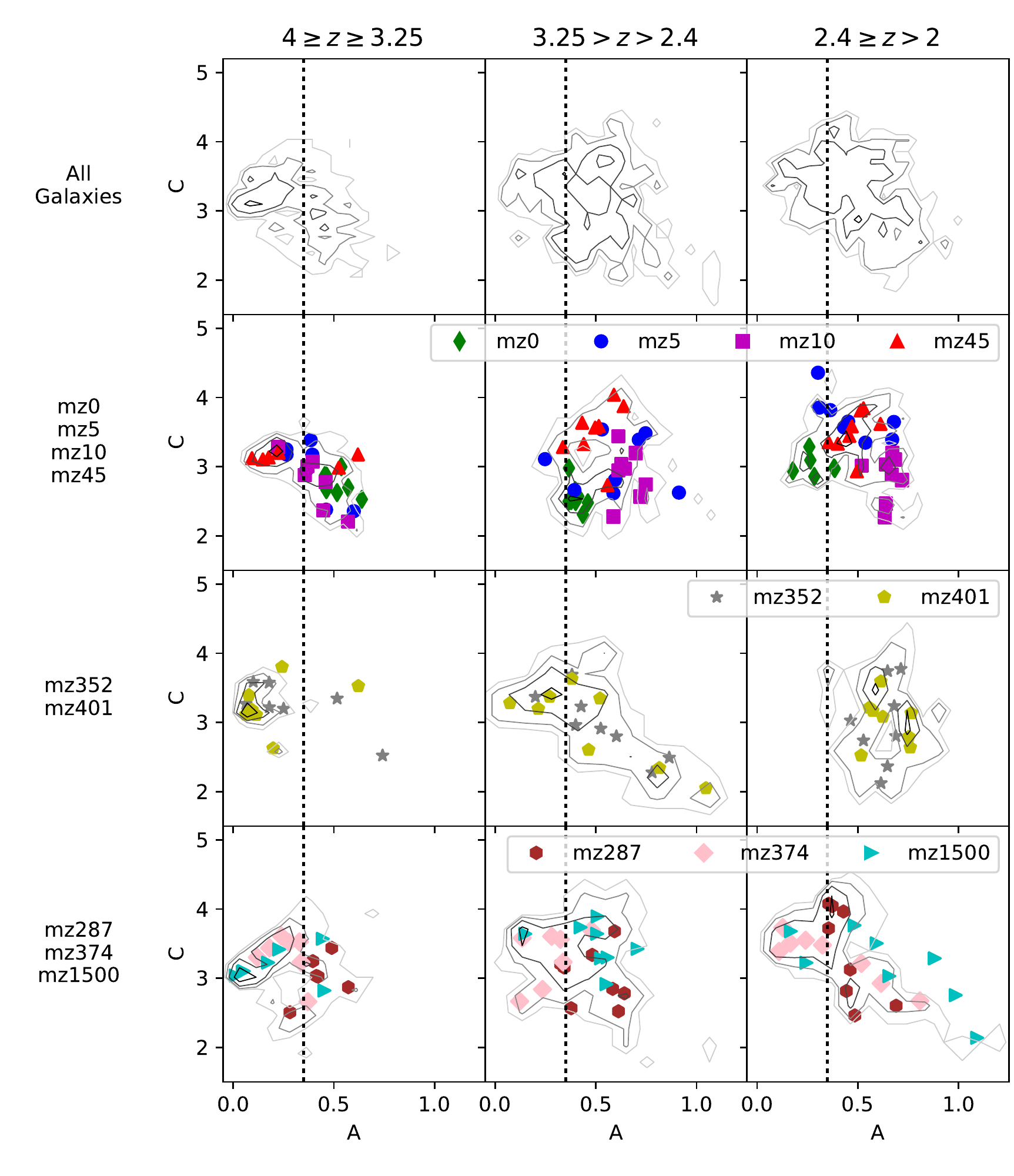}
% Give the caption for the Figure here. 
\caption{\label{figure:median_conc_asym} Location in $C-A$ space of
  the central galaxy in each simulation. Akin to
  Figure~\ref{figure:median_gm20}, the contours are spaced logarithmically 
  from the second smallest non-zero number density value to the second 
  largest value. Likewise, the markers are the sightline medians (over the 
  unique viewing angles) and the different shapes correspond to the
  different simulations. The dashed line indicates a division in $A$
  values used to identify mergers in local galaxies
  \citep{conselice03a}; points to the right of the vertical dashed
  line typically indicate galaxy mergers.}
\end{figure*}

\subsection{$G-M_{20}$ and $C-A$ Space} \label{section:evolution_param_space}

After having built our intuition in \S~\ref{section:evo}, we now
consider our model galaxies in $G-M_{20}$ and $C-A$ space.  In
Figure~\ref{figure:median_gm20}, we present the location of our model
galaxies in $G-M_{20}$ space, and in
Figure~\ref{figure:median_conc_asym}, we show the same in $C-A$ space.
In the top row of each, we show all of our model galaxies in three
redshift bins, and in the subsequent rows, we bin the galaxies by mass
in $G-M_{20}$ and $C-A$ space.  We additionally label the plot with
the traditional Hubble morphological classification associated with
particular regions in these plots \citep{lotz08a,conselice03a}.  The
points shown are sightline medians.  The generic trends discussed in
\S~\ref{section:evo} become more apparent in $G-M_{20}$ and $C-A$
space.

%In Figure~\ref{figure:median_gm20}, the median $G$ value of mz374 below 
%$G\sim0.3$ can be attributed to the small number of measurements that meet 
%our selection criteria at high $z$. Specifically that value comes from a 
%snapshot where all but measurements of $G-M_{20}$ were discarded.

On average, galaxies move from the non-merger
region of $G-M_{20}$ and $C-A$ space to the merger region as they
evolve over cosmic time.  The most massive galaxies arrive in the
merger region first due to the contribution of bright satellites at
early times.  Lower mass galaxies reside in the non-merger region of
$G-M_{20}$ and $C-A$ due to satellites at early times being too faint
to be detected in the segmentation map. At later times, as the
accretion rate increases and the contribution of satellites becomes
more significant, the lower mass galaxies move into the merger
regions.

\begin{figure}
% Make sure the figure is centered:
\centering
\includegraphics[width = 3.15 in]{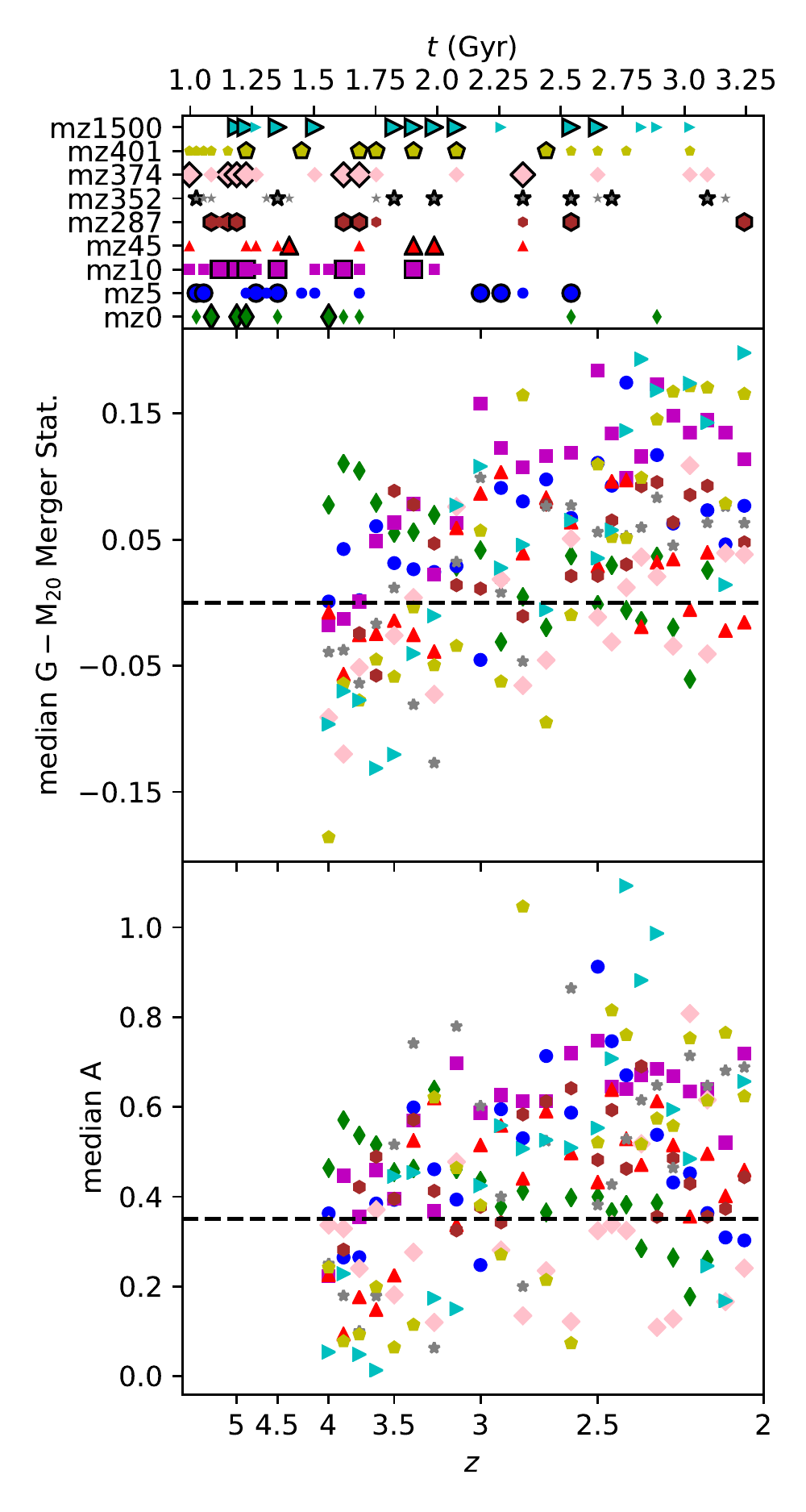}
% Give the caption for the Figure here. 
\caption{\label{figure:median_gm20_stat} Evolution of merger diagnostics across $z$ with indications of
  galaxy mergers. The upper panel illustrates mass increases
  indicative of galaxy merger. Larger markers with black outline
  correspond to a mass increase indicative of a major merger
  ($R\ge0.25$), whereas smaller markers without an outline indicate
  mass increases indicative of minor mergers ($0.10\le R <0.25$).  The
  middle panel shows the median $G-M_{20}$ merger statistic vs. $z$
  for the central galaxy in each simulation. Likewise, the bottom
  panel shows the median $A$ over $z$. The merger indicators were
  obtained by taking the median over all valid measurements at unique
  viewing angles at each $z$ for each simulation. The different
  combinations of shapes and colours represent data from different
  simulations.}
\end{figure}

In Figure~\ref{figure:median_gm20_stat}, we quantify these trends with
mass, and plot the evolution of the median $G-M_{20}$ merger
statistic, and $A$ with redshift.  The $G-M_{20}$ merger statistic
\citep[also referred to in the literature as ``mergyness'',
  e.g.][]{thompson15b} is defined as the perpendicular distance from
the canonical $G-M_{20}$ merger line (i.e. $G = -0.14M_{20}+0.33$),
with more positive values being further in the merger region.
Similarly, we plot the median $A$ value, noting the canonical $A=0.35$
line that defines mergers.  We show, at the top, the location of major
and minor mergers for each model.  Nearly all galaxies show more signs
of mergyness at lower redshift, though (as seen repeatedly now), more
massive galaxies move to the merger regions of the $G-M_{20}$ and
$C-A$ spaces earlier.

\begin{table*}
\caption{Descriptions of the $M_\ast$ galaxy bins. Specifically, 
we list the number of snapshots between $4\leq z <2$ where there is a merger and at least one measurement of a diagnostic, and the total number of snapshots for which we make at least one measurement of a diagnostic.}
\label{tab:smass_bins}
\begin{tabular}{lcccc}
\hline
$\rm{log}_{10}(M_\ast/M_\odot)$ Bins & $G-M_{20}$ mergers & $A$ mergers & $G-M_{20}$ measurements & $A$ measurements\\
\hline
$[7.33, 8.55)$ & 12 & 13 & 45 & 48\\
$[8.55,9.77)$ & 10 & 10 & 64 & 65\\
$[9.77,10.99]$ & 8 & 8 & 85 & 87\\
\hline
\end{tabular}
\end{table*}

\begin{figure*}
% Make sure the figure is centered:
\centering
\includegraphics[width = 6.9 in]{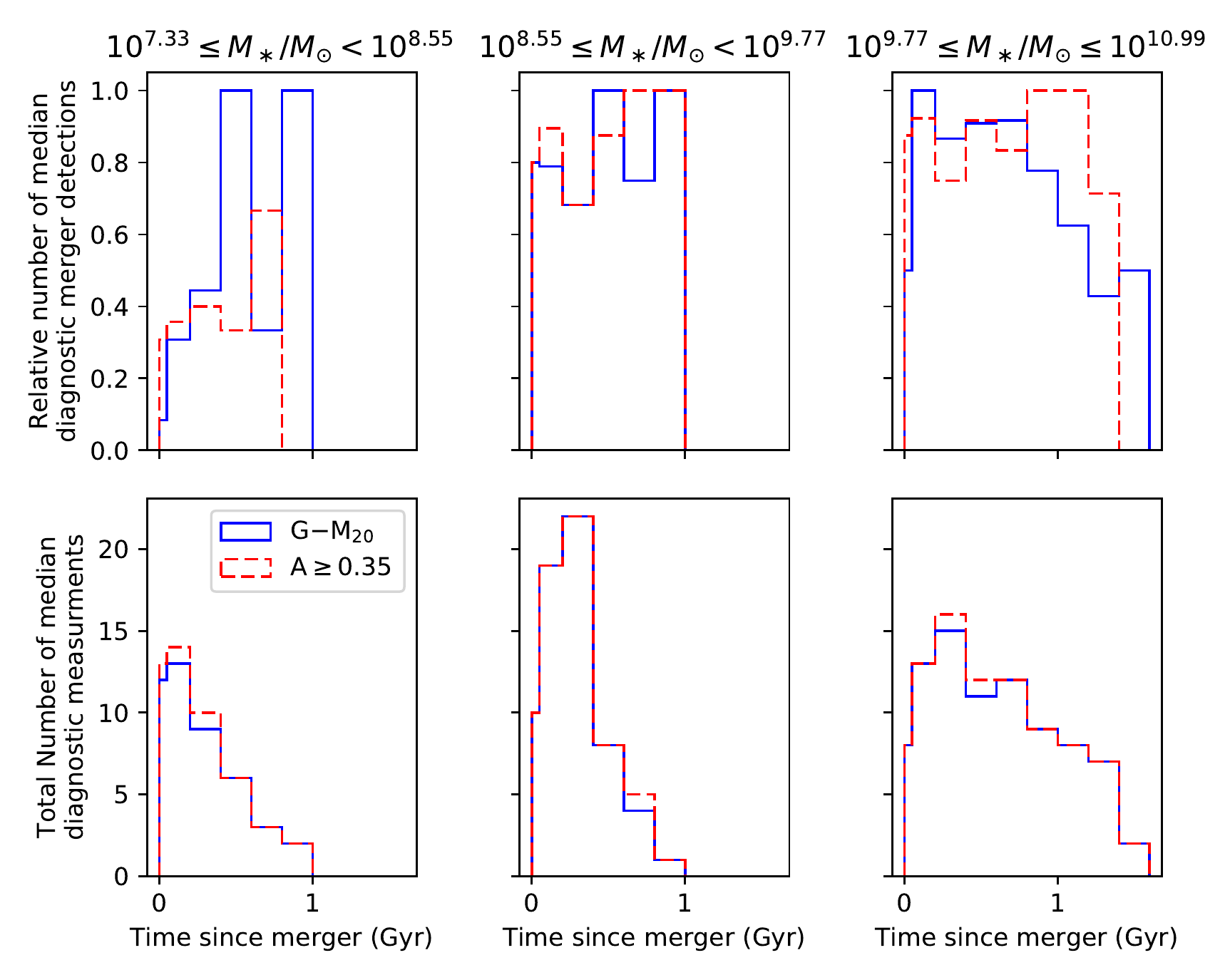}
% Give the caption for the Figure here. 
\caption{\label{figure:delta_t_hist} Fraction of median diagnostic 
  merger detections (top panel) and total number of median diagnostic
  measurements (bottom panel) as a function of $\Delta t$, or the 
  amount of time the diagnostic is measured after a mass increase 
  indicative of major merger ($R\ge 0.25$).  Our model galaxies are binned by stellar mass, with each column corresponding to a different mass bin.  $G-M_{20}$ is shown in solid 
  blue lines while $A\geq0.35$ is indicated by 
  dotted red lines. Each bin has a width of $0.2$ Gyr except for the first two bins which 
  extend from $0$ Gyr to $0.05$ Gyr and $0.05$ Gyr to $0.20$ Gyr. 
  Note that the minimum amount of time between snapshots is 
  $\approx0.0513$ Gyr and therefore the first bin only contains median 
  diagnostic measurements at $-\Delta t = 0$ Gyr.  Except for G-M20 with high smass galaxies, neither diagnostic's sensitivity to merger detections discriminates between a galaxy's proximity in time to a merger.}
\end{figure*}

%\begin{figure*}
%% Make sure the figure is centered:
%\centering
%\includegraphics[width = 6.9 in]{figures/delta_t_histogram_smass_A_200Myr.pdf}
% Give the caption for the Figure here. 
%\caption{\label{figure:delta_t_hist_A} Same as the top panel of 
%  \ref{figure:delta_t_hist} except it shows the performance of 
%  $A\geq 0.5$ and $A\geq0.65$. The absolute number of median diagnostic 
%  measurements are omitted from this figure because they are identical to 
%  the absolute number of $A\geq0.35$ measurements displayed in the bottom 
%  row of figure \ref{figure:delta_t_hist}.}
%\end{figure*}

\begin{figure*}
% Make sure the figure is centered:
\includegraphics[width = 6.9 in]{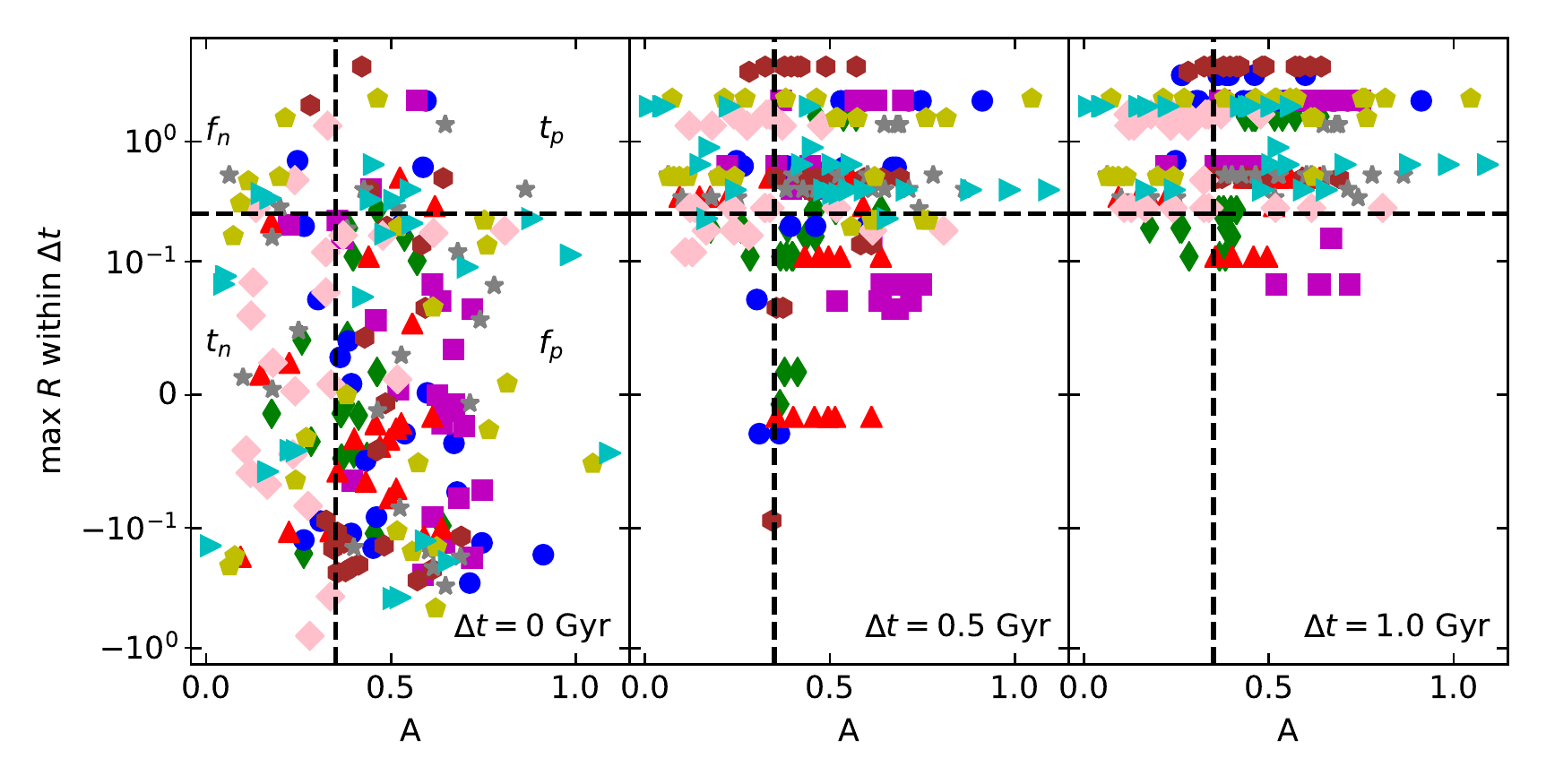}
% Give the caption for the Figure here. 
\caption{\label{figure:ratio_asym} The maximum baryonic mass 
  increase, $R$, at the time when the observation is measured 
  or any earlier time within $\Delta t$, as a function of the
  sightline-median $A$ for all simulations; $\Delta t$ 
  represents a potential timescale after a merger during which 
  $A\geq0.35$ might show heightened to detecting the merger. 
  Each panel is labelled with the employed value $\Delta t$ in 
  the bottom right corner. The grid on each panel illustrate 
  boundaries used to classify indications of galaxy mergers by 
  $A$, using a threshold of $A\geq0.35$ (the conventional 
  threshold at low-$z$), for major mergers ($R>0.25$), as true 
  positives ($t_p$, upper right), false positives ($f_p$, 
  lower right), false negatives ($f_n$, upper left) and true 
  negatives ($t_n$, lower left).}
\end{figure*}

\begin{figure}
% Make sure the figure is centered:
\includegraphics[width = 3.5 in]{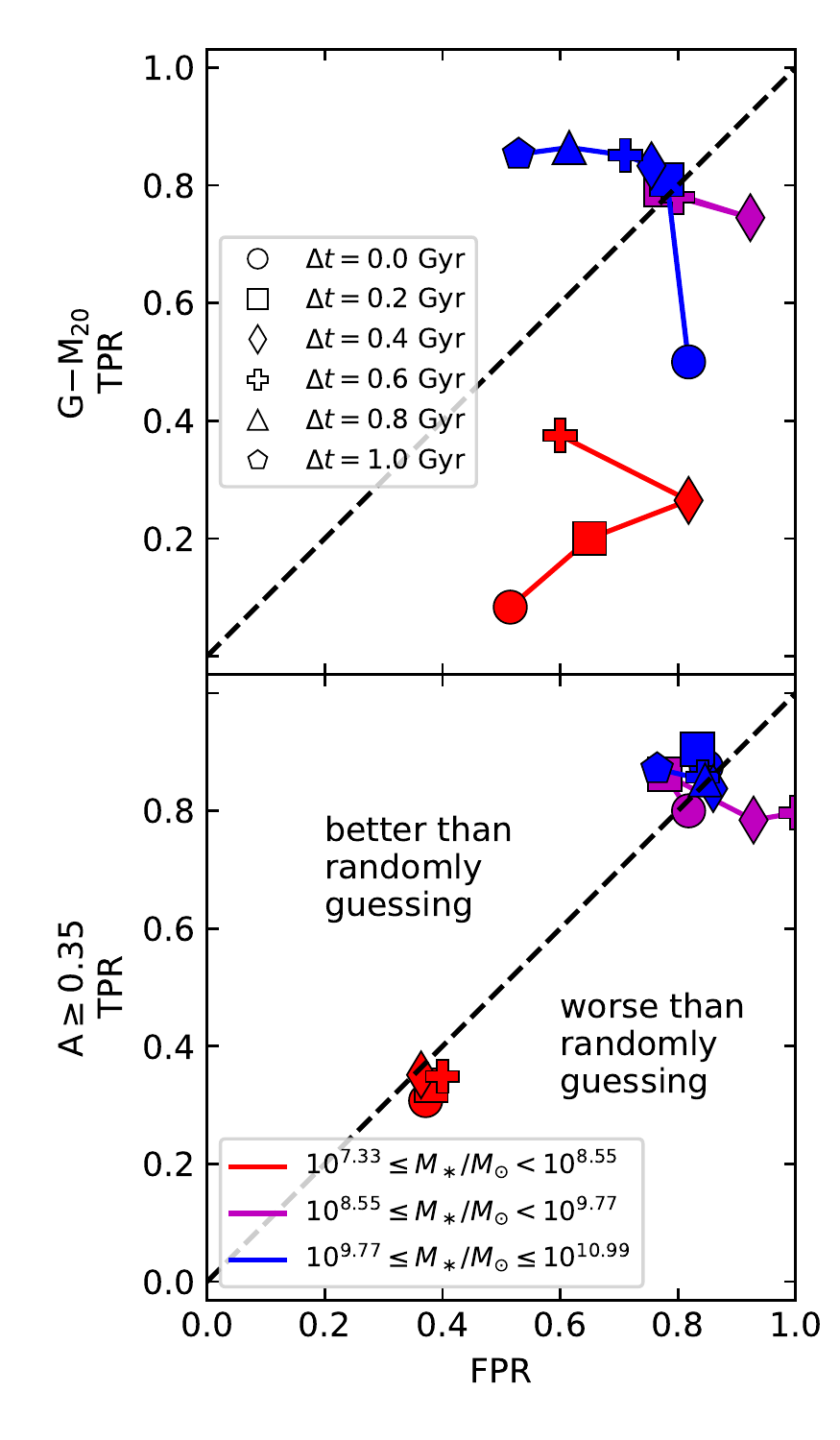}
\caption{\label{figure:ROC_major} Utility of $G-M_{20}$ and $C-A$ as
  major merger indicators for high-redshift galaxies. 
We show this by computing
  ROC (Receiver Operating Characteristic) curves, or the median True 
  Positive Rate (TPR) and average False Positive Rate (FPR) curve for 
  sightline-median values of $G-M_{20}$ (top panel) and $A\geq0.35$ 
  (bottom panel)for identifying major galaxy mergers as functions of 
  possible observability timescales $\Delta t$. The red, magenta, and 
  blue curves are constructed from observations of low, intermediate 
  and high \smass\ galaxies. Each point in a curve corresponds to the 
  FPR and TPR at a given $\Delta t$ which varies from 0 Gyr to 1.0 Gyr 
  in 0.2 Gyr increments. We omit values for the low and intermediate \smass\ bins
  for $\Delta t \geq 0.8$ due to the dearth of observations of 
  galaxies in those \smass\ bin more than 0.6 Gyr after the most 
  recent major merger. Note that the $G-M_{20}$ point for galaxies in 
  the intermediate \smass\ bin with $\Delta t=0$ Gyr is almost directly 
  below the point for galaxies in the intermediate \smass\ bin with 
  $\Delta t=0.2$ Gyr. In both panel, a black dashed 
  runs along TPR$=$FPR. At a given point on line, $(x,x)$, the point 
  represents the performance of randomly classifying $x$ per cent of 
  galaxies as undergoing a merger.  {\bf Except for the most massive 
  galaxies and the longest post merger time scales, $\Delta t$, 
  $G-M_{20}$ and $A$ tend to perform comparably to randomly guessing.}}
\end{figure}

\begin{figure}
% Make sure the figure is centered:
\includegraphics[width = 3.5 in]{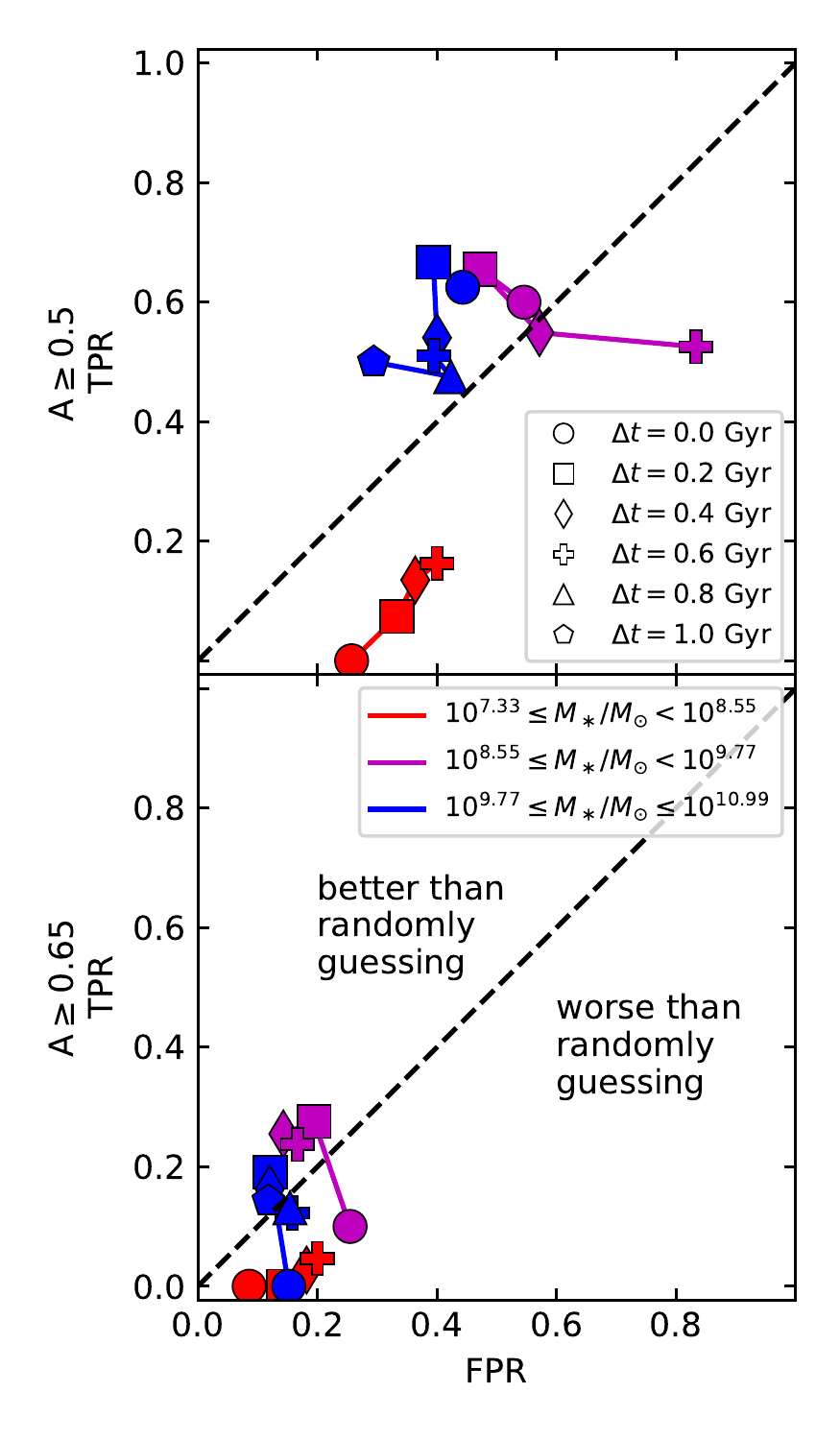}
\caption{\label{figure:ROC_major_A} Like Figure \ref{figure:ROC_major} 
  except it illustrates the utility of $A\geq0.5$ and $A\geq0.65$ as 
  merger indicators for high-redshift major galaxy mergers. Other than 
  $A\geq0.5$ for intermediate and high \smass\ galaxies at selective 
  time scales $\Delta t$, both $A$ criteria perform comparably to 
  randomly guessing.}
\end{figure}

%\begin{figure*}
% Make sure the figure is centered:
%\includegraphics[width = 6.9 in]{figures/all_sens_delay_time.pdf}
% Give the caption for the Figure here. 
%\caption{\label{figure:ind_sens_all} The same as Figure
%  \ref{figure:ind_sens}, but depicting the True Positive Rate and
%  False Positive Rate for all mergers ($R\ge0.10$).}
%\end{figure*}

\begin{figure*}
% Make sure the figure is centered:
\centering
\includegraphics[width = 6.9 in]{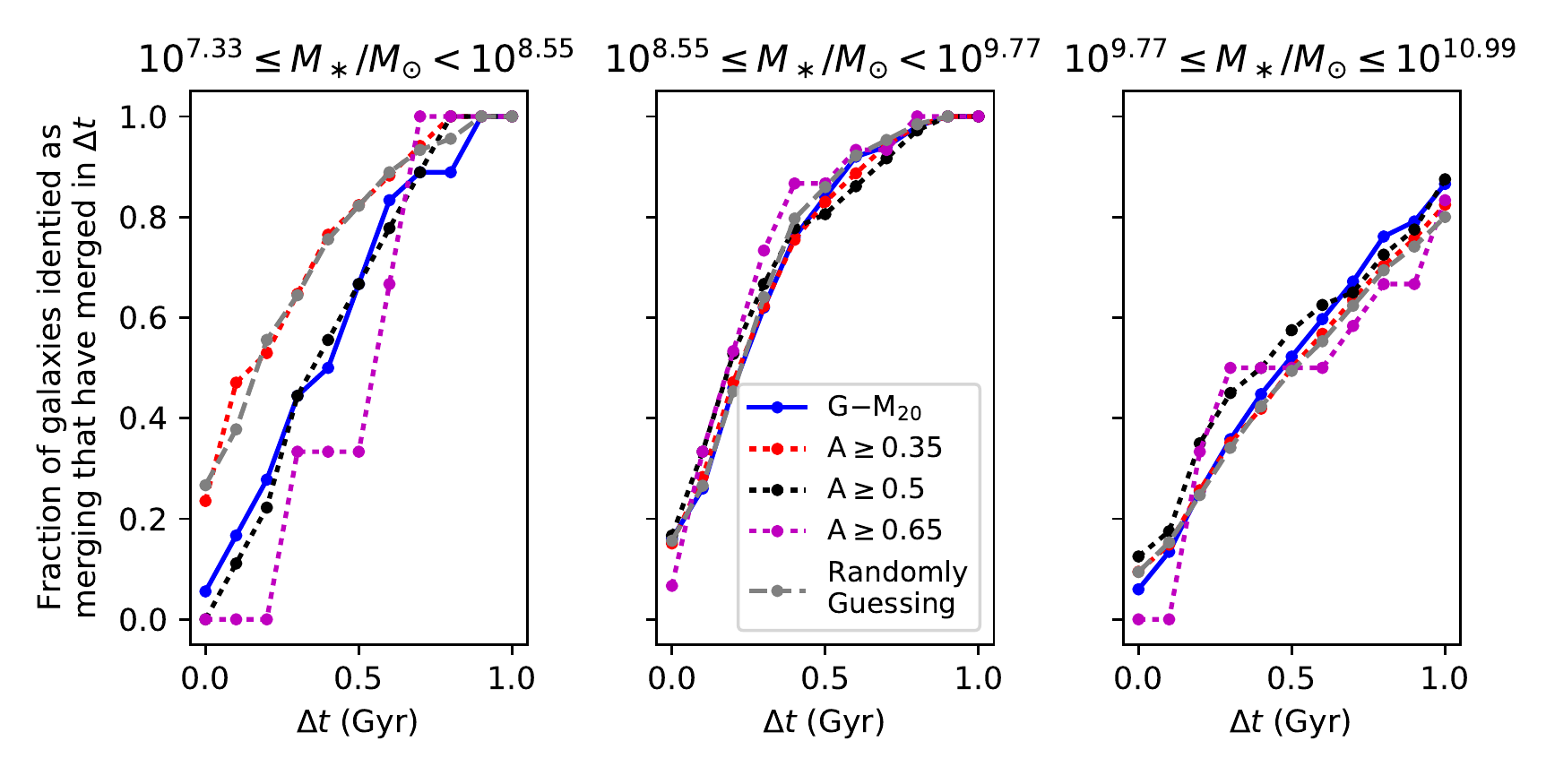}
% Give the caption for the Figure here. 
\caption{\label{figure:precision_array} Fraction of systems 
identified as mergers that are actually mergers as a function 
of $\Delta t = 0$, or time after a merger. The blue solid 
line represents data for $G-M_{20}$ using the standard 
merger criteria. The dotted lines 
represent data for $A$ while the different colours represent 
different thresholds; red represents $A\geq 0.35$ (standard), 
black represents $A\geq0.5$ and magenta represents 
$A\geq 0.65$. The grey dashed line shows the fraction of 
systems identified as mergers if one randomly classifies 
$x\%$ of all galaxies as merging such that $0<x<100$. 
Each panel uses data for observations in which the central 
galaxy lies in a different \smass\ bin. There is no case 
where any diagnostic performs appreciably better than 
randomly guessing.}
\end{figure*}

\begin{figure*}
% Make sure the figure is centered:
\centering
\includegraphics[width = 6.9 in]{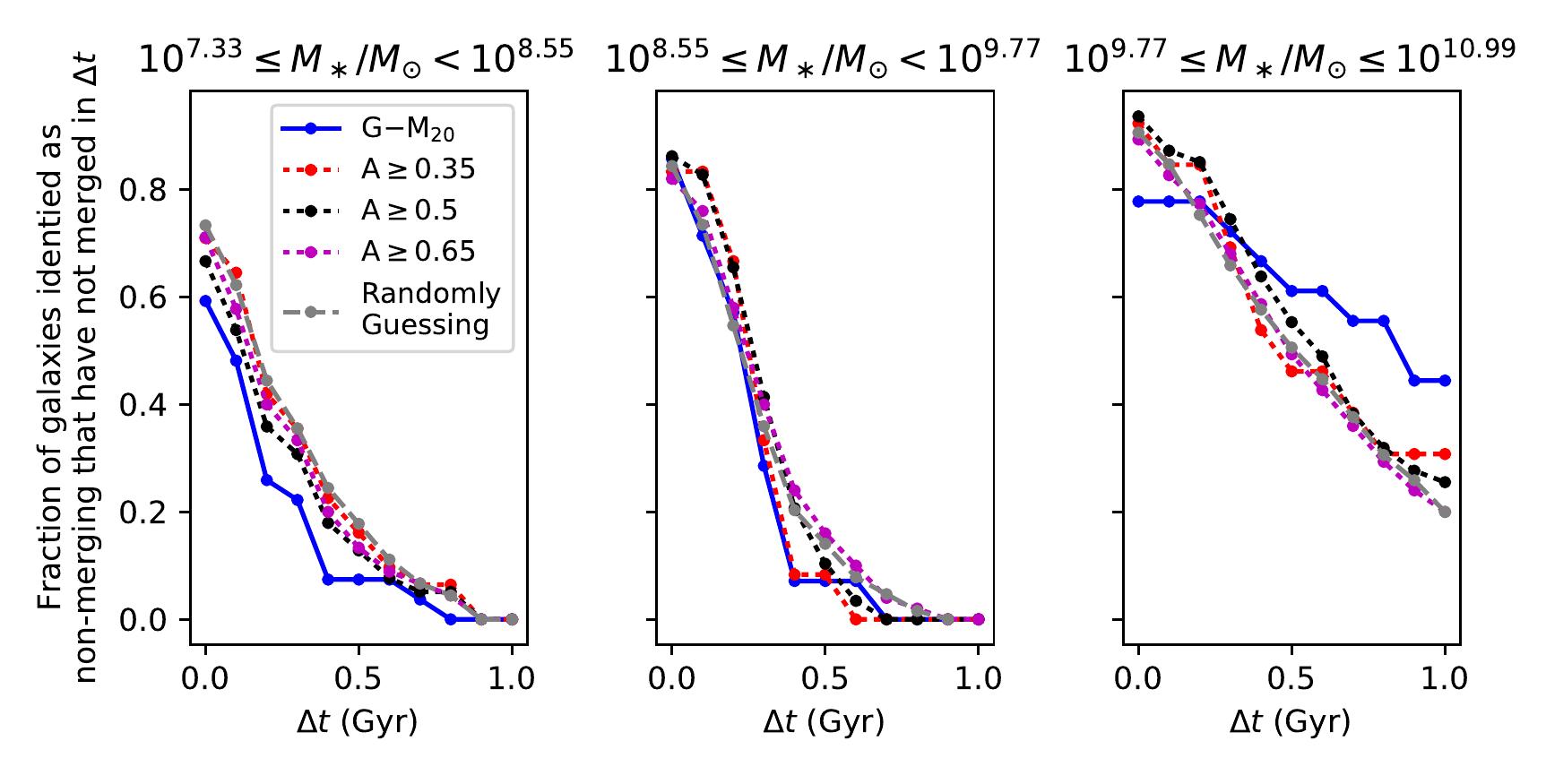}
% Give the caption for the Figure here. 
\caption{\label{figure:npv_array} NPV (negative predictive value) of 
diagnostics as functions of $\Delta t$, or time after a merger. 
Diagnostics and diagnostic thresholds are represented by the same 
combinations of colours and lines as in 
Figure~\ref{figure:precision_array}. The only instance where any 
diagnostic achieve considerably better performance than randomly 
guessing is $G-M_{20}$ for the galaxies in most massive \smass\ 
bin at a time scale $\Delta t$.}
\end{figure*}

\section{Discussion}
\label{section:discussion}
\subsection{Utility of $G-M_{20}$ and $A$ as diagnostics for 
mergers at $z\sim 2-4$} 
\label{section:utility_gm20_a}

We begin our discussion with an extended study into the utility of $G-M_{20}$ and $A$ in identifying 
galaxy mergers amongst massive galaxies at high-redshift.   Fundamentally, we want to answer:  how 
often are $G-M_{20}$ and $A$ correctly reporting mergers, and how often do they incorrectly identify 
a galaxy as merging when it is not?  Relatedly, we want to understand the 
fraction of time mergers may be missed by these metrics.  Throughout 
this discussion we will keep returning to the same fundamental point: the 
complex environments surrounding massive galaxies at high-redshift force 
$G-M_{20}$ and $A$ to be elevated for the bulk of the galaxy's life, thus 
rendering the metrics unable to capture mergers on short time scales.

To understand this, the first item we should address is: how often are our galaxies actually merging?  From the top panel of 
Figure~\ref{figure:median_gm20_stat}, it is clear that both major and minor mergers are relatively 
frequent in galaxies of all masses, except the most massive halos (mz0,mz5).   Quantitatively, every galaxy in our model mass 
range undergoes a merger \textit{at least} every $\sim$Gyr.  We show this exactly in the bottom panel
of Figure~\ref{figure:delta_t_hist}, where we bin our model galaxies into three mass bins, and plot the 
timescales after the last major merger for all of our model galaxies.   Only for the most massive bin 
(i.e. the rarest galaxies) are there a few systems that go $\sim 1.5$ Gyr between mergers.

Understanding the typical cadence of major mergers is important because it informs an ideal feature of 
non-parametric galaxy mergers for galaxies at high ($z \ga 2$) redshift:
they must be able to reliably detect mergers on time scales $t_{\rm merger} << 1 $ Gyr.  Once a galaxy 
approaches timescales $\sim 1$ Gyr since the last merger, it is almost guaranteed to undergo another 
merger again.  Any merger classifier that only works on time scales comparable to the time between 
mergers in effect is then just tracing the entire galaxy population, and is thus of limited use.
  
  \subsubsection{The Utility of $G-M_{20}$ and $A\geq0.35$: Post Merger Timescales}
  \label{section:utility_timescales}
  
We first aim to understand the utility of non-parametric morphological 
metrics in terms of the time scale ($\Delta t$) since a major merger.  
To do this, we return to Figure~\ref{figure:delta_t_hist}, where we 
now highlight the top row.   As a reminder, the galaxies are binned in 
three stellar mass bins whose bounds were chosen to chosen to distribute 
the snapshots where a mass increase indicating a merger as evenly as 
possible, with details of the bins in Table~\ref{tab:smass_bins}. We 
purposely omit 3 galaxies with \smass $\la10^{7.26}\ M_\odot$ because 
their inclusion would have made the distribution of galaxies more 
unequal. Note, unlike the previous section where we refer to the sizes 
of all galaxies in a simulation based on the mass of the central 
galaxy in the simulation at $z\sim2$, in this section we bin the 
galaxies based on the \smass\ it has at a given snapshot, entirely 
independent of which simulation it comes from.

The morphological statistics are sightline averages.  The top row illustrates the fraction of snapshots 
in a given time bin that register a merger diagnostic, while the bottom row shows the total number of 
galaxy $G-M_{20}$ and $A$ measurements in a given time bin.    Note, in the case of multiple mergers 
in rapid succession, we consider these as individual merger events.  For example, if a given galaxy has 
a merger at time $T=0$ Gyr, and then another one at $T=0.3$ Gyr, then we will consider these as two 
separate merger events.    The blue line denotes every time a galaxy would be classified as a merger 
via $G-M_{20}$, whereas the red line shows when it would be classified via $A\geq0.35$.

Two salient points arise from Figure~\ref{figure:delta_t_hist}.  First, at the lowest 
stellar masses, neither non-parametric method appears to work particularly well.  
Both $G-M_{20}$ and $A\geq0.35$ detect $\sim 20-100\%$ of the mergers during a given 
$\Delta t$ time bin, and these values are independent of the time since merger.  
As we will show quantitatively shortly, if a galaxy is merging, within $\sim 1$ Gyr, 
both methods work roughly as well as randomly guessing. This trend may be partially 
explained by the compact size of the lowest \smass\ galaxies; typically these galaxies 
consist of a small, bright nucleus surrounded by a relatively uniform brightness envelope 
of light. %This structure naturally lends itself towards having low $A$, $G$, and $M_{20}$. 
Several of the panels during $z\sim4-3.5$ in Figures~\ref{figure:mz352_stamps}-\ref{figure:mz1500_stamps}, illustrate how \caesar 's FOF finder frequently associates 
under-dense outlying regions of stellar density with these galaxies, which do not appear in 
the visual morphology. Mergers may be % (certainly not the case for all mergers)
detected when the outlying regions of the merging galaxies overlap, but before the 
densest regions interact. In other words, there is a delay between the merger 
and morphological disturbance. Once the main galaxies interact, their compact size 
cause them to coalesce relatively quickly. Because of the sensitivity of the 
diagnostics to small irregularities, minor mergers could also lead to detections of 
mergers. Additionally, this trend may also be partially explained by the incompleteness of our sample of low \smass\ galaxies.

Second, in the intermediate and larger mass bins (\smass $>10^{8.5}$), the relative number of median diagnostic merger detections stays uniformly large ($\ga
80\%$) during the entire $\Delta t = 1$ Gyr timescale.  This is despite the fact that the number of 
galaxies with large time lags since the last merger drops dramatically as $\Delta t\rightarrow 1$ Gyr. What this means is that both $G-M_{20}$ and $A\geq$ register nearly all mergers up to $\sim 1 $ Gyr after the merger.   At the same time, nearly every galaxy undergoes mergers on time scales $t < 1$ Gyr (bottom row of Figure~\ref{figure:delta_t_hist}).  In other words, effectively {\it all galaxies at all times are within $\Delta t = 1 $ Gyr after a merger}, and neither method is able to discriminate on the time since a merger.  Therefore, that $G-M_{20}$ and $A\geq0.35$ nearly always register galaxies as mergers within $1 $ Gyr means that they in effect are simply tracing all massive galaxies at high-redshift.  This suggests a limited utility in these metrics.

Why do massive galaxies register as mergers for the majority of their lives? 
As discussed previously (c.f. \S\ref{section:evo}) this is due
to the fact that more massive galaxies have larger optical morphologies.   When galaxies have larger 
optical morphologies, the source extraction algorithm is less likely
to separately identify the central galaxy and satellites.  As a
result, the final segmentation maps are less likely to mask out
infalling galaxies, and therefore include both merging galaxies.  Both
$G-M_{20}$ and $A\geq0.35$ therefore register the merger more easily in more
massive systems.  Concurrently, more massive halos tend to have
multiple ongoing mergers, and therefore are more likely to register as
a merger via $G-M_{20}$ and $A\geq0.35$ over multiple sightlines.

\subsubsection{The Utility of $G-M_{20}$ and $A\geq0.35$: True and False Positive Rates}
\label{section:utility_tprfpr}
In the previous section, we demonstrated that both $G-M_{20}$ and $A\geq0.35$ register mergers for significant periods of time following an actual merger event (up to $\sim 1$ Gyr).   We now quantify the expected true and false positive rates of these morphological statistics.

We define two quantities. The True Positive Rate (TPR) 
is the fraction of all mergers that are detected by either $G-M_{20}$ or 
$A\geq0.35$ within $\Delta t$.  The False Positive Rate (FPR) is the fraction of
snapshots within $\Delta t$ that are incorrectly classified as mergers
by the non-parametric measures. In other words:
\begin{eqnarray}
  \rm {TPR} = \frac{t_p}{t_p+f_n}\\
  \rm {FPR} = \frac{f_p}{f_p+t_n}
\end{eqnarray}
where $t_p$ and $f_p$ are the number of true and false positives,
respectively, and $t_n$ and $f_n$ are the number of true and false
negatives, respectively.  Note that $\mathrm{TPR+FPR}$ does not
necessarily equal 1.  The TPR answers the question, "if a system is 
a merger, how often does a statistic identify it as a merger 
(within a time $\Delta t$)?"

We compute both fractions, for major mergers identified by 
median $G-M_{20}$ and $A$ taken over individual sightlines 
of each snapshot, separately, using different values of 
$\Delta t$. It is important to note that each value of 
$\Delta t$ fundamentally changes the definitions of what we 
consider a $t_p$, $f_p$, $t_n$ and $f_n$. For example, if we 
increase $\Delta t$, we expect a diagnostic to detect a 
merger in a larger fraction of our mock observations, thereby 
converting some $f_p$ into $t_p$, and some $t_n$ into $f_n$.

To compute the number of true and false positives, we define $D_i (t)$ as the 
relative fraction of galaxies, in \smass\ bin $i$, identified as mergers 
$\Delta t$ Gyr after a merger. We also define $G_i(t)$ as the 
distribution of galaxies in \smass\ bin $i$ with the respect to the 
amount of time to the most recent major merger. 

We can then compute the number of true and false positives directly from 
$G_i(t)$ and $D_i(t)$. Assuming that one observes 
$N$ galaxies in \smass\ bin $i$ and define the merger time scale 
as $\tau$, then the $t_p$ and $f_p$ are given by
\begin{eqnarray}
  t_p = N\int_0^\tau G_i(t) D_i(t) dt\\
  f_p = N\int_\tau^\infty G_i(t) D_i(t) dt
\end{eqnarray}
The equations for $f_n$ and $t_n$ are found by replacing 
$D_i(t)$ with $(1-D_i(t))$ in the equations for $t_p$ and 
$f_p$, respectively.

To build some intuition, in Figure~\ref{figure:ratio_asym}, we 
show the distribution of our model galaxies that have true 
positive, true negative, false positive and false negative 
measurements.  We only consider galaxies that undergo major 
mergers for simplicity, and for illustrative purposes, couch 
this in terms of $A\geq0.35$ measurements. In 
Figure~\ref{figure:ratio_asym}, each panel represents a 
different $\Delta t$ value\footnote{Note that this figure was 
only constructed with data back to $z\sim5.6$, and 
consequently there is not a full Gyr of data before every 
observation; this effect is insignificant as there is a major 
merger in every simulation during $z=5.6-4$.}. The ideal 
scenario is to have most 
of the model galaxy points in the top right (true positive; 
$t_p$) or bottom left (true negative; $t_n$) quadrants. 
Figure~\ref{figure:ratio_asym} clearly exhibits the expected 
trend we previously mentioned that as we increase $\Delta t$ 
(i.e. as we move toward right most panel), galaxies tend to 
be converted from $f_p$ into $t_p$, and $t_n$ into $f_n$ (i.e.
they move to the top of each panel towards the $t_p$ and 
$f_n$ quadrants).

Figure~\ref{figure:ratio_asym} also demonstrates that $A$ does not clearly correlate in any way with the underlying merger ratio.  If $A$ represented some measure of whether an object has recently undergone a merger, one might expect that it would be positively correlated with $R$.  However, for no $\Delta t$ does such a correlation appear.  Although we do not show it here, a similar lack of correlation is seen for the $G-M_{20}$ merger statistic with $R$.  This foreshadows that identifying mergers with quantitative morphological measures will not be easy.  We next quantify this more precisely.

% I DON'T NEED TO CITE ROC as a name - I recently saw the same name 
% used in the DES DR1 paper without citation
We now use the $t_p$ and $f_p$ metrics to ask, over what time scales are 
$G-M_{20}$ and $A\geq0.35$ likely to produce true and false positives for a given 
mass bin.  In Figure~\ref{figure:ROC_major}, we show this via a Receiver 
Operating Characteristic Curve.  In the ROC, the axes are TPR and FPR, and 
the 1:1 line is plotted.  The 1:1 line essentially signifies the division 
at which randomly guessing if a galaxy is merging would be better:   above 
this line, the non-parametric measure performs better than randomly 
guessing.  A point along that line, $(x,x)$ indicates the 
performance randomly classifying $x$ per cent of galaxies as 
undergoing a merger.   For example, if a diagnostic has a TPR of $0.75$ and an FPR of $0.75$, the exact same result can be achieved by randomly classifying 75\% of all observed galaxies as undergoing a merger.
Within the ROC curves, we show three mass bins (signified by the 
three different colours), as well as six different time 
scales following a merger event.

In short, The ROC curves in Figure~\ref{figure:ROC_major} 
demonstrate that at best, for most masses and most time 
scales $\Delta t$, $G-M_{20}$ and $A \geq 0.35$ perform 
comparably to randomly guessing.   Only for the most massive 
time bin, and for the longest time scales do these metrics 
beat guessing.  This is evident in 
Figure~\ref{figure:delta_t_hist}, where for the most massive 
time bin, the likelihood of $G-M_{20}$ detecting a merger a 
significant time ($\sim 1$ Gyr) after a merger is relatively 
small.  An elevated asymmetry cut of $A\geq0.5$ performs a bit 
better than $A\geq0.5$ for massive systems, as does $A\geq0.65$. 

While it is not shown here, we also examined the effectiveness of 
the statistics by combining $\Delta t$ with $-\Delta t$ values of 
0.1 Gyr and 0.2 Gyr (The inclusion of $-\Delta t$ in the calculations 
allowed us to ask if a given non-parametric morphology statistic 
can identify mergers {\it prior} to the snapshot where the stellar 
mass increase indicating a merger is measured). We ultimately found 
that this had neutral or negative effects on each of the ROC curve 
and have thus omitted it from our figures.

%In Figure~\ref{figure:ROC_major}, $G-M_{20}$ for the most massive 
%\smass\ bin starts below the 1-to-1 line and increase vertically 
%with increasing $\Delta t$. For $\Delta t\sim 0.2 - 0.5$ the value 
%hovers around the 1-to-1 line and then it moves leftward with 
%increasing $\Delta t$. $G-M_{20}$'s behaviour for the highest mass 
%\smass\ bin is the only case in figure~\ref{figure:ROC_major} shows 
%diagnostic perform considerably better than randomly guessing. At 
%$0.2\la \Delta t\la 1.0$ Gyr $G-M_{20}$ correctly identifies 
%$\sim 80$ per cent of mergers and incorrectly identifies 
%$\sim 50 - 80$ per cent of non-merging galaxies as merging. Despite, 
%this high degree of confusion at large enough $\Delta t$, $G-M_{20}$ 
%performs better than guessing. It's performance is maximized at 
%$\Delta t\sim 1.0$ Gyr. Figure ~\ref{figure:ROC_major} shows that 
%$G-M_{20}$ for all other \smass\ bins and $A\geq 0.35$ for all 
%\smass\ bins each perform comparably to or worse than randomly 
%guessing.

At face value, then,
Figure~\ref{figure:ROC_major} suggests
that the usage of quantitative morphology measures at high-redshift
(especially in the intermediate to low \smass\ regime) is complicated 
by the significant false positive rates in comparison to the true 
positive rates; with the exception of $G-M_{20}$ for the largest 
\smass\ bin large at large $\Delta t$, the diagnostics are unable to 
discriminate between merging and non-merging galaxies.  This is simply a recasting of the results seen in Figure~\ref{figure:delta_t_hist} -- on average, except for the most massive galaxies and longest time scales, one may as well randomly guess if a galaxy is merging or not.

It is worth noting, however, that there is a significant uncertainty
in our calculated false positive rate.  As a reminder, we have
discarded all sightlines that either have poor SNR ($\left<SNR\right>
< 20$), or non-contiguous segmentation maps.  In particular, while
relatively few sightlines for $C-A$ data are discarded, $G-M_{20}$
values that come from rather abnormal morphologies can sometimes
result in either background pixels that lower the signal to noise
ratio, or non-contiguous segmentation maps.  As a result, $G-M_{20}$
values from observations following a merger and observation of very 
compact galaxies are particularly likely to be discarded, thus 
lowering the total number of true positives
registered in Figure~\ref{figure:ROC_major}. 

In Appendix
\ref{section:optimisic_tpr/fpr} we illustrate the uncertainty
associated with our measurements by regenerating Figure
\ref{figure:ROC_major} and at every timescale, 
assume that all of the discarded measurements have values that are 
the most optimistic for each diagnostic's performance.  Figure~\ref{figure:ROC_major_opt} demonstrates that despite this, the results for $A>0.35$ are unchanged.   The results for $G-M_{20}$, however, can be greatly improved by including sightlines that have non-contiguous segmentation maps or low SNR detections.  We suggest, however, that this is a somewhat unrealistic scenario as it requires including data that would otherwise be discarded for quality issues.  In summary, $G-M_{20}$ performs best at diagnosing merging large \smass\ galaxies at 
$\Delta t\sim1.0$ Gyr. The is the only case in which a conventional diagnostic appears 
to definitively work significantly better than randomly guessing. $A$ works comparably to randomly guessing for most masses and post-merger timescales.

%We conclude, therefore, that $G-M_{20}$ has a very uncertain true positive %rate 
%and false positive rate for galaxies smallest \smass\ bin, which are $\sim %10-80$
%per cent and $\sim 25-70$ per cent. For the other two \smass\ bins, there %is less 
%uncertainty in the true positive rates of $G-M_{20}$. Considering our %uncertainties, 
%for intermediate \smass\ galaxies, $G-M_{20}$ correctly identifies $\sim %80-90$ per cent 
%of mergers and incorrectly identifies $\sim25-90$ per cent of non-merging %galaxies. 
%though has a false positive rate that is somewhat uncertain. Finally, for %the highest 
%\smass\ galaxies, at the optimal timescale, $+\Delta t\sim1.0$ Gyr, $G-%M_{20}$ 
%has a true positive rate of $\sim 80$ per cent and false positive rate of %$\sim 40 - 50$.

\subsubsection{Utility of alternative $A$ criteria}\label{section:other_A_thresh}

While examining the full $G-M_{20}$ space for potentially better merger criteria is 
outside the scope of this work, we briefly examine a small number of more 
stringent Asymmetry criteria in Figure~\ref{figure:ROC_major_A}. In 
Figure~\ref{figure:ROC_major_A} we illustrate ROC curves for $A\geq0.5$ and 
$A\geq0.65$ at a variety of $\Delta t$ for each of the three \smass\ bins.

Essentially, the ROC curves in Figure~\ref{figure:ROC_major_A} illustrates 
how for most time scales and masses both $A$ criteria performs comparably 
to randomly guessing, at best. The only cases in which either criteria is 
substantially superior to randomly guessing is $A\geq0.5$ for intermediate 
to most massive \smass\ bin with a timescale $\Delta t\sim 0.2$ Gyr.

Similar to $G-M_{20}$ for high \smass\ galaxies with $\Delta t=1.0$ Gyr, 
$A\geq0.5$ for both intermediate to high \smass\ galaxies with 
$\Delta t = 0.2$ Gyr, has a relatively high probability of detecting mergers 
at immediately after a galaxy merges that decreases substantially around its 
optimal $\Delta t$. However, for galaxies in the intermediate (high) \smass\ 
bin, $A\geq0.5$ shows a spike in the probability of detecting mergers greater 
than (comparable to) the probability of detecting mergers less than 0.2 Gyr 
after the merger. This spike occurs for galaxies that have not merged in 
0.6-1.0 Gyr (0.6-0.8 Gyr after which it decreases again). Because this spike 
in probability occurs in time bins that contain $\sim17\%$ ($\sim14\%$) of 
all galaxies in a given bin that have not merged in at least 0.2 Gyr, the 
average probability of $A\geq0.5$ identifying a galaxy less than 0.2 Gyr 
after a merger is greater than that for a galaxy more than 0.2 Gyr after a 
merger.  

\subsection{Questions of Interest} 
\label{section:targeted_questions}
Thus far, we have discussed the performance of $G-M_{20}$, $A\geq0.35$, 
$A\geq0.5$, $A\geq0.65$ as merger diagnostics. We have concluded that there 
are only two cases where a diagnostic performs substantially better than 
randomly guessing:
\begin{enumerate}
	\item $G-M_{20}$ on timescales of $\Delta t\sim 1$ Gyr for 
    high \smass\ galaxies
    \item $A\geq0.5$ on timescales of $\Delta t\sim 0.2$ Gyr for 
    intermediate to high \smass\ galaxies.
\end{enumerate}

We will use the results we have developed so far to answer questions 
targeted questions that may be of use to observational surveys.

\subsubsection{Is my observed galaxy undergoing a merger right now?}
What is the probability that any given observed galaxy is undergoing a galaxy merger, based on the results of $G-M_{20}$ or $A$ diagnostics?
To answer this, we measure the probability 
that a galaxy with a merger diagnostic indicating a merger comes from 
a snapshot where $R\geq0.25$ (we effectively measuring the probability 
for $D_i(t=0)$).

The specific metric with which we can measure this probability is 
called PPV (positive predictive value), or precision, which 
is given by
\begin{equation}
\rm{PPV}= \frac{t_p}{t_p+f_p}.
\end{equation}
In other words, PPV measures the ratio of true positives to all positive signals.  A PPV = $100\%$ is ideal.   In Figure~\ref{figure:precision_array}, we plot the PPV as a 
function of $\Delta t$ for each \smass\ bin using the median diagnostic 
value taken over all unique lines of sight for a given snapshot. 
To answer this question we only care about the values of PPV when 
$\Delta t = 0$.

The values of PPV are sensitive to the distribution of galaxies as a 
function in time after the merger. To illustrate the effects of the 
underlying distribution of galaxies on the PPV, we include a curve in 
Figure~\ref{figure:precision_array} that shows the PPV for randomly 
identifying galaxies with probability $x$. One can show, that as 
long as $x$ satisfied $0<x<1.0$ its PPV is entirely determined by 
the underlying distribution of galaxies.

From Figure~\ref{figure:precision_array}, we find that each of the $A$ 
thresholds and $G-M_{20}$ have PPVs ranging from $0$ to $\sim 0.20$ at 
$\Delta t=0$. In other words there is less than a 20 per cent chance 
that a galaxy is undergoing a merger right now. In nearly all cases, 
one can identify a larger or comparable fraction of merging galaxies 
by randomly guessing.

\subsubsection{Has my observed galaxy undergone a merger within a characteristic timescale, $\Delta t$?}

We again turn to Figure~\ref{figure:precision_array} to 
determine this answer.  We see that as $\Delta t$ increases, 
so does the precision of each diagnostic.  While there is more 
dispersion in the PPV of the diagnostic at lower $\Delta t$, 
the precision appears to converge at $\geq0.6$ Gyr. 
%As 
%stated previously for the ROC curves, the PPV data for the intermediate 
%\smass\ bin must be treated with skepticism at $\Delta t \ga 0.6$. 
That said, it is important to note that the upward trend with 
$\Delta t$ is expected. As we increase the $\Delta t$, 
$f_p$ are converted into $t_p$ (e.g. Figure~\ref{figure:ROC_major}). 
Therefore, as $\Delta t$ 
is increased the precision can only get better or remain the same.

We find that for galaxies in the largest \smass\ bin, about 
$85$ per cent of galaxies identified as merging by  
$G-M_{20}$ have merged within a $\Delta t \sim 1.0$ Gyr. For 
galaxies in the intermediate(high) \smass\ 
bin, about $55$ ($35$) per cent of galaxies with $A\geq 0.5$ have 
merged within $\sim0.2$ Gyr.

\subsubsection{Is a galaxy identified as non-merging actually not merging?}

This question is quantitatively answered by NPV (negative predictive 
value) which is defined as 
\begin{equation}
\rm{NPV}= \frac{t_n}{t_n+f_n}.
\end{equation}
In other words, NPV is the fraction of all negative signals 
(i.e. when $G-M_{20}$ and $A$ diagnostics say that a galaxy is 
not merging) that are truly not merging.  Figure~\ref{figure:npv_array} 
illustrates NPV as a function of $\Delta t$ since the most 
recent major galaxy for each \smass\ bin. It is critical to 
consider NPV alongside PPV. Like PPV, NPV is also highly dependent 
on the underlying distribution of galaxies with time after a merger.
To illustrate this dependence, we include the NPV in 
Figure~\ref{figure:npv_array} for randomly identifying $x$ per cent 
of galaxies as merging, such that $0<x<100$.

From Figure~\ref{figure:npv_array}, we find that NPV decreases 
as $\Delta t$ increases. This is expected because as we 
increase $\Delta t$, $t_n$ are converted into $f_n$ 
(e.g. Figure~\ref{figure:ROC_major}; consequentially NPV can 
only get worse or remain the same. Additionally, 
Figure~\ref{figure:npv_array} shows that within a  
\smass\ bin, all of the $A$ merger criteria evolve similarly. 
We also find while $G-M_{20}$ evolves relatively similarly to 
$A$ in the low and intermediate \smass\ bin, in the most 
massive \smass\ bin, the NPV of $G-M_{20}$ decreases with a 
much shallower slope that that of $A$. 

Figure~\ref{figure:npv_array} indicates that $\sim45$ per cent of 
galaxies in the most massive \smass\ bin identified by $G-M_{20}$ 
as non-merger have actually merged in the last Gyr. The figure 
also demonstrates how $\sim65$ ($\sim 85$) per cent of galaxies 
in the intermediate (high) \smass\ bin with $A<0.5$ have not merged 
within the previous $\sim 0.2$ Gyr.

Considering both the PPV and NPV together, we conclude that $A>0.5$ 
is not a particularly useful diagnostic for identifying galaxy 
mergers at high $z$. We find that $A>0.5$ does not achieve considerably 
better performance than randomly guessing in terms of PPV and 
NPV for galaxies in both the intermediate and highest \smass\ 
bins. The fact that $A\geq0.5$ has a larger NPV than PPV suggests 
that it may be used to remove galaxies from a sample that have not 
merged within the last 0.2 Gyr. Then in principle, other methods could 
be employed to identify which of the remaining sample of galaxies 
actually merged. Unfortunately, $A\geq0.5$'s utility in screening out 
galaxies that have not merged in 0.2 Gyr is limited by its TPR of 
$\sim0.65$ (see Figure~\ref{figure:ROC_major_A}); about $35\%$ of 
galaxies that merged in the last 0.2 Gyr would be screened out in the 
process.

On the other hand, for the largest \smass\ galaxies at the longest 
time scale, $\Delta t=1.0$ Gyr, $G-M_{20}$ has a NPV approximately 
double that of randomly guessing and a PPV comparable to that of 
randomly guessing. Due to its long time scale of activity, the 
main use of $G-M_{20}$ would be to identify high \smass\ 
galaxies that have not merged in the last Gyr. The absolute value 
$G-M_{20}$'s NPV ($\sim45\%$) considerably limits the utility of 
$G-M_{20}$ (despite being twice as large as that of guessing); more 
than half of all high \smass\ galaxies that do not meet the $G-M_{20}$ 
merger criterion have undergone a merger in the last Gyr. While an 
argument could be made for using $G-M_{20}$ to screen out merging 
galaxies, this is not particularly relevant to how well $G-M_{20}$ 
identifies mergers.

\subsection{Comparison to other Theoretical Studies}

In the last decade, several authors have sought to utilize numerical
simulations to study quantitative morphology measures.  These studies
span a diverse range of methods, ranging from (i) studying the
morphologies idealised galaxy merger simulations
\citep[e.g.][]{lotz08b,lotz10a, lotz10b}; (ii) studying the
morphological measures in of galaxies from a collection of
cosmological zoom simulations \citep[e.g.][]{hambleton11a,snyder15a};
and (iii) studying the morphological measures for a statistically
large sample of mock observations generated from coarser resolution
large-box cosmological simulations
\citep[e.g.][]{snyder15b,bignone17a}. In all studies from the first
two categories, the mock observations were generated with full Monte
Carlo dust radiative transfer simulations \citep{lotz08b,lotz10a,
  lotz10b, snyder15a}, while studies in the third classification use
mock observations from the Illustris Project \citep{snyder15b,
  bignone17a}, which were generated with radiative transfer
simulations that omit the effects of dust \citep{torrey15a}
emission. Because our work falls into the second category and has many
similarities to works in the first category, we primarily focus our
comparison works in these categories.

\citet{lotz08b} were the first to study morphological measures in
numerical galaxy formation simulations. These authours utilised a
combination of {\sc Sunrise} dust radiative transfer
\citep{Jonsson2006,Jonsson2010a,Jonsson2010b} with idealised galaxy
merger simulations in order to generate mock observations of galaxies.
Lotz et al.  studied the dependence of $G-M_{20}$, $CAS$ on a variety
of factors for equal mass binary mergers including merger orbital 
parameters and orientation, viewing angle, dust, image resolution, gas
fraction, scale length, and different models of supernova feedback. In
\citet{lotz10a} and \citet{lotz10b}, the same group employed similar
methods to study the dependence of morphological measures on the mass
ratio and gas fractions of the merging galaxies, respectively. In each
of these papers they analysed the average observability timescales on 
which $G-M_{20}$ and $A$ identify mergers of local galaxies.  Their 
observability time scale measures the line-of-sight averaged total 
amount of time that a diagnostic indicates disturbed morphologies 
for {\bf both} pre-merging galaxies and the post-merger system; 
our merger time scale, $\Delta t$ measures a fudamentally different 
quantity. While a direct comparison of our models to theirs is not 
straightforward, owing to the messy complex environments that 
surround high-$z$ massive galaxies in
cosmological simulations (as compared to the relatively cleaner
environments of idealised binary mergers), in Appendix
\ref{section:idealized_comparison}, we apply our methods to an
idealised binary galaxy merger simulation and find comparable results
to these previous works. 

%\citet{Lotz2011} summarizes that the $G-M_{20}$ timescales are 
%somewhat dependent on the mass ratios and 
%weakly correlated with gas fractions, whereas $A$ is dependent on 
%both mass ratios and gas fractions. Furthermore, \citet{Lotz2011} 
%concludes that $G-M_{20}$ can detect at least 1:10 galaxy mergers and 
%$A$ can detect at least 1:4 mergers when the combined gas fraction of 
%both merging is 0.2 and at least 1:10 mergers for higher gas fractions.
%At the lowest combined gas fraction they simulated, 0.2, both $G-M_{20}$ 
%and $A$ have similar average timescales of below 500 Myr for all merger 
%ratios \citep{Lotz2010b}. At higher combined gas fractions, 0.4 and 
%$\sim0.5$, $G-M_{20}$ has a relatively constant timescale while $A$ 
%has timescales of 500 Myr and larger \citep{Lotz2010b}. 

\citet{hambleton11a} studied the $CAS$ morphology of a collection of
$z=0$ simulated galaxies in zoom simulations, and compared the
morphology trends to that of the \citet{Frei1996} catalogue of local
galaxies. Using \sunrise, they generated mock observations at $z=0$
for a collection of 15 galaxies simulated with
\gasoline\ \citep{Wadsley2004} cosmological zoom simulations. This
collection contained 12 galaxies with masses similar to that of the
Milky Way and 3 galaxies with masses similar to that of the masses of
dwarf Galaxies.  These authours, concentrating
on understanding the $C-A$ diagnostic, found galaxies that typically
found comparable concentration indices as observed local galaxies,
though model galaxies that have significantly higher $A$ values.  

% Hambleton made observations for 6 galaxies along 5 lines of sight
% and observations along 2 or 3 lines of sight for the remaining 
% galaxies.

\citet{snyder15a} used \sunrise\ to generate mock observations of galaxies, 
from \art\ \citep{Kravtsov1997,Kravtsov2003} cosmological zoom 
simulations, that closely mimic the resolution, depth, filters, and 
noise of the \hst\ observations from the CANDELS-Wide survey 
\citep{Grogin2011}. Unlike in our analysis where we study the 
quantitative morphological measures computed from the 
observations in the rest-frame $B$ filter with $\left<SNR\right>\ge 20$ 
along 13 lines of sight, they examined the morphologies in the closest 
\hst\ filter to the rest-frame $B$ filter with a selection criteria 
of $H<24.5$ along 5 lines of sight.  \citet{snyder15a} studied 
the evolution of the morphology of 10 simulated galaxies, with 
$10^{9.2}\le M_{*}/M_{\odot}\le 10^{10.3}$ (at $z\sim2$), from $z\sim3.5$ to 
$z\sim0.7$. Additionally, they also examine the 
timescales on which $G-M_{20}$ and $MID$ are sensitive to a mergers for 
3 galaxy simulations in which a single major or minor merger occurs at 
$z\la 2.2$ \citep{snyder15a}.

Our simulation sample generally spans larger masses at high-redshift
than the \citet{snyder15a} work, though there is some overlap in our
low-mass regime and their high-mass end.  Similar to these authours,
we see an increase in $r_p$ with decreasing redshift for our model
galaxies.  We additionally see a similar dynamic range in measured $G$
values, with particularly good agreement from mz287 and mz374 at $z\sim2$.
However, our model $M_{20}$ extends over a much larger range of
values (and, broadly, larger values).  This discrepancy forces our
model galaxies, on average, to reside in the merger region of
$G-M_{20}$ while those from the \citet{snyder15a} work typically are
not at the same redshift.

This difference may be physical.  As demonstrated in
Appendix~\ref{section:idealized_comparison}, our radiative transfer and
source detection algorithms result in similar results as comparison
models when performing apples-to-apples tests.  At the same time,
while the low mass models in our simulation sample tend toward merger
regions in $G-M_{20}$ and $C-A$ space at late ($z\approx 2$) times,
these galaxies also have relatively larger merger rates
(c.f. Figure~\ref{figure:param_evo}).  It may be that our model
galaxies that share an overlapping mass range with those studied in
\citet{snyder15a} undergo a richer merger history during the redshift
range of interest.

\section{Conclusions}
\label{section:conclusions}

We have analysed the effectiveness of quantitative morphological
measures $G, M_{20}, C$ and $A$ in galaxies at high-redshift ($z=2-4$)
by combining a series of cosmological zoom simulations of galaxy
formation with dust radiative transfer models in order to create mock
observations of galaxies.  Our results focus on galaxies ranging from
proto-Milky Way mass through relatively high ($M_{\rm halo} \approx
10^{13} M_\odot$ at $z=2$).  Our primary results follow.

\begin{enumerate}

  \item Galaxies tend to move from the ``non-merger'' (e.g. Sb/Sc/Ir)
    region of $G-M_{20}$ space to the `''merger'' region with cosmic
    time.  Higher mass galaxies transition from the non-merger
    $\rightarrow$ merger region at earlier times, followed by lower
    mass galaxies at lower redshifts.  A similar effect is true in
    $C-A$ space.

  \item Generally, $G-M_{20}$ and $A\geq0.35$ tend to accurately identify
  	most major mergers at high-redshift in the intermediate to high \smass\ 
    galaxies within our modeled mass range.  Galaxies in the most massive 
    have a large merger rate at early times ($z\ga4$), while the galaxies 
    in the less massive simulations have increased merger rates at later 
    times.  These are both reflected in the $G-M_{20}$ and $A\geq0.35$ 
    values.

  \item At the same time, there is a significant false positive
    rate, relative to the true positive rate in both $G-M_{20}$ 
    and $C-A$ space, complicating the interpretation of these 
    metrics. These false positive rates are due to the highly complex 
    environments characteristic of massive galaxies at
    high-redshift.  Subhalos surrounding the central galaxy
    distort the final segmentation maps, causing increased $G$ and
    $A$ values, even when a galaxy is not actively merging.

  \item On average, both $G-M_{20}$ and $A\geq0.35$ perform 
    comparably to randomly guessing if a galaxy is merging or not. 
    The main exception to this is the most massive galaxies 
    (\smass $\sim 10^{10}\msun$) on the longest post-merger 
    time scales ($\Delta t \sim 1$ Gyr).

  \item In most cases other $A$ merger thresholds, $A\geq0.5$ and 
    $A\geq0.65$, achieve slightly better performance than 
    $A\geq0.35$. However the only cases where either threshold 
    achieves significantly better performance than randomly 
    guessing whether a galaxy is merger is $A\geq0.5$ on a 
    post-merger timescale of $\Delta t\sim0.2$Gyr for 
    intermediate mass (\smass\ $\sim10^9\msun$) and the most 
    massive galaxies.
\end{enumerate}

These effects taken together suggest that the application of traditional non-parametric galaxy morphology measures for galaxy mergers to high-redshift galaxies may simply trace the complex
environments of massive halos. These systems will typically eventually merge, but it is not straightforward to interpret these measures as reflective of ongoing active mergers.

\section*{Acknowledgements}
The authors are grateful to J. Lotz for providing the code used to
compute morphological measures for idealize galaxy mergers
\citep{lotz08b,lotz10a,lotz10b} and to L. Bignone for providing data
from his recent paper in digital format \citep{bignone17a}.  We thank
Greg Snyder and Paul Torrey for helpful conversations.  The code used to analyse data
for this paper used \numpy\ \citep{vanderWalt2011},
\cython\ \citep{Behnel2011}, and
\matplotlib\ \citep{Hunter2007}. M.A. acknowledges support from the
Haverford College Koshland Integrated Natural Sciences Center.
D.N. was partially supported by NSF AST-1724864, AST-1715206, and HST
AR-15043.0001.  The simulations for this paper were conducted on the
Fock cluster at Haverford College, and the HiPerGator2.0 facility at
the University of Florida.  M.A. and D.N. thank Joe Cammisa at Haverford College for his efforts in maintaining the Fock cluster.

\bibliographystyle{mnras}
\bibliography{allref}

\appendix
\section{Direct Method Comparison}\label{section:diggss}

As a first test of our modeling methods, we examine the g3iso
simulation from the ``Dusty Interacting Galaxy GADGET-SUNRISE
Simulations'' (DIGGSS) simulation
series\footnote{\url{http://archive.stsci.edu/prepds/diggss}}, 
and compared them with the reported values available with the 
mock observations of the simulations.

Here, we start with the sky subtracted observations and detection
segmentation maps supplied with the catalogues. Like in the procedure
outlined in \citet{lotz08b}, we find the centre by minimizing the
second-order moment of the central galaxy's pixels, and determined the
properties of the best fit ellipse using the algorithm implemented in
the IDL task FIT\textunderscore ELLIPSE \citep{Fanning2002}. After
this point we use the exact same procedure outlined earlier to compute
$r_p$, $a_p$, $C$, $A$, $G$, and $M_{20}$ are explained in
\ref{section:image_analysis}.

\begin{figure*}
% Make sure the figure is centered:
\includegraphics[width = 6.9in]{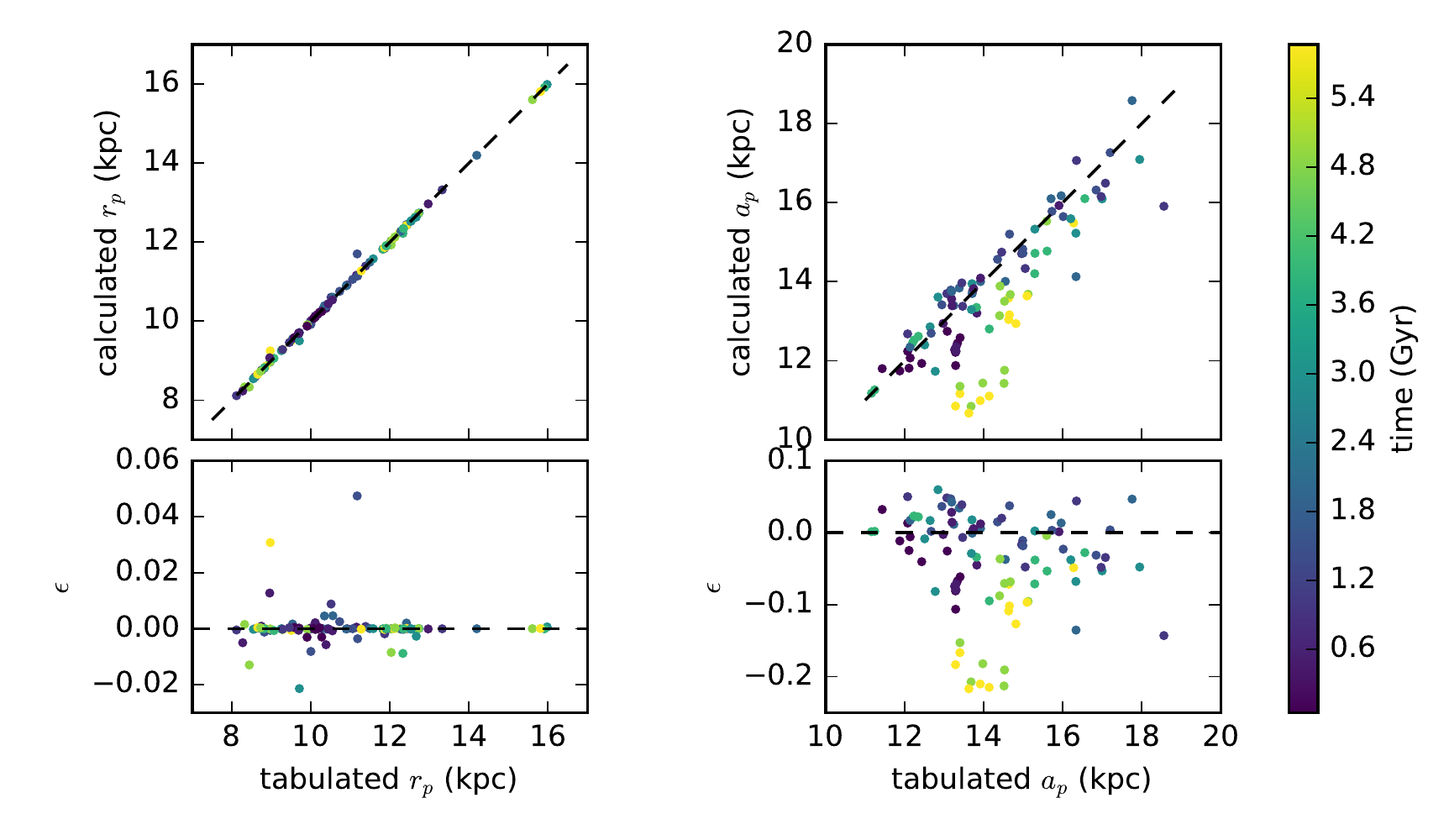}
% Give the caption for the Figure here. 
\caption{\label{figure:lotz_petro} Comparison of the calculated $r_p$ 
  and $a_p$ using methods described earlier in this paper with the 
  tabulated values for the DIGGSS g3iso galaxy. In the lower panels, 
  $\epsilon$ is the relative from the tabulated values.  The symbol colours denote the simulation time.}
\end{figure*}

\begin{figure*}
% Make sure the figure is centered:
\centering
\includegraphics[width = 6.9in]{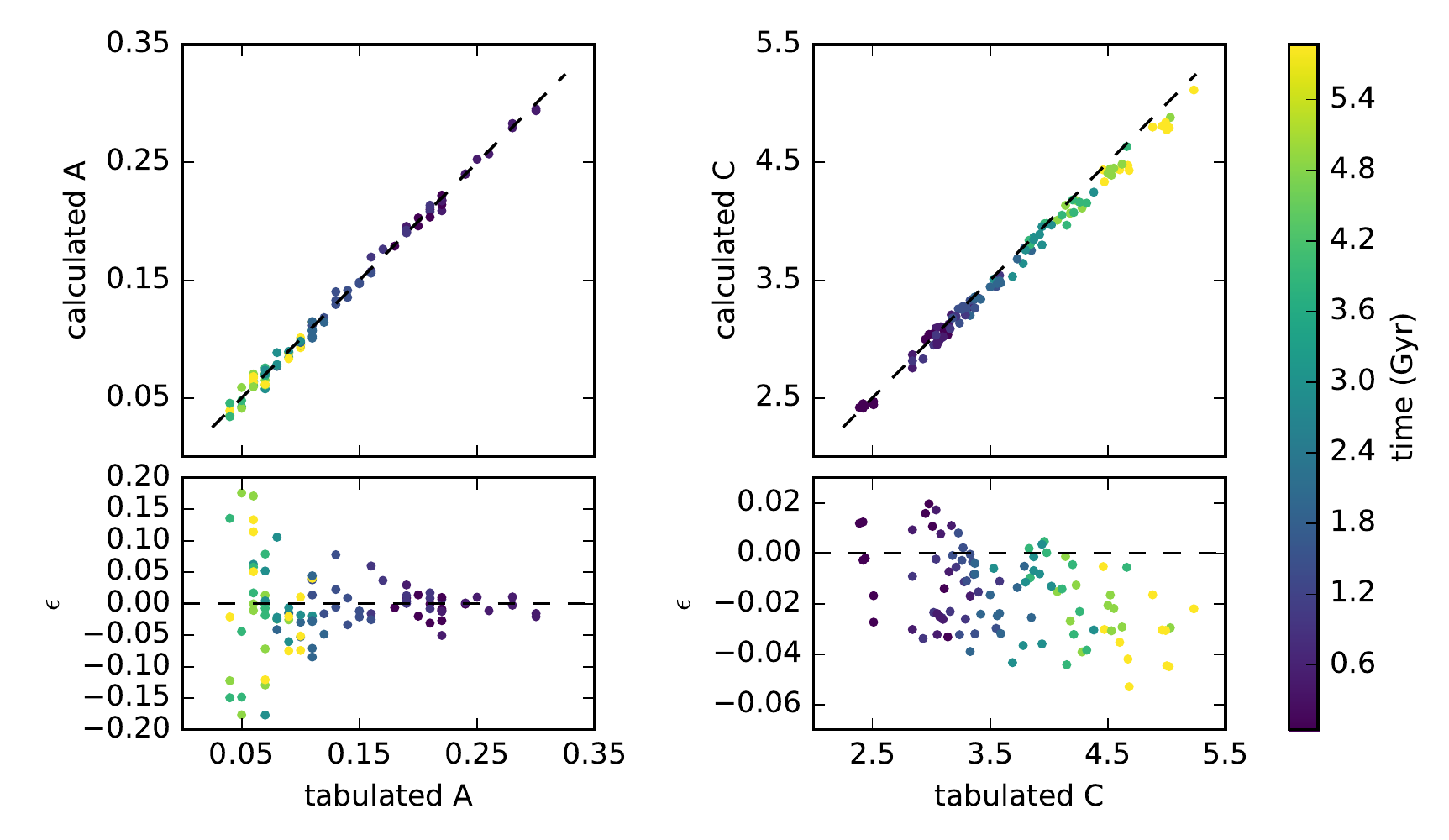}
% Give the caption for the Figure here. 
\caption{\label{figure:lotz_cas} Comparison of the calculated $A$ and
  $C$ using methods described earlier in this paper with the tabulated
  values for the DIGGSS g3iso galaxy. In the lower panels, $\epsilon$
  is the relative from the tabulated values.  Symbol colours denote
  the simulation time.}
\end{figure*}

\begin{figure*}
% Make sure the figure is centered:
\centering
\includegraphics[width = 6.9in]{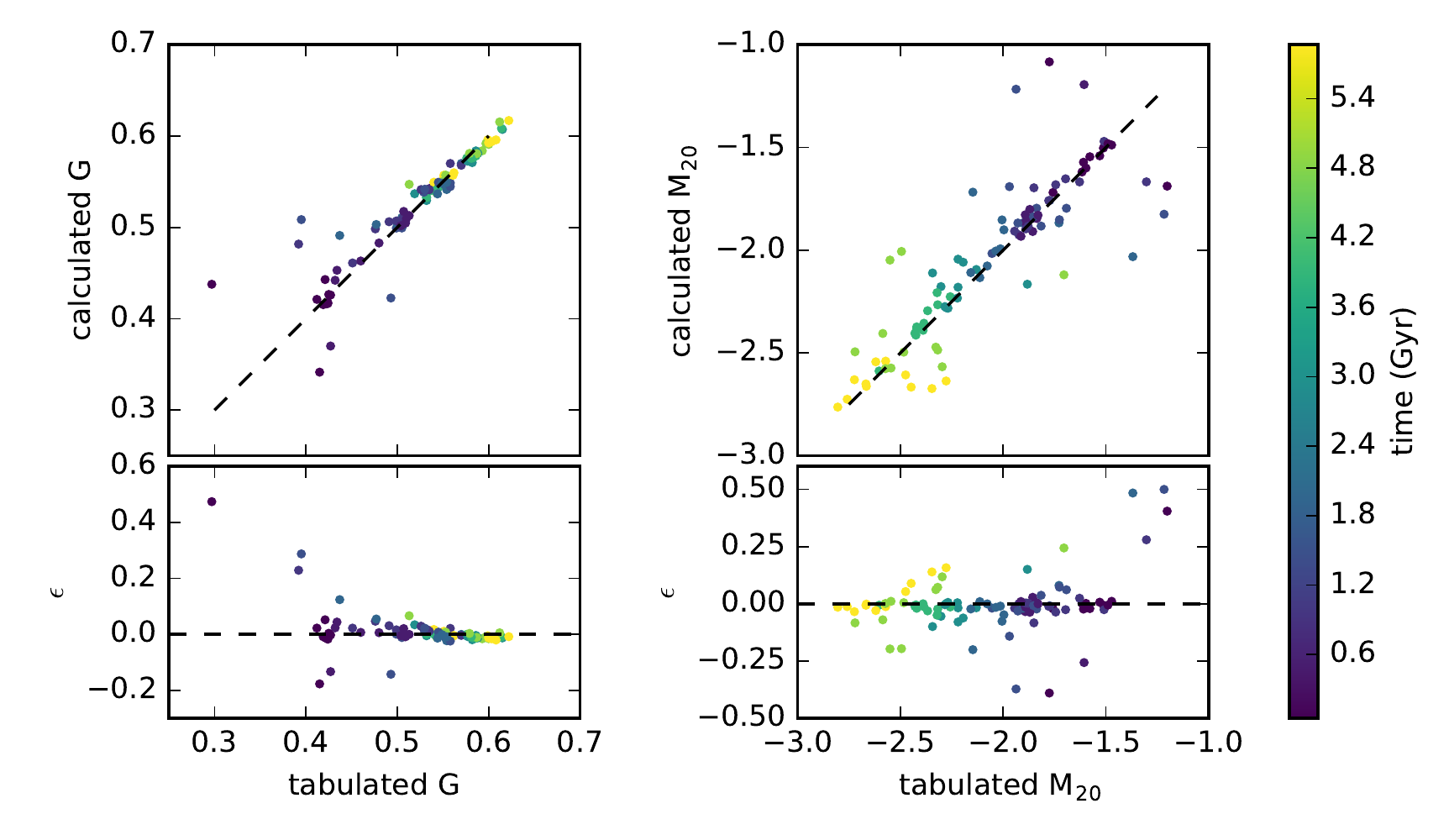}
% Give the caption for the Figure here. 
\caption{\label{figure:lotz_gm20} Comparison of the calculated $G$ 
  and $M_{20}$ using methods described earlier in this paper with the 
  tabulated values for the DIGGSS g3iso galaxy. In the lower panels, 
  $\epsilon$ is the relative from the tabulated values.  Symbol colours denote
  the simulation time.}
\end{figure*}

Figures \ref{figure:lotz_petro} and \ref{figure:lotz_cas} illustrate
the excellent agreement between the calculated and reported values of
$r_p$, $C$, and $A$. In figures \ref{figure:lotz_petro} and
\ref{figure:lotz_gm20}, slightly worse agreement is demonstrated for
$a_p$, $G$, and $M_{20}$. Recall that we employ a different method to
compute the value of $a_p$ than was used to compute the tabulated
values in the catalogue. As we will explain in \S~\ref{ap
  explanation}, we conclude that our utilised method determines more
accurate measurements of $a_p$ for lower resolution
observations. Therefore, we expect some modest deviations in the value of $a_p$,
$G$, and $M_{20}$.

%Because our current method to compute $a_p$ only results in moderately 
%poorer agreement in $M_{20}$ than the alternative algorithm and we 
%conclude that it actually returns more accurate measurements of $a_p$, 
%we conclude that our algorithm determines satisfactory measurements 
% $a_p$, $G$, and $M_{20}$.

\section{Idealized Merger Comparison} \label{section:idealized_comparison}

\begin{figure*}
% Make sure the figure is centered:
\includegraphics[width = 6.9in]{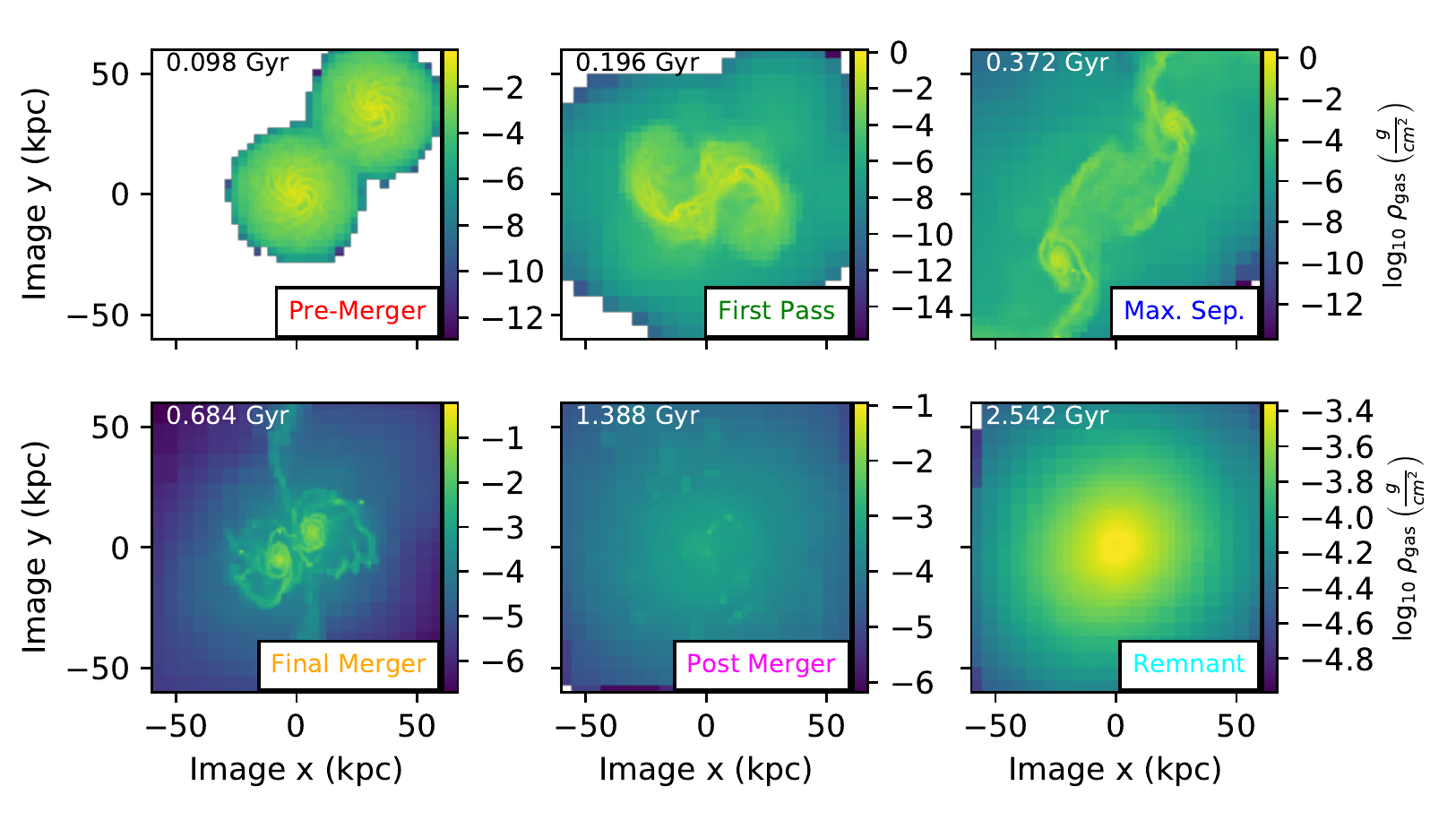}
% Give the caption for the Figure here. 
\caption{\label{figure:ideal_stages} Gas surface densities of snapshots 
  from each merger stage. Each panel is labelled with its associated 
  merger stage in the lower right corner and the time since the simulation 
  started in the upper left corner. The time periods for each merger stage 
  are given in Table \ref{tab:stages}. }
\end{figure*}

%\begin{figure}
% Make sure the figure is centered:
%\centering
%\includegraphics[width = \columnwidth]{figures/idealized_comp.pdf}
%\includegraphics[width = 3.15 in]{figures/idealized_comp.pdf}
% Give the caption for the Figure here. 
%\caption{\label{figure:ideal_gm20} $G-M_{20}$ plot for all unique inclinations in select snapshots of the d4e idealized simulation. Points are coloured by the merger stage. The stages pre-merger, first pass, maximal separation, final merger, post merger, and remnant are coloured red, green, blue, orange, magenta, and cyan.}
% This label can be used in the text to refer to the figure by number, if
% desired; see example \ref command in the text where this figure is
% referred to. 
%\end{figure}

\begin{figure}
% Make sure the figure is centered:
\centering
\includegraphics[width = 3.15 in]{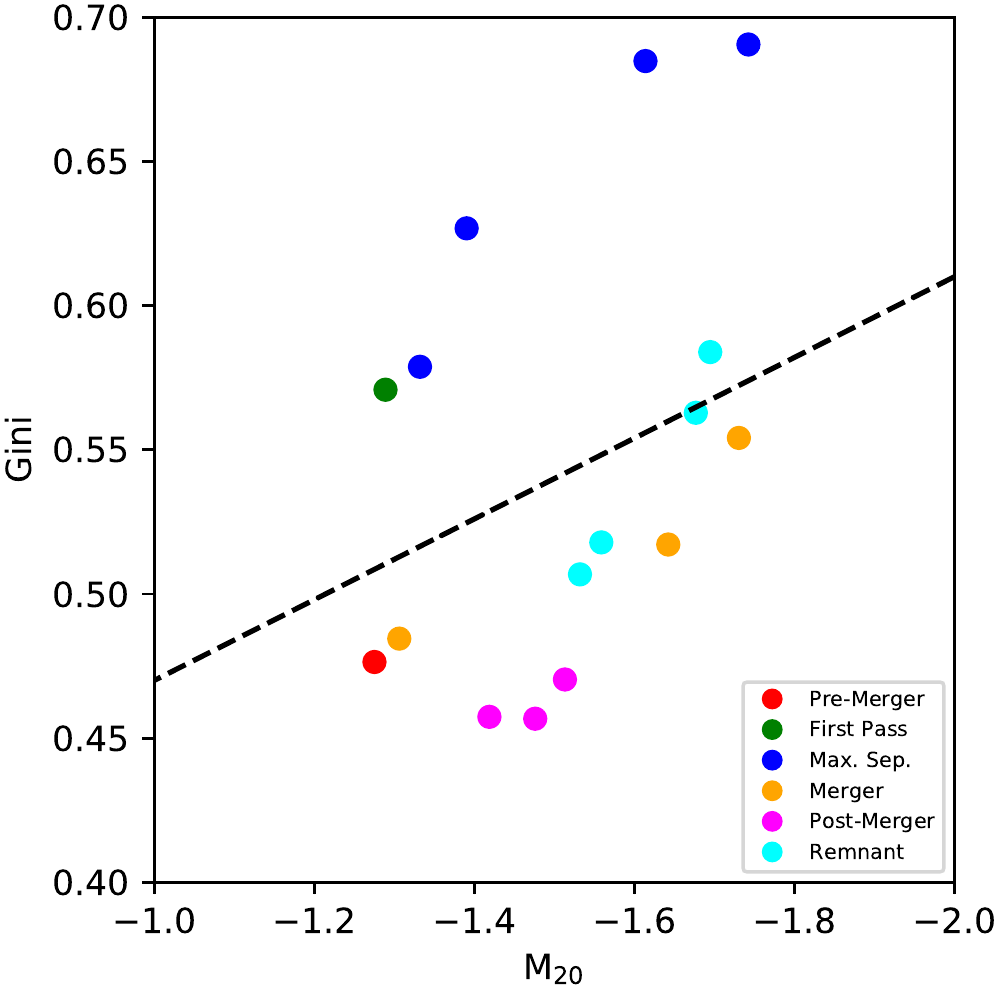}
% Give the caption for the Figure here. 
\caption{\label{figure:ideal_gm20_avg} $G-M_{20}$ plot for select snapshots of the d4e idealized simulation. The points are the average $G$ and $M_{20}$ over all unique inclinations in a given snapshot. Points are coloured by the merger stage. The stages pre-merger, first pass, maximal separation, final merger, post merger, and remnant are coloured red, green, blue, orange, magenta, and cyan}
% This label can be used in the text to refer to the figure by number, if
% desired; see example \ref command in the text where this figure is
% referred to. 
\end{figure}

\begin{figure}
% Make sure the figure is centered:
\centering
\includegraphics[width = 3.15 in]{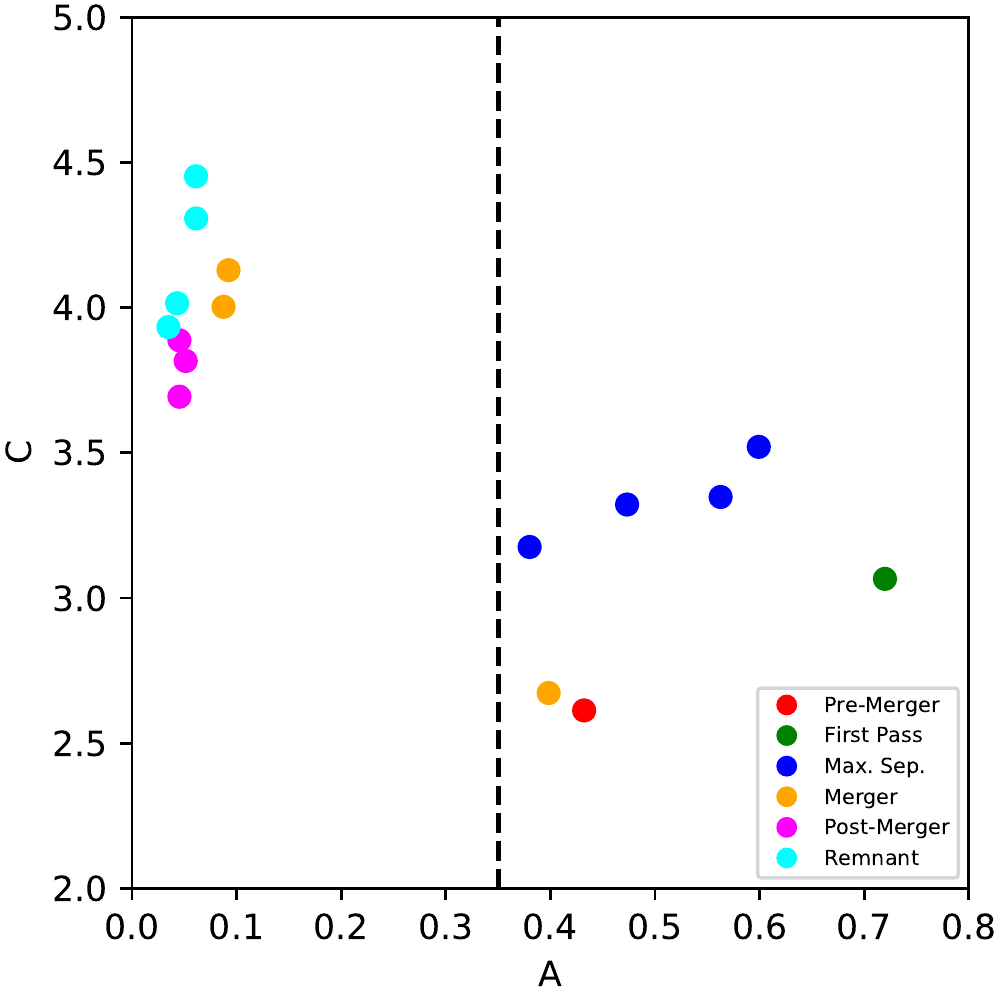}
% Give the caption for the Figure here. 
\caption{\label{figure:ideal_ca_avg} $C-A$ plot for select snapshots of the d4e idealized simulation. The points are the average $C$ and $A$ over all unique inclinations in a given snapshot. Like Figure \ref{figure:ideal_ca_avg}, points are coloured by the merger stage.}
% This label can be used in the text to refer to the figure by number, if
% desired; see example \ref command in the text where this figure is
% referred to. 
\end{figure}

As a second check on the validity of the morphological measures, we 
compare the morphological measures of a galaxy merger in idealized 
simulation to the results found by \citet{lotz08b}. In particular, 
we make use of the d4e simulation suite developed originally by \citet{cox06a,
cox06b} and \citet{robertson06a,robertson06b}.

It is important to note a few minor differences between the mock
observations generated for this comparison, and those for the
cosmological zoom simulations that comprise our main study.  Unlike
the cosmological zoom simulations, these mock observations have been
simulated at $\lambda = 4686$ \angstrom\ to simulate the SDSS $g$
filter \citep[in order to best compare with][]{lotz08b}, and noise was
added such that all pixels containing any projected stellar mass
belonging to any galaxy had an \avgSNR $\sim 25$.  Additionally, the
binding lengths used for FOF halo and galaxy finding were slightly
different from those used for the cosmological simulation. We set them
both equal to 0.10 times the mean inter-particle distance as we found
this greatly improved the ability of \caesar \ to distinguish between
the two galaxies in early merger stages.

These mock observations are simulated at an angular diameter distance
of a galaxy at $z\sim 2$ where $0.05'' \sim 430$ pc. They have a pixel
scale of $\sim 0.05'' \mathrm{pixel}^{-1}$ and are convolved with a
Gaussian of $FWHM \sim 0.15''$.  Here, 0.05'' $\sim$ 430
pc. For comparison, the mock observations produced by \citet{lotz08b},
have noise added slightly differently and the observations are
simulated for SDSS at a distance where $0.396'' \sim 105$ pc. Those
observations have a pixel scale $\sim 0.396'' \mathrm{pixel}^{-1}$ and
are convolved with a Gaussian of $FWHM \sim 1.5''$.  One final
difference from the analysis of the mock observations of the
cosmological simulations is that we use a 9x9 tophat filter with a
radius of 4.5 pixels during source detection; we found that this
yielded better detection segmentation maps than those yielded by the
5x5 tophat filter with a radius of 2.5 pixels for the idealized
simulation.

\begin{table}
	\centering
	\caption{The definitions for the times at which each merger 
          stage begins are listed below. The actual calculated times 
          for this idealized simulation are also listed. Each stage 
          other than the ``remnant'' Stage includes all events from 
          its starting time to the starting time time of the next 
          stage. The ``remnant'' stage includes all events after its 
          start time.}
	\label{tab:stages}
	\begin{tabular}{lcc}
		\hline
		Merger Stage & Definition & Time (Gyr)\\
		\hline
		Pre-Merger & $0$ & 0 \\
		First Pass & $0.5t_{fp}$ & 0.098 \\ 
		Max. Sep. & $0.5\left(t_{fp}+t_{max}\right)$ & 0.284 \\
		Merger & $0.5\left(t_{max}+t_{merg}\right)$ & 0.601 \\
		Post-Merger & $t_{merg}+0.5$ Gyr & 1.331 \\
		Remnant &$t_{merg}+1.0$ Gyr & 1.831 \\
		\hline
	\end{tabular}
\end{table}

We classify the merger stages in a similar way to \citet{lotz08b}.  We
determine $t_{fp}$, the time at which the galaxies are closest during
their first pass, $t_{max}$, the time when the galaxies are the most
separated and $t_{merge}$, the time at which the nuclei are merged
(functionally, we determine this when the nuclear black hole sink
particles in the two progenitors have merged).  within 1 kpc of one
another.  Using these time-steps, we classify a galaxy's merger stage
into the following categories: 'pre-merger', 'first pass', 'maximal
separation', 'merger', 'post-merger', and 'remnant'. See Table
\ref{tab:stages} for the definitions of the merger stages and Figure
\ref{figure:ideal_stages} for sample gas surface density plots in each
merger stage.

Figures \ref{figure:ideal_gm20_avg} and \ref{figure:ideal_ca_avg} show
the average $G-M_{20}$ and $C-A$ values for select snapshots of the
d4e simulation.  The $G-M_{20}$ and $C-A$ values have been averaged 
over all unique lines of sight at every snapshot. Note that no 
$\left<SNR\right>$ cuts were made. Comparison of these
results to those of \citep{lotz08b} demonstrate that our 
image analysis methods yield comparable results to other image
analysis methods.

\section{Additional Postage Stamps}\label{section:sim_stamps}

Here, we include series of postage stamps analogous to Figure
\ref{figure:mz0_stamps} for mz5, mz10, mz45, mz287, mz352, mz374 and
mz401. In each Figure, we include snapshots close to $z\sim$ 5, 4,
3.5, 3, 2.5 and 2 that all have valid $A$ and $G-M_{20}$ along a
consistent line of sight.

\begin{figure*}
% Make sure the figure is centered:
\centering
\includegraphics[width = 6.8 in]{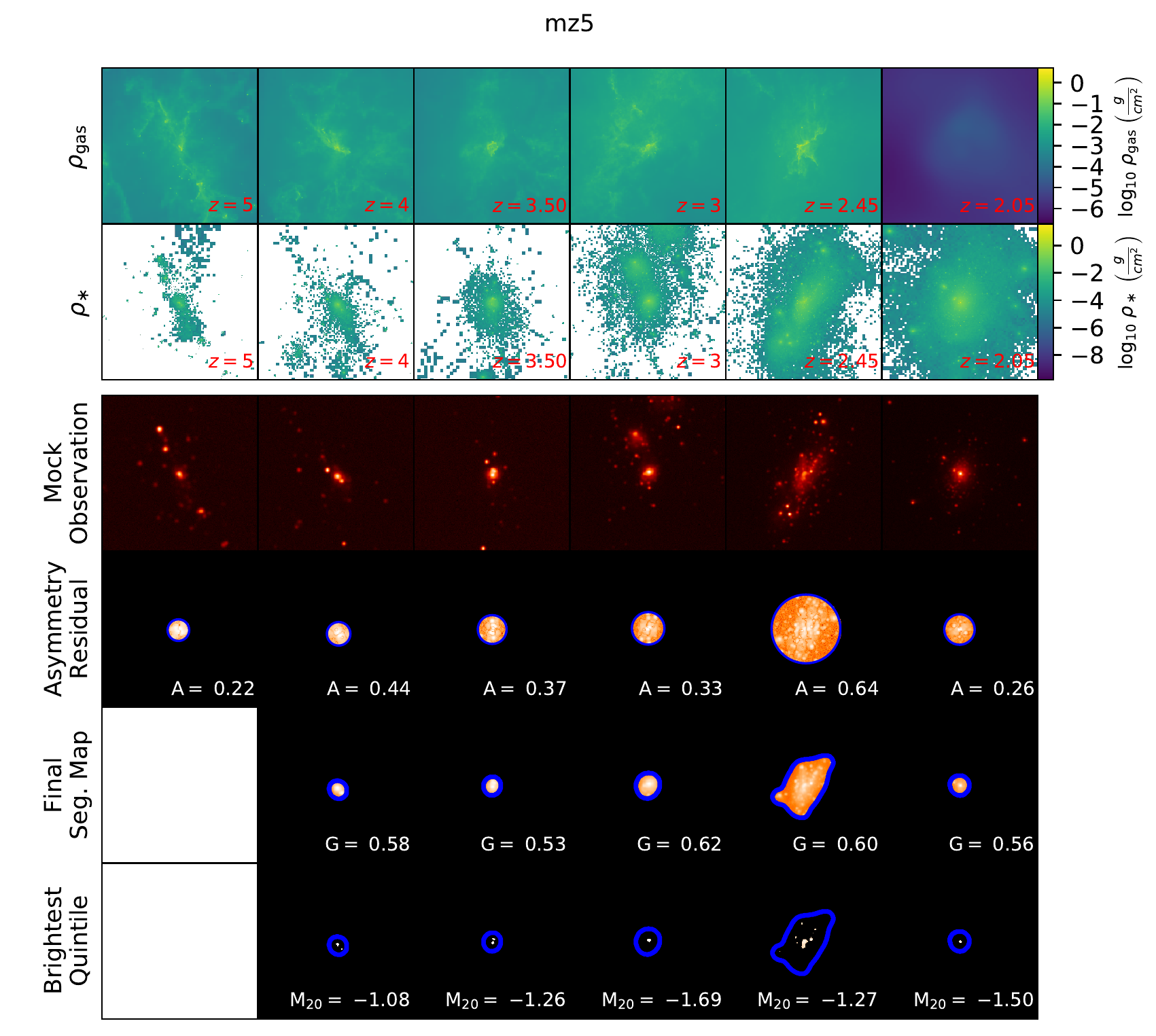}
\caption{\label{figure:mz5_stamps} The same as
  Figure~\ref{figure:mz0_stamps} except that the panels have
  correspond to the central galaxy of mz5, rather than that of mz0,
  and the panels are generated at $z\approx$ $5$, $4$, $3.5$, $3$,
  $2.6$ and $2.05$.  The final segmentation map and brightest
  quintile has been omitted at the $z$ when the final segmentation map
  has \avgSNR $< 20$ or is not contiguous.}
\end{figure*}

\begin{figure*}
% Make sure the figure is centered:
\centering
\includegraphics[width = 6.8 in]{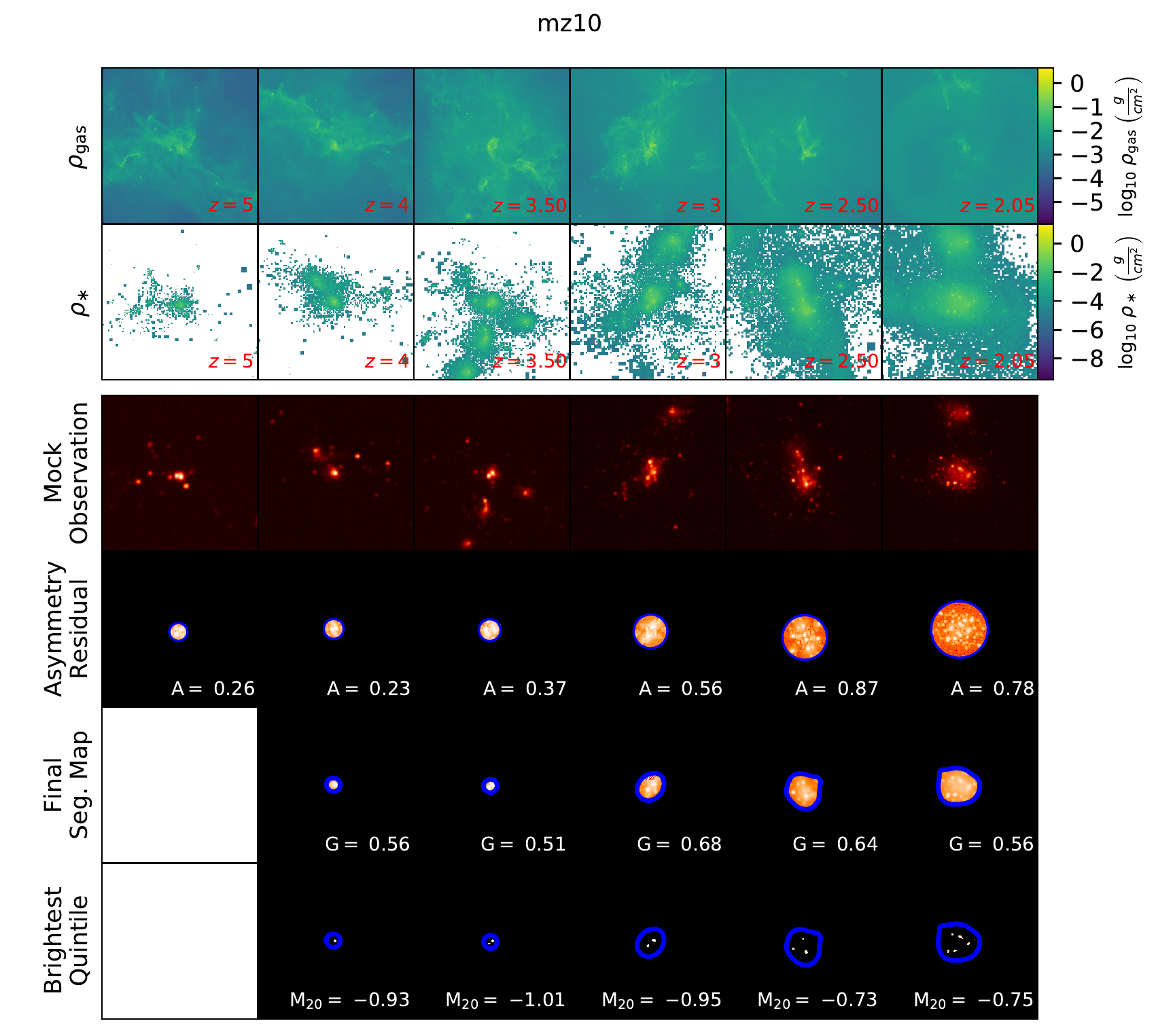}
\caption{\label{figure:mz10_stamps} The same as Figure~\ref{figure:mz0_stamps} except 
that the panels have correspond to the central galaxy of mz10, rather than that 
of mz0, and the panels are generated at $z\approx$ $5$, $4$, $3.5$, $3$, $2.5$ 
and $2.05$.   The final segmentation map and brightest quintile has been omitted at the $z$ when the final segmentation map has \avgSNR $< 20$ or is not contiguous.}
\end{figure*}

\begin{figure*}
% Make sure the figure is centered:
\centering
\includegraphics[width = 6.8 in]{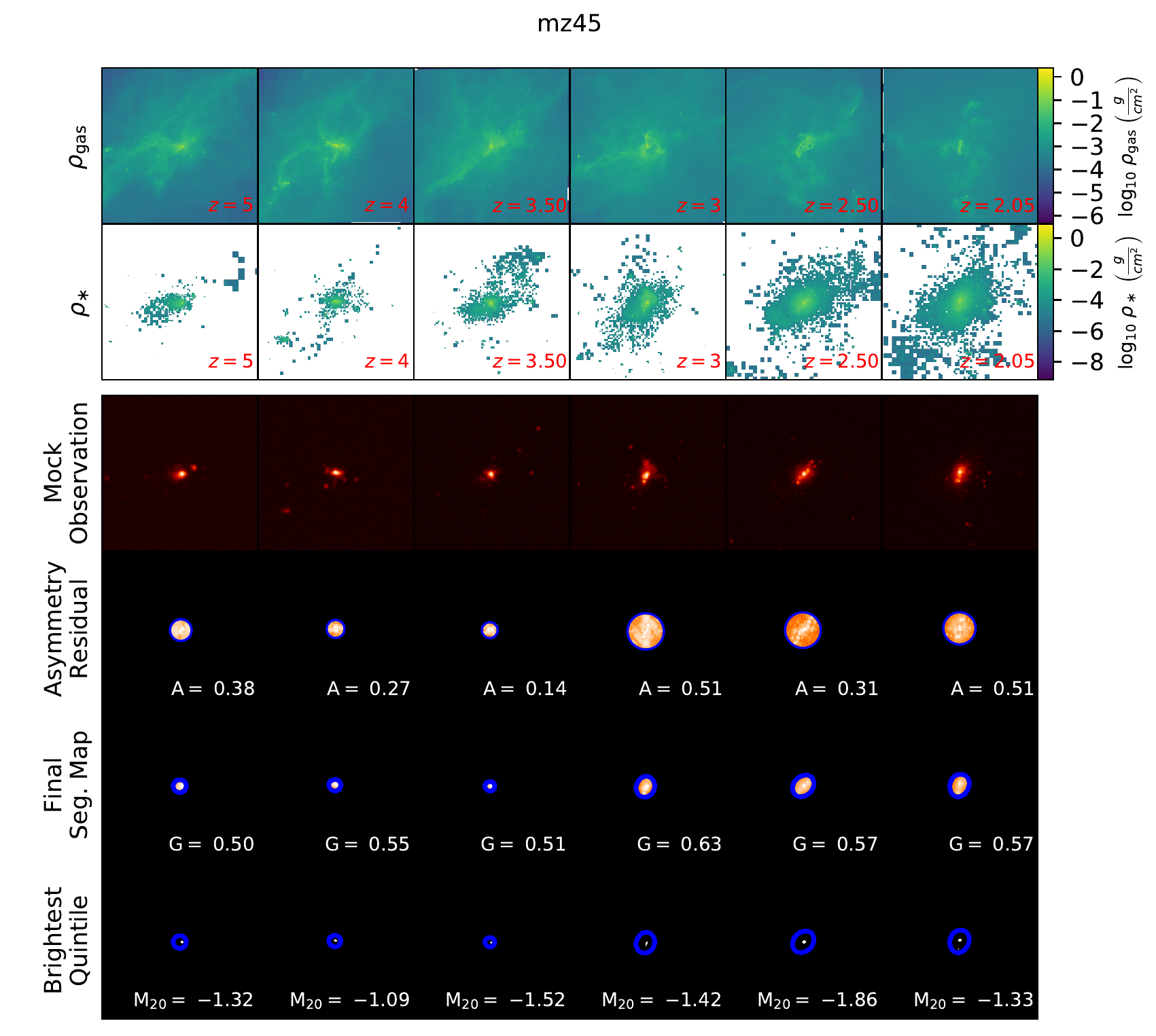}
\caption{\label{figure:mz45_stamps} The same as
  Figure~\ref{figure:mz0_stamps} except that the panels have
  correspond to the central galaxy of mz45, rather than that of mz0,
  and the panels are generated at $z\approx$ $5$, $4$, $3.5$, $3$,
  $2.5$ and $2.05$.  The final segmentation map and brightest quintile
  has been omitted at the $z$ when the final segmentation map has
  \avgSNR $< 20$ or is not contiguous.}
\end{figure*}

\begin{figure*}
% Make sure the figure is centered:
\centering
\includegraphics[width = 6.8 in]{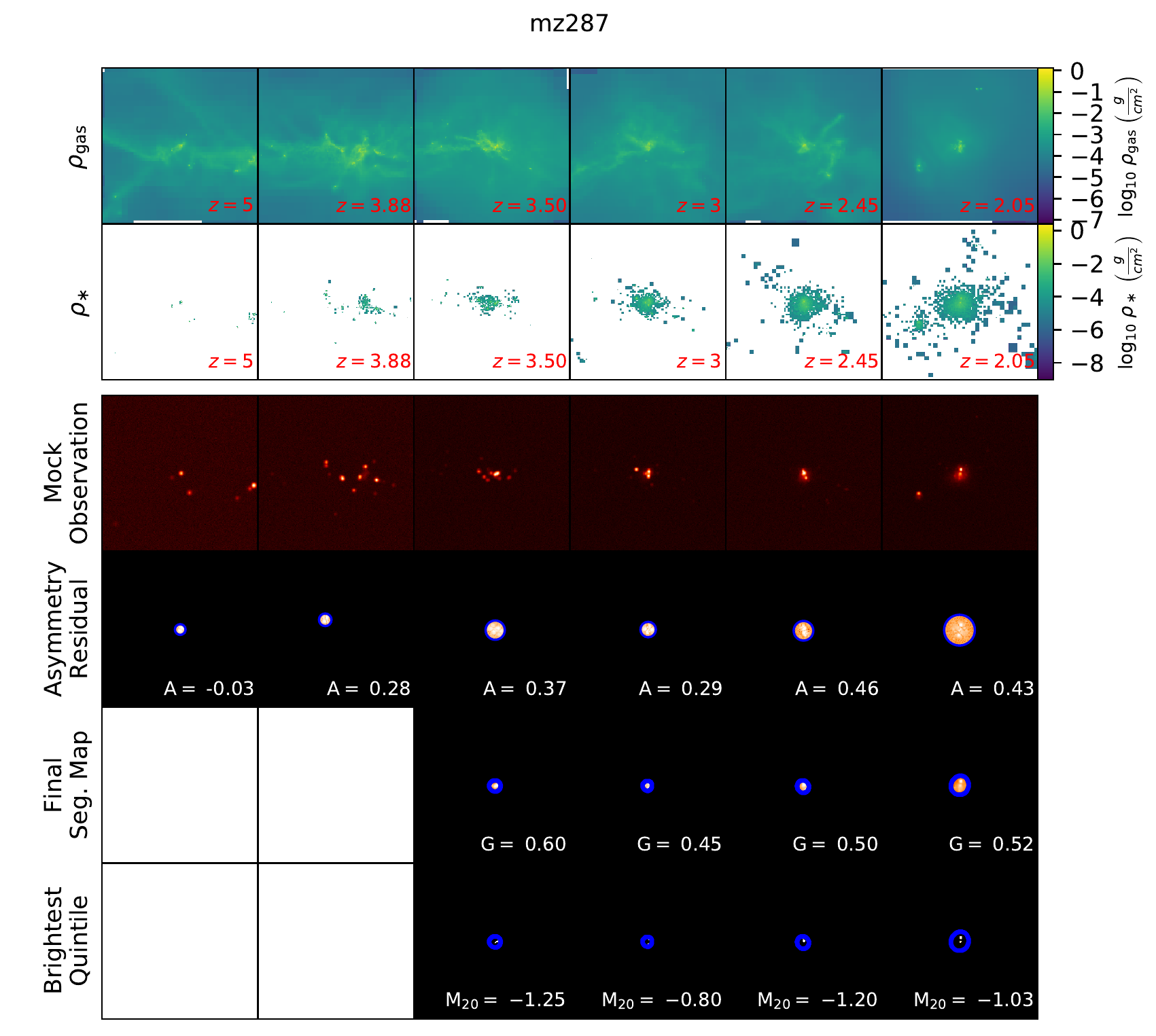}
\caption{\label{figure:mz287_stamps} The same as
  Figure~\ref{figure:mz0_stamps} except that the panels have
  correspond to the central galaxy of mz287, rather than that of mz0,
  and the panels are generated at $z\approx$ $5$, $4$, $3.5$, $3$,
  $2.5$ and $2.05$.  The final segmentation map and brightest quintile
  has been omitted at the $z$ when the final segmentation map has
  \avgSNR $< 20$ or is not contiguous.}
\end{figure*}

\begin{figure*}
% Make sure the figure is centered:
\centering
\includegraphics[width = 6.8 in]{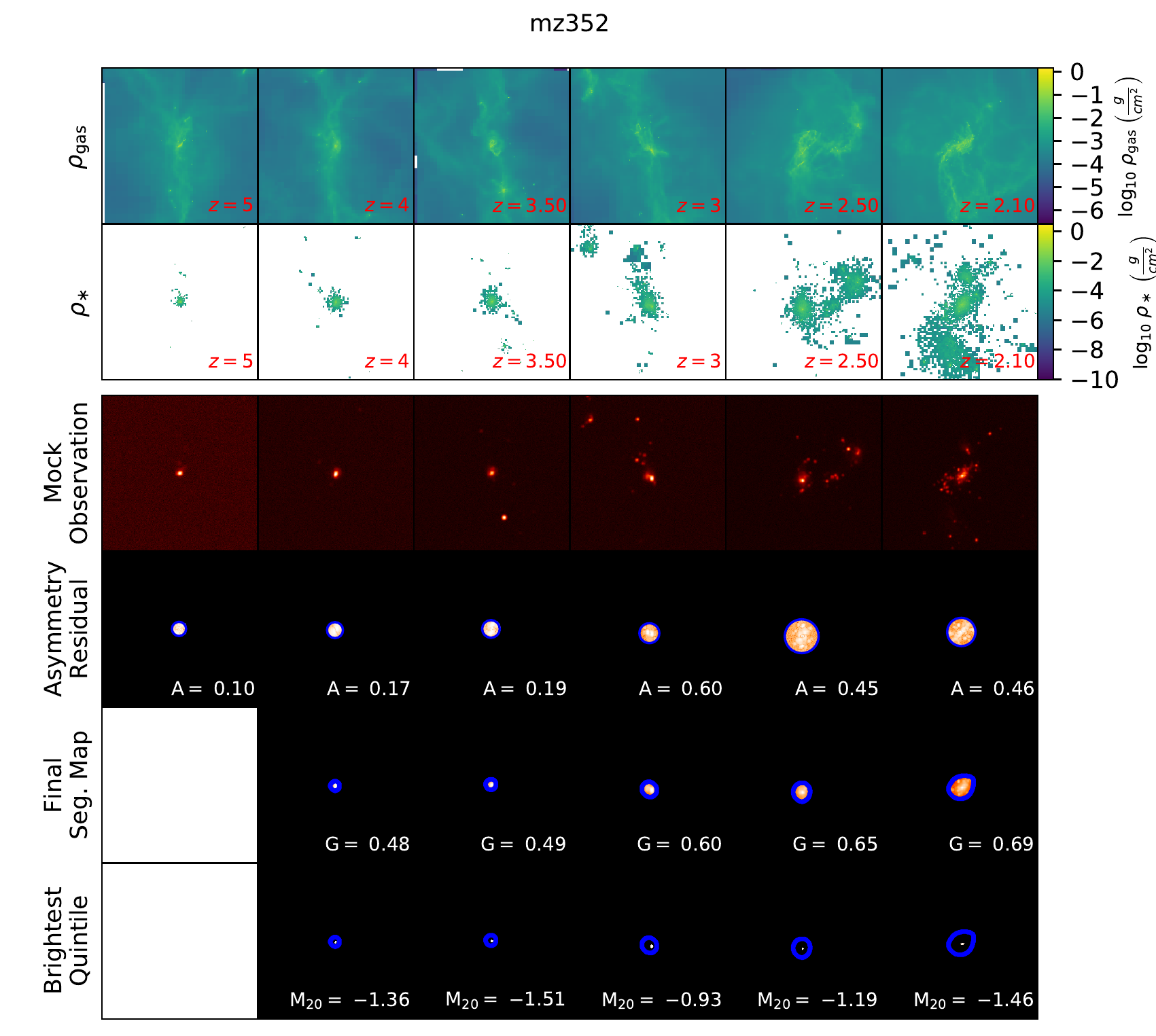}
\caption{\label{figure:mz352_stamps} The same as
  Figure~\ref{figure:mz0_stamps} except that the panels have
  correspond to the central galaxy of mz352, rather than that of mz0,
  and the panels are generated at $z\approx$ $5$, $4$, $3.625$, $2.9$,
  $2.45$ and $2.15$.  The final segmentation map and brightest
  quintile has been omitted at the $z$ when the final segmentation map
  has \avgSNR $< 20$ or is not contiguous.}
\end{figure*}

\begin{figure*}
% Make sure the figure is centered:
\centering
\includegraphics[width = 6.8 in]{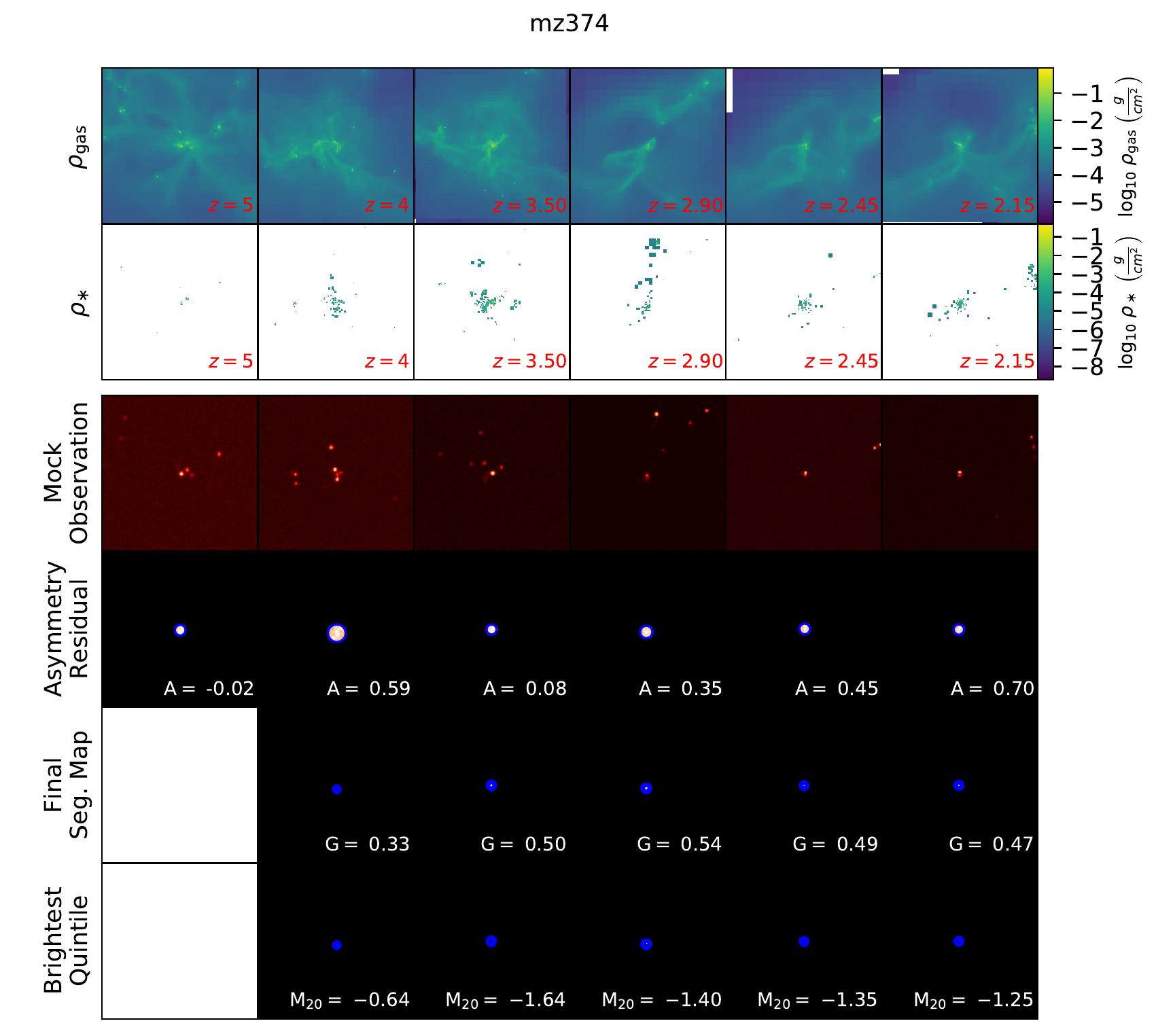}
\caption{\label{figure:mz374_stamps} The same as
  Figure~\ref{figure:mz0_stamps} except that the panels have
  correspond to the central galaxy of mz374, rather than that of mz0,
  and the panels are generated at $z\approx$ $5$, $4$, $3.5$, $3$,
  $2.45$ and $2.05$.  The final segmentation map and brightest
  quintile has been omitted at the $z$ when the final segmentation map
  has \avgSNR $< 20$ or is not contiguous.}
\end{figure*}

\begin{figure*}
% Make sure the figure is centered:
\centering
\includegraphics[width = 6.8 in]{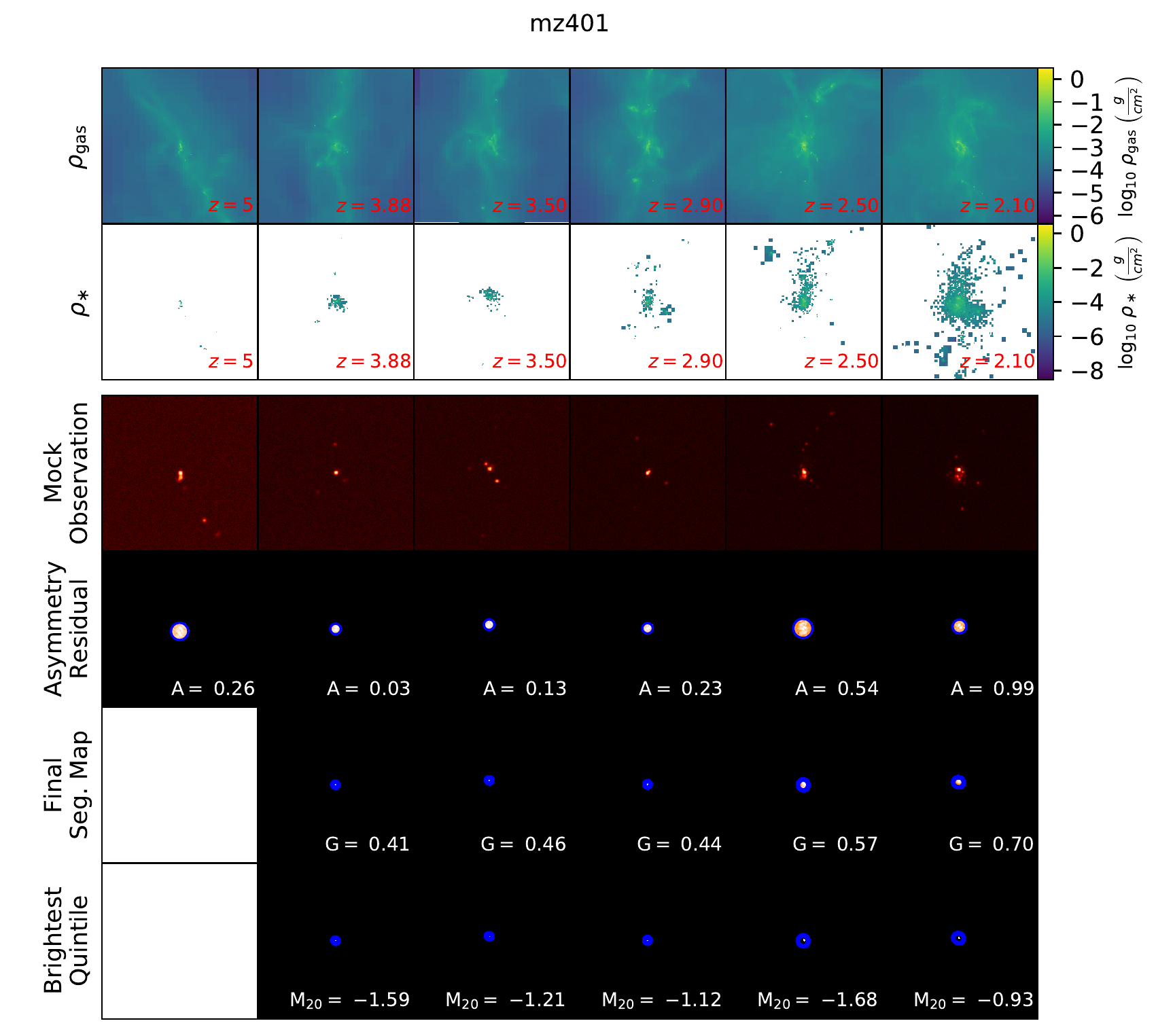}
\caption{\label{figure:mz401_stamps} The same as
  Figure~\ref{figure:mz0_stamps} except that the panels have
  correspond to the central galaxy of mz401, rather than that of mz0,
  and the panels are generated at $z\approx$ $5$, $3.875$, $3.5$,
  $2.90$, $2.5$ and $2.10$.  The final segmentation map and brightest
  quintile has been omitted at the $z$ when the final segmentation map
  has \avgSNR $< 20$ or is not contiguous.}
\end{figure*}

\begin{figure*}
% Make sure the figure is centered:
\centering
\includegraphics[width = 6.8 in]{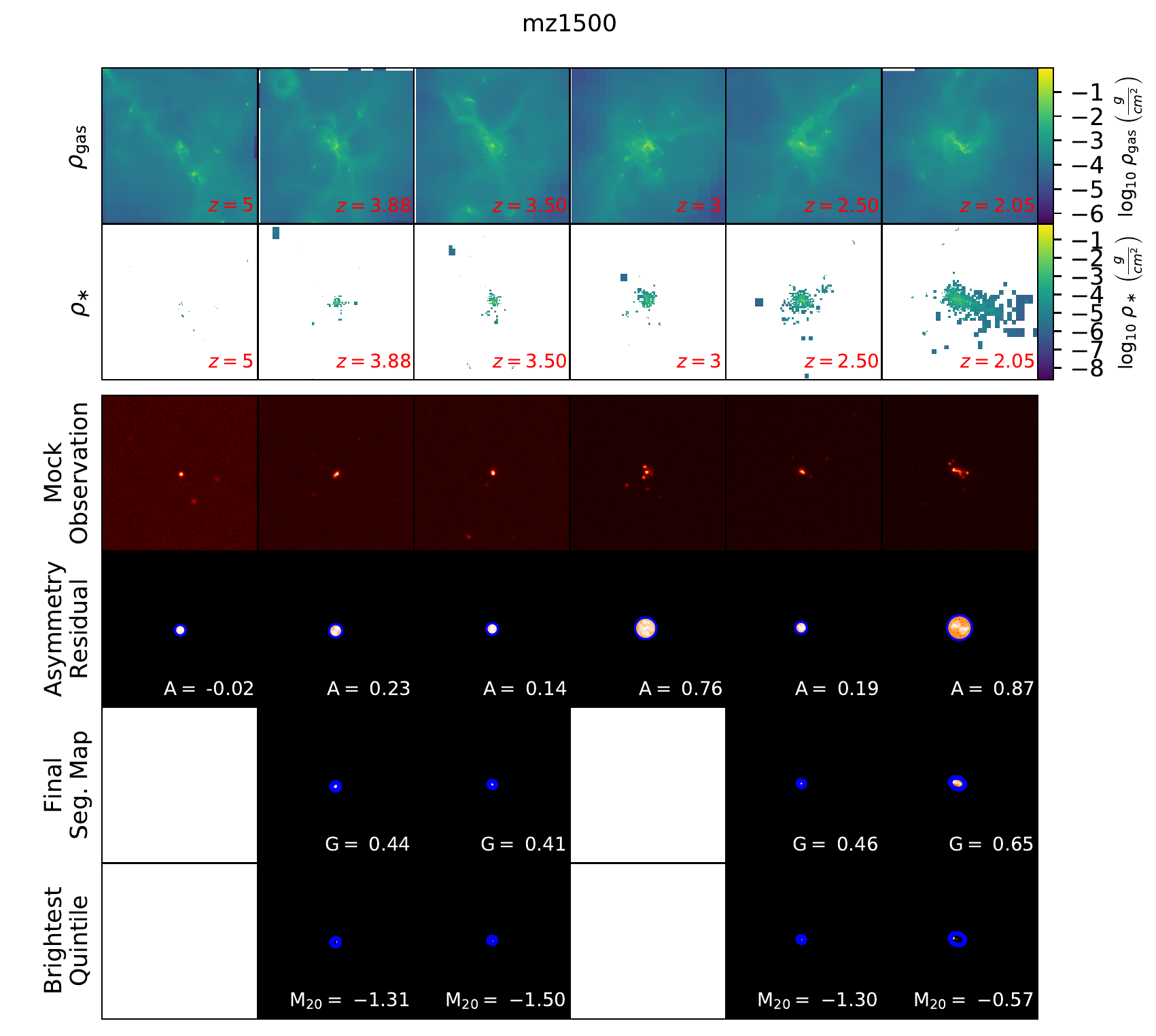}
\caption{\label{figure:mz1500_stamps} The same as
  Figure~\ref{figure:mz0_stamps} except that the panels have
  correspond to the central galaxy of mz1500, rather than that of mz0,
  and the panels are generated at $z\approx$ $5$, $3.875$, $3.5$,
  $3$, $2.5$ and $2.05$.  The final segmentation map and brightest
  quintile has been omitted at the $z$ when the final segmentation map
  has \avgSNR $< 20$ or is not contiguous.}
\end{figure*}

\section{Effects of Selection Criteria on Uncertainty of Merger Diagnostics Performance} \label{section:optimisic_tpr/fpr}

\begin{figure}
% Make sure the figure is centered:
\includegraphics[width = 3.5 in]{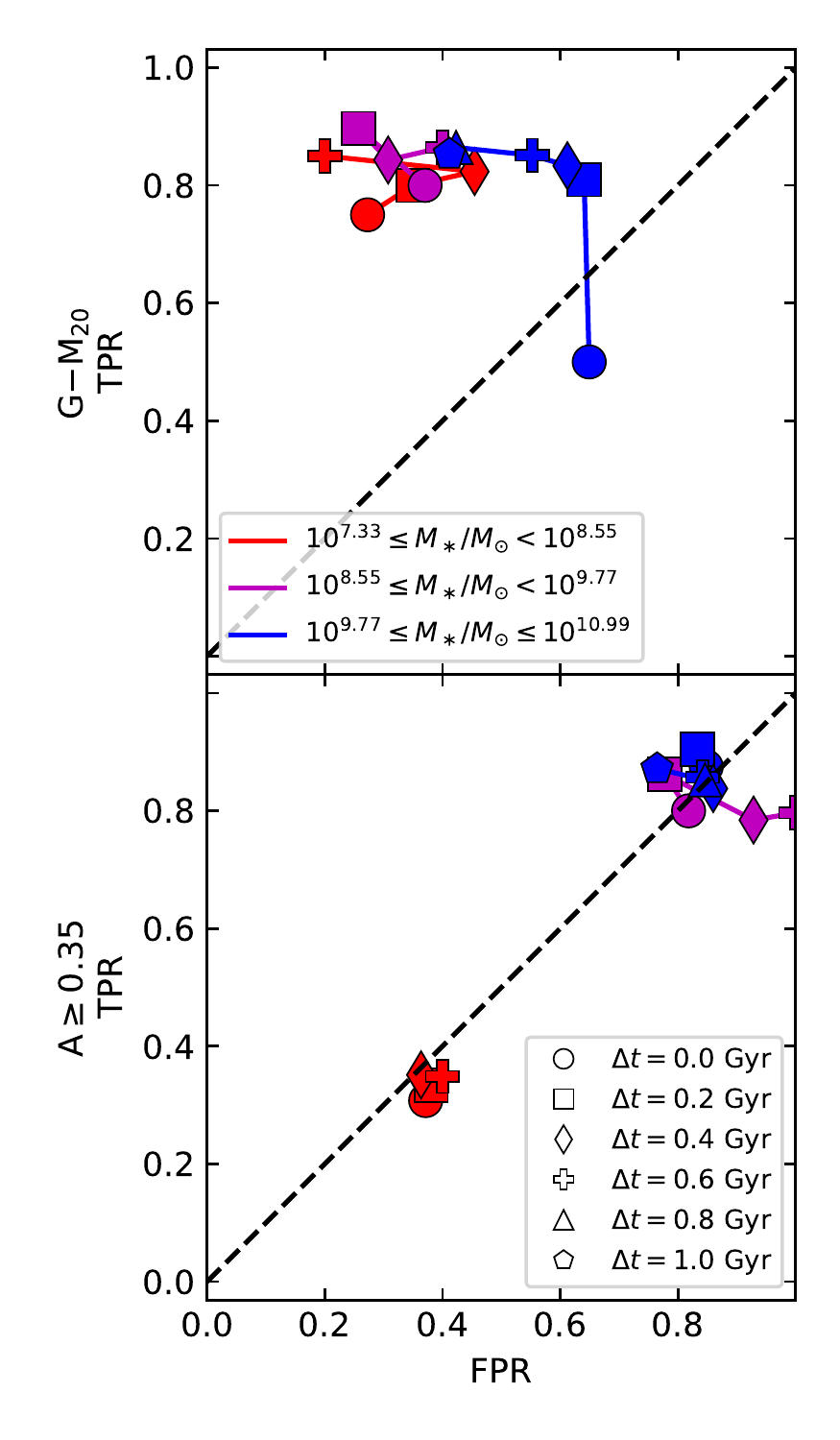}
% Give the caption for the Figure here. 
\caption{\label{figure:ROC_major_opt} The same as Figure \ref{figure:ROC_major}, 
but instead of ignoring the discarded data, at each possible $\Delta t$, 
$-\Delta t$ timescale we assume that all discarded data have values that 
maximize TPR and minimize FPR values.}
\end{figure}

Figure~\ref{figure:ROC_major_opt} illustrates the 
uncertainty in the performance of the merger diagnostics that results 
from our sample selection. The figure shows the most optimistic 
hypothetical true positive rates and false positive rates at any given  
potential timescale that the merger diagnostics could have if all of 
our data met our selection criteria.

\defcitealias{Jones2001}{Jones~et~al.~2001}

\section{Petrosian Semi-Major Axis}\label{ap explanation}

Here, we discuss our methodology of computing $a_p$, and compare it to
the method employed by \citet{lotz08b}.  As discussed in \S
\ref{section:pet_semi-major}, the method we use to compute $a_p$ draws
heavy inspiration from the method employed in the photometric pipeline
of SDSS to compute $r_p$.  For the sake of this discussion, we will
refer to method we currently use as Method A and the method used by
\citet{lotz08b} as Method B.  As we will discuss, at our model
resolution Method A returns somewhat more accurate results than method
B.  This owes to:   (i) 
the exactness of the photometry and (ii) the spacing of the 
points on the light curves constructed by each method.
%Our algorithm to compute $r_p$, as 
%described in \S \ref{section:pet_radius}, closely follows the 
%method used by \citet{lotz08b} to compute $r_p$ and it 
%therefore is very similar to Method B.

%Method A uses more exact photometry and constructs a light 
%curve from exponentially spaced points, whereas Method B 
%uses less exact photometry and constructs a light curve 
%from points spaced linearly. 

The photometry we use in Method A 
subsamples the flux of pixels partially enclosed by an 
annulus. For comparison, in the photometry employed in Method 
B subsampling is not used. Instead, the entire flux of the pixel 
is included or not included based on where the annulus passes 
through the centre of the pixel. In higher resolution images, 
where the features of galaxies are distributed over more pixels, we expect 
there to be minimal difference in the photometry. However in lower 
resolution images we expect the differences in the photometry to be 
more significant.

Likewise, we expect the differences in the spacing of the points on 
the light profiles constructed by each method to have similar impacts on 
the accuracy of the recovered $a_p$ at different image resolutions. In 
Method A the light profile is measured at exponentially spaced points. 
Near the centre of the galaxy, where features have a larger impact on the 
light profile (because they make up a larger fraction of the 
total enclosed flux), the light profile has sub-pixel spacing and further 
from the centre points on are spaced by more than a single pixel. In 
Method B, the light profile is measured at points with constant 1 pixel 
spacing. For low resolution images, we expect Method A 
to recover more accurate measurements as the features of the galaxy are 
condensed over a smaller number of pixels. However, for higher resolution 
images, we expect either minimal differences in the accuracy or 
that Method B might have better accuracy as it has constructs the light 
profile with more finely spaced points at large distances.

\begin{figure}
% Make sure the figure is centered:
\centering
\includegraphics[width = 3.15 in]{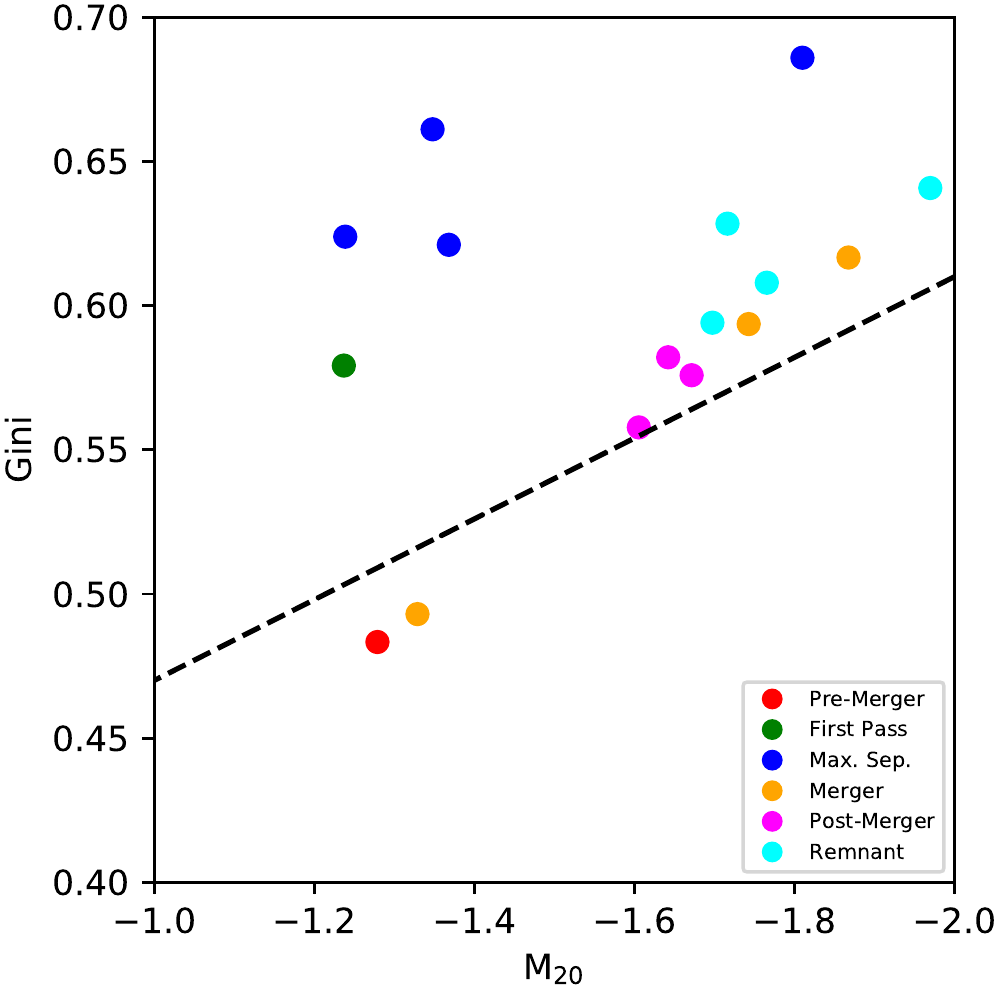}
% Give the caption for the Figure here. 
\caption{\label{figure:ideal_gm20_avg_lotz_ap} $G-M_{20}$ plot for 
  select snapshots of the d4e idealized simulation produced when 
  $a_p$ is computed using the algorithm from \citet{lotz08b} 
  (Method B). The data illustrated here can be directly compared 
  against that featured in Figure \ref{figure:ideal_gm20_avg}.}
\end{figure}

\begin{figure*}
% Make sure the figure is centered:
\centering
\includegraphics[width = 6.9in]{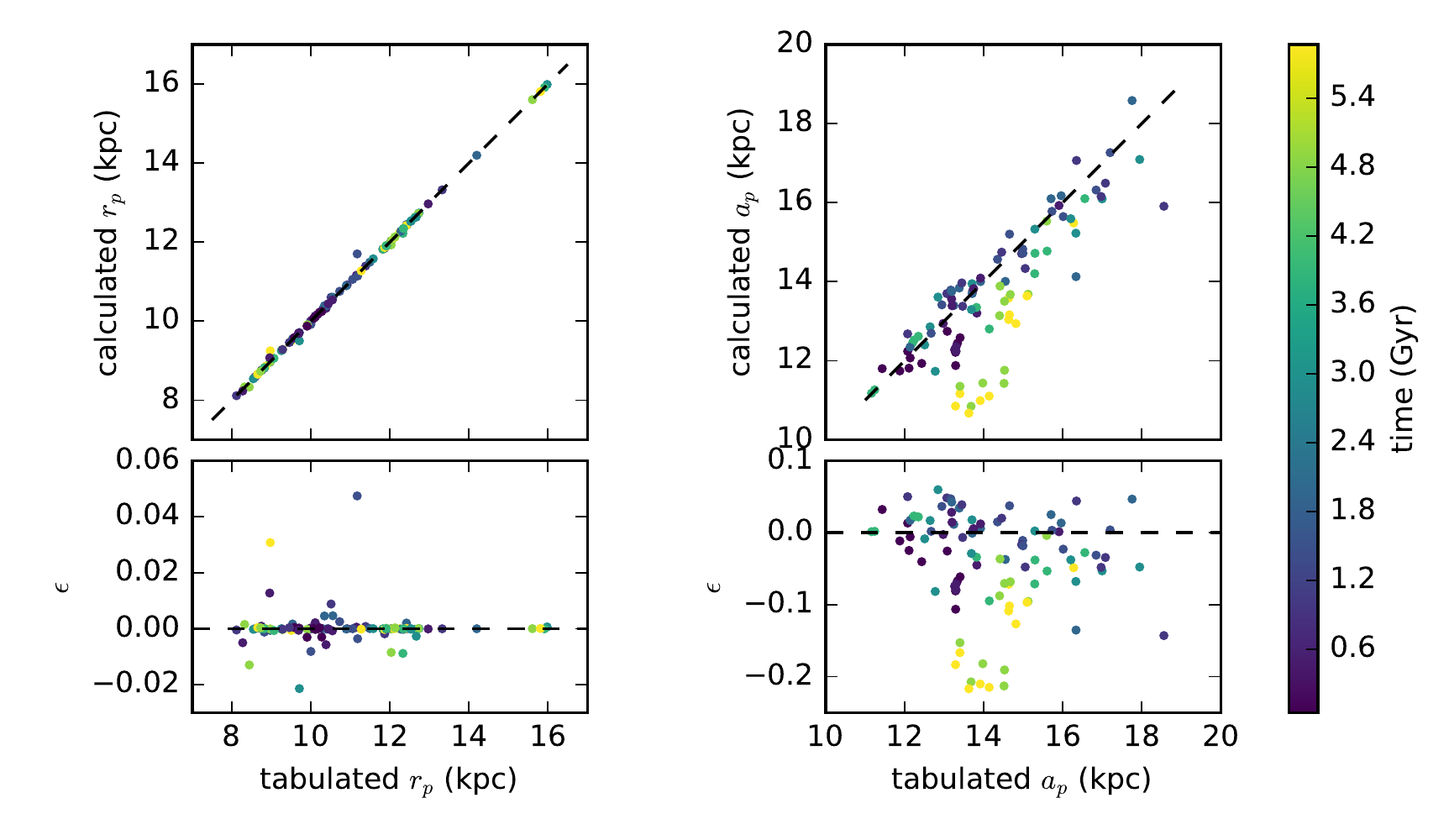}
% Give the caption for the Figure here. 
\caption{\label{figure:lotz_petro_lm} This is the same as figure 
  \ref{figure:lotz_petro} except that $a_p$ values are computed 
  with the algorithm used in \citet{lotz08b} (Method B). }
\end{figure*}

\begin{figure*}
% Make sure the figure is centered:
\centering
\includegraphics[width = 6.9in]{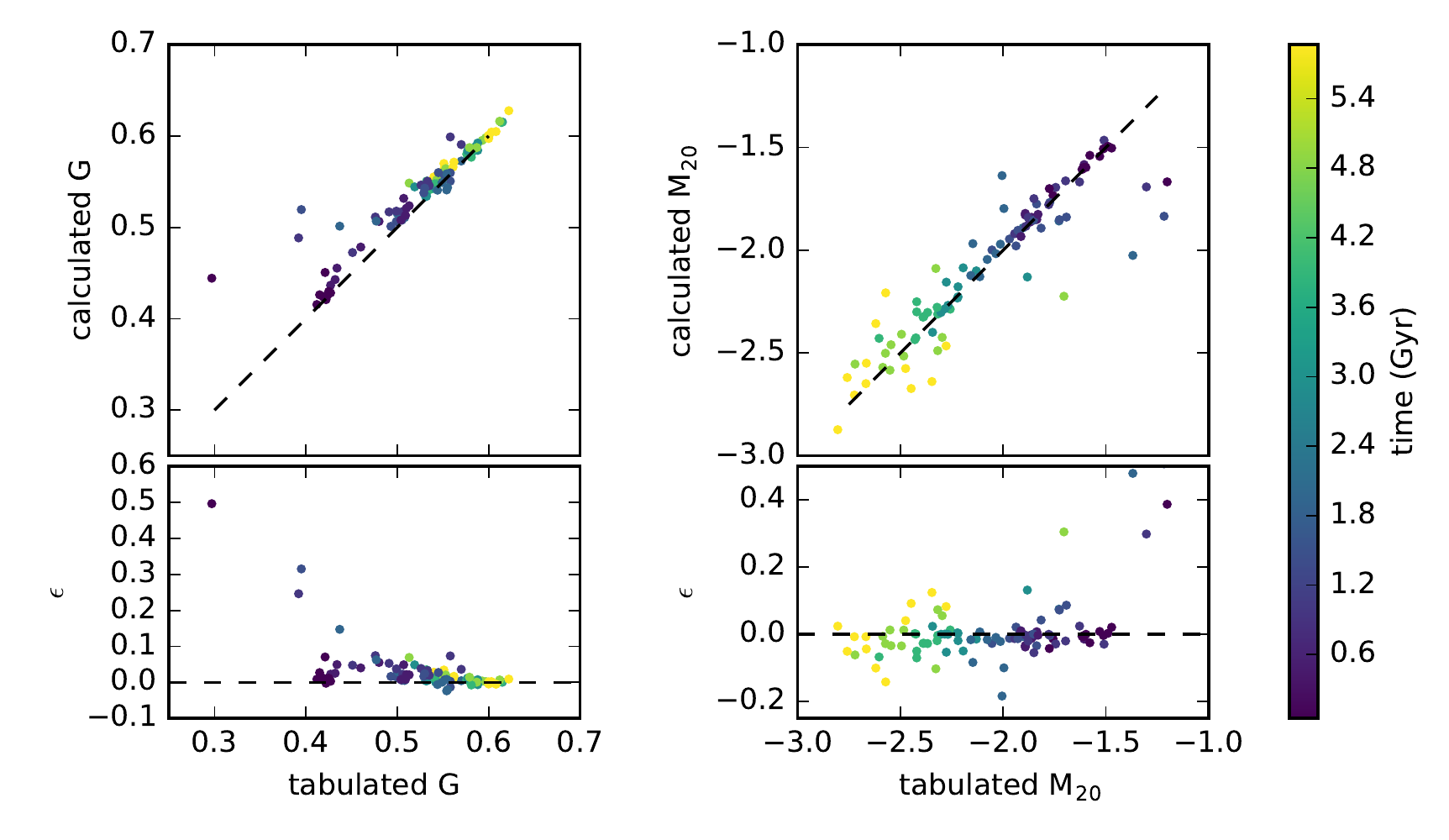}
% Give the caption for the Figure here. 
\caption{\label{figure:lotz_gm20_lm} Comparison of the calculated $G$ 
  and $M_{20}$, using the values of $a_p$ calculated with the algorithm 
  employed in \citet{lotz08b} (Method B), with the tabulated values for 
  the DIGGSS g3iso galaxy. In the lower panels, $\epsilon$ is the 
  relative from the tabulated values. All points have been coloured by 
  the time since the simulation started. }
\end{figure*}

In order to compare the performance of the methods, we used the exact
\citet{lotz08b} methodology, and re-computed the $a_p$, $G$, and
$M_{20}$ for the the d4e idealized simulation and the g3iso simulation
from the DIGGSS simulation series.  As a reminder, the values computed
for the simulations using Method A, are described in appendices
\ref{section:idealized_comparison} and \ref{section:diggss},
respectively. In Figure~\ref{figure:ideal_gm20_avg_lotz_ap} we
illustrate the average $G-M_{20}$ values calculated for the idealized
d4e simulation calculated with Method B. Comparing this to
Figure~\ref{figure:ideal_gm20_avg}, it is evident that Method A
performs slightly better regarding the location of post-merger
remnants in $G-M_{20}$ space.  Similarly, Figures
\ref{figure:lotz_petro_lm} and \ref{figure:lotz_gm20_lm} illustrate
the comparison of the values of $a_p$, $G$, and $M_{20}$, calculated
using Method B, to the tabulated values for the g3iso DIGGSS
simulation\footnote{As an aside, the deviation in
  these values likely arises from differences in the implementation of
  Powell's method for minimization between the \scipy\ function
  fmin\textunderscore powell \citepalias{Jones2001} and the IDL
  procedure POWELL. This difference likely causes slightly different
  centre to be determined while minimizing $A$ and because this centre
  serves as an initial guess for using Powell's method to minimize
  $M_{tot}$, the effect is compounded for the $G$ and $M_{20}$
  values.}. For comparison, figures \ref{figure:lotz_petro} and
\ref{figure:lotz_gm20} illustrate the deviations of the measurements
calculated with Method A from the tabulated measurements. It is
apparent that the measurements made with Method B are moderately
closer to the tabulated values. However, because Method A recovers
more accurate measurements at lower resolutions, we consider its
deviations tolerable.

%We conclude that for the resolution of the mock observations created 
%for this paper, Method A is superior to Method B. We do not change 
%the method of computing $r_p$ from algorithm employed by \citet{lotz08b}, 
%because that algorithm does an excellent job of reproducing the 
%g3iso $r_p$ measurements, and it yields reasonable measurements 
%of $A$ and $C$ for the d4e simulations.

\end{document}